\documentclass[twocolumn]{aastex63}

\usepackage{amsfonts,amsmath,graphicx,natbib,subfigure,color,gensymb,longtable,verbatim,graphicx,float,soul}

\received{\today}
\revised{\today}
\accepted{\today}

\submitjournal{ApJ}

\shorttitle{SN\,2014C}
\shortauthors{Daniel Brethauer}

\graphicspath{{./}{figures/}}

\begin{document}

\title{Seven years of coordinated \emph{Chandra-NuSTAR} observations of SN\,2014C unfold the extreme mass-loss history of its stellar progenitor
}

\author[0000-0001-6415-0903]{D. Brethauer}
\affiliation{Department of Astronomy, University of California, Berkeley, CA 94720-3411, USA}
\affiliation{Center for Interdisciplinary Exploration and Research in Astrophysics (CIERA) and Department of Physics and Astronomy, Northwestern University, Evanston, IL 60208}

\author[0000-0003-4768-7586]{R. Margutti}
\affiliation{Department of Astronomy, University of California, Berkeley, CA 94720-3411, USA}
\affiliation{Department of Physics, University of California, 366 Physics North MC 7300,
Berkeley, CA 94720, USA}

\author[0000-0002-0763-3885]{D. Milisavljevic}
\affiliation{Purdue University, Department of Physics and Astronomy, 525 Northwestern Ave, West Lafayette, IN 47907 }
\affiliation{Integrative Data Science Initiative, Purdue University, West Lafayette, IN 47907, USA}

\author[0000-0002-0592-4152]{Michael F. Bietenholz} 
\affiliation{Department of Physics and Astronomy, York University, 4700 Keele St.,
Toronto, M3J 1P3, Ontario, Canada}

\author[0000-0002-7706-5668]{R. Chornock} 
\affiliation{Department of Astronomy, University of California, Berkeley, CA 94720-3411, USA}

\author[0000-0001-5126-6237]{D.L. Coppejans} 
\affiliation{Department of Physics, University of Warwick, Coventry CV4 7AL, UK}

\author[0000-0002-3137-4633]{Fabio De Colle} 
\affiliation{Instituto de Ciencias Nucleares, Universidad Nacional Autonoma de México, A.P. 70-543 04510, Mexico City, Mexico}

\author[0000-0003-2349-101X ]{Aprajita Hajela} 
\affiliation{Center for Interdisciplinary Exploration and Research in Astrophysics (CIERA) and Department of Physics and Astronomy, Northwestern University, Evanston, IL 60208}

\author[0000-0003-0794-5982]{Giacomo Terreran} 
\affiliation{Las Cumbres Observatory, 6740 Cortona Drive, Suite 102, Goleta, CA 93117-5575, USA}
\affiliation{Department of Physics, University of California, Santa Barbara, CA 93106-9530, USA}

\author[0000-0001-5518-9689]{Felipe Vargas} 
\affiliation{Instituto de Ciencias Nucleares, Universidad Nacional Autonoma de México, A.P. 70-543 04510, Mexico City, Mexico}

\author[0000-0003-4587-2366]{Lindsay DeMarchi} 
\affiliation{Center for Interdisciplinary Exploration and Research in Astrophysics (CIERA) and Department of Physics and Astronomy, Northwestern University, Evanston, IL 60208}

\author{Chelsea Harris} 
\affiliation{Department of Physics and Astronomy, Michigan State University, East Lansing, MI 48824, USA}

\author[0000-0002-3934-2644]{W.~V.~Jacobson-Gal\'{a}n} 
\affiliation{Department of Astronomy, University of California, Berkeley, CA 94720-3411, USA}

\author{Atish Kamble}
\affiliation{Center for Astrophysics | Harvard \& Smithsonian, 60 Garden St, Cambridge, MA 02138}

\author[0000-0002-7507-8115]{Daniel Patnaude}
\affiliation{Center for Astrophysics | Harvard \& Smithsonian, 60 Garden St, Cambridge, MA 02138}

\author[0000-0002-3019-4577]{Michael C. Stroh} 
\affiliation{Center for Interdisciplinary Exploration and Research in Astrophysics (CIERA) and Department of Physics and Astronomy, Northwestern University, Evanston, IL 60208}

\begin{abstract}
We present the results from our seven-year long broad-band X-ray observing campaign of SN\,2014C with \emph{Chandra} and \emph{NuSTAR}. These coordinated observations represent the first look at the evolution of a young extragalactic SN in the 0.3-80 keV energy range in the years after core collapse. We find that the spectroscopic metamorphosis of SN\,2014C from an ordinary type Ib SN into an interacting SN with copious hydrogen emission is accompanied by luminous X-rays reaching $L_x\approx 5.6\times10^{40}\, \rm{erg\,s^{-1}}$ (0.3--100 keV) at $\sim 1000$ days post explosion  and declining as $L_x\propto t^{-1}$ afterwards. The broad-band X-ray spectrum is of thermal origin and shows clear evidence for cooling after peak, with $T(t)\approx 20 \,{\rm keV}(t/t_{\rm pk})^{-0.5}$. Soft X-rays of sub-keV energy suffer from large photoelectric absorption originating from the local SN environment with  $NH_{\rm int}(t)\approx3\times 10^{22}(t/400 \,\rm{days})^{-1.4}\,\rm{cm^{-2}}$. We interpret these findings as the result of the interaction of the SN shock with a dense ($n\approx 10^{5}-10^{6}\,\rm{cm^{-3}}$), H-rich disk-like circumstellar medium (CSM) with inner radius $\sim2\times 10^{16}$ cm and extending to $\sim 10^{17}$ cm. Based on the declining $NH_{\rm int}(t)$ and X-ray luminosity evolution,  we infer a CSM mass of $\sim(1.2\,f$--2.0$\sqrt{f}) \rm{M_{\odot}}$, where $f$ is the volume filling factor. Finally, we place SN\,2014C in the context of 119 core-collapse SNe with evidence for strong shock interaction with a thick circumstellar medium and we highlight the challenges that the current mass-loss theories (including wave-driven mass loss, binary interaction and line-driven winds) face when interpreting the wide dynamic ranges of CSM parameters inferred from observations. 

\end{abstract}

\keywords{supernovae: specific (SN\,2014C)}

\section{Introduction}
\label{Sec:Intro}
Observational studies of evolved massive stars are starting to reveal an eventful history of mass loss as these stars approach core collapse. Observational evidence has been accumulating from a variety of independent lines, including: the direct detection of pre-supernova outbursts from across the mass spectrum of exploding stars (e.g., \citealt{Pastorello07,Pastorello13,Pastorello18,Margutti14,Ofek14,Strotjohann21,Jacobson-Galan22}); bright UV emission in type IIP SNe at early times (e.g., \citealt{Morozova18,Morozova20,Bostroem19,Dessart22}); narrow spectral lines originating from a dense circumstellar medium ionized by the explosion's shock (as in type IIn, type Ibn, and type Icn SNe, e.g., \citealt{Schlegel90,Filippenko97,Pastorello08,Perley22}); luminous X-ray and radio emission powered by efficient conversion of shock kinetic energy into radiation as the SN shock is decelerated in the environment by mass lost by the star before stellar demise (e.g., \citealt{Chevalier06,Soderberg062003bg,Dwarkadas10,Chevalier17,Stroh21}). Here we
present the results from a coordinated  campaign on SN\,2014C with the Chandra X-ray Observatory (\emph{CXO}) and the Nuclear Spectroscopic Telescope Array (\emph{NuSTAR}) during the first seven years after core collapse, and  
update the analysis by \cite{Margutti2017}.
These observations map the evolution of the emission from the interaction of a H-stripped SN shock with a H-rich medium in the soft and hard X-rays for the first time. 

SN\,2014C originally attracted the attention of the SN community because of its highly unusual spectroscopic metamorphosis from an ordinary H-stripped core-collapse SN of type Ib into a SN with clear signs of interaction with an H-rich medium, as  documented in \cite{Milisavljevic15} (see \citealt{Mauerhan18} for a follow-up study). The optical spectroscopic metamorphosis was accompanied by rising luminous radio and X-ray emission \citep{Anderson16,Margutti2017,Brethauer20,Bietenholz20a, Bietenholz21,Thomas22}, as well as luminous infrared emission \citep{Tinyanont16,Tinyanont19}. Despite different modeling assumptions in those papers, and in theoretical investigations such as \cite{Harris20} and \cite{Vargas21}, a concordant picture emerged that associates the SN\,2014C phenomenology with the presence of dense H-rich circumstellar material (CSM) in the immediate vicinity of a H-poor SN. Furthermore, the analysis of archival pre-explosion images pointed at a low-mass star of $M_{\rm{ZAMS}} \approx 11\,\rm{M_{\odot}}$ as progenitor   \citep{Milisavljevic15,Sun20}. 

Clear signs of H-rich CSM interaction have occurred in other stripped envelope explosion types; SNe 2001em \citep{Chugai06}, 2004dk \citep{Mauerhan18}, 2017dio \citep{Kuncarayakti18}, and 2019yvr \citep{Kilpatrick21}, as well as Super Luminous SNe iPTF13ehe, iPTF15esb, iPTF16bad \citep{Yan17}, and 2017ens \citep{Andrews19} all showed a delayed appearance of H$\alpha$ emission. Similarly, late-time radio emission that is believed to be connected to this phenomenology has been observed in SNe 2001em \citep{Schinzel09}, 2003gk \citep{Bietenholz14}, 
2004C \citep{DeMarchi22}, 2007bg \citep{Salas13}, PTF11qcj \citep{Corsi14}, VT J121001+495647 \citep{Dong21}, and others identified in \cite{Stroh21}. 

Several explanations have been proposed to explain these observations, ranging from Luminous Blue Variable-like eruptions \citep[e.g.,][]{Smith06,Smith14}, to stellar H-envelope ejection as a result of  binary interaction \citep[e.g.,][]{Podsiadlowski92}, to internal gravity wave driven mass ejections proposed in \cite{Quataert12}, to nuclear burning instabilities \citep{Smith14b}. By assembling a large sample of 119 core-collapse SNe with signatures of interaction with a thick CSM, we show how each of these mechanisms might naturally explain only a portion of the mass-loss parameter space of evolved massive stars.

The paper is organized as follows. In \S\ref{Sec:Obs}, we present the analysis of our broad-band X-ray campaign of SN\,2014C with the \emph{CXO} and \emph{NuSTAR}. In \S\ref{Sec:Inf} we model these observations and derive our inferences on the SN shock and circumstellar environment. We discuss our findings within the broader context of a large sample of 119 core-collapse SNe with observational signatures of interaction with a dense medium in \S\ref{Sec:Disc}. Finally, we discuss  different mass-loss scenarios, including line-driven winds (\ref{SubSubSec:windwind}), interaction with a stellar companion (\ref{SubSubSec:binary}) and internal gravity-wave driven mass loss (\ref{SubSubSec:nuclear}). Conclusions are drawn in \S\ref{Sec:Conc}. In Appendix \ref{Appendix} we provide a detailed accounting of each individual interacting SN in our sample, including classification, modeled properties from other literature, and our methods for filling in information that was not explicitly provided.

Following \cite{Freedman01} we adopt a distance of 14.7 $\pm 0.6$ Mpc for the host galaxy of SN\,2014C (NGC 7331). Times are referred to the time of first light from \cite{Margutti2017}, which is December 30th of 2013 $\pm$ 1 day (corresponding to MJD 56656 $\pm 1$).   Uncertainties are quoted at 1$\sigma$ confidence level and upper limits at the 3$\sigma$ confidence level unless stated otherwise.

\begin{figure*}
    \centering
  \includegraphics[width=0.97\textwidth]{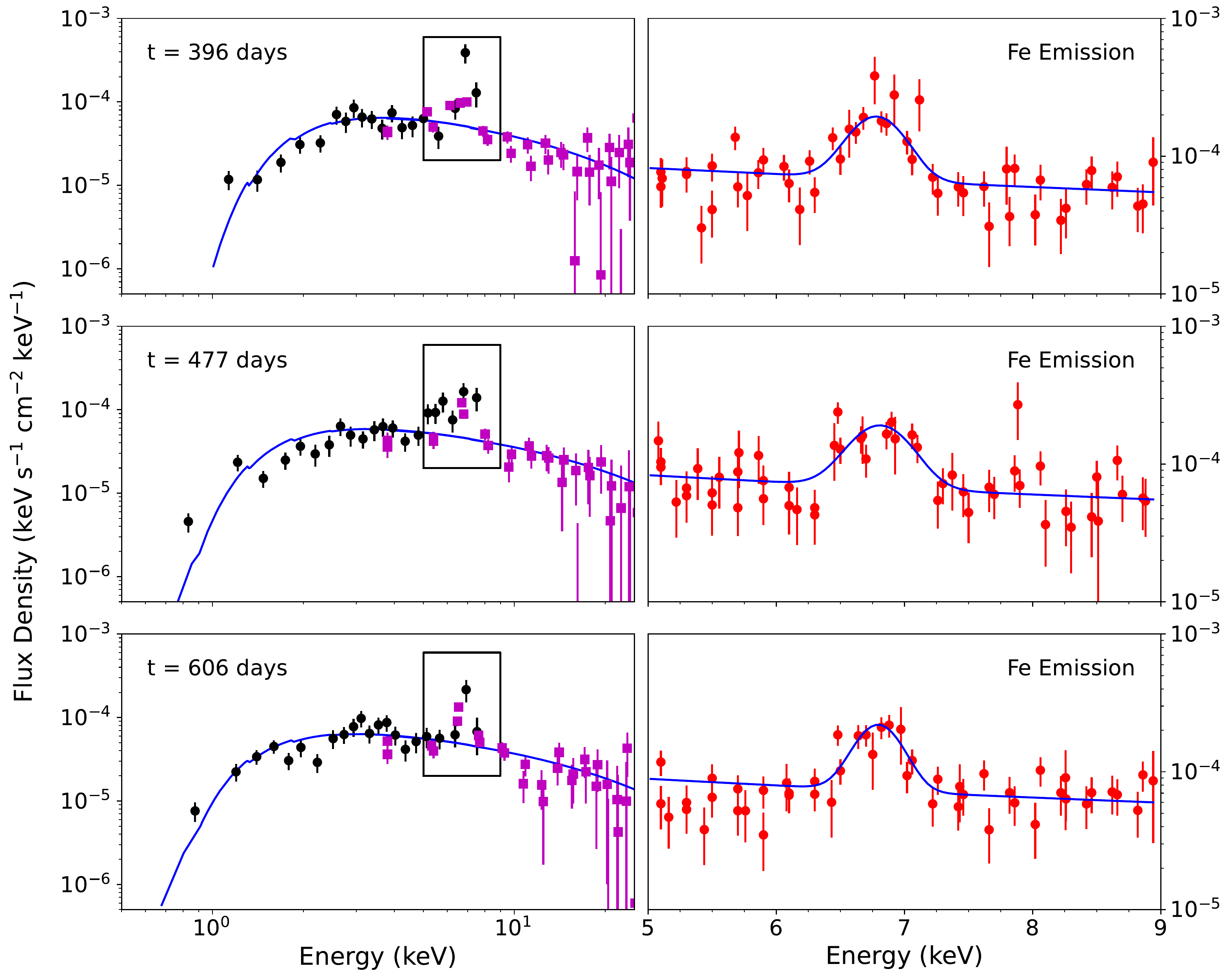}
    \caption{\emph{Left Panels:} Broadband (0.5-40 keV) X-ray  spectrum of SN\,2014C acquired with coordinated \emph{CXO} (black filled circles) and \emph{NuSTAR} (magenta squares) observations at $\delta t=396,477, \rm{and}\, 606$ days since explosion. In each panel a thick blue line marks the best fitting absorbed bremsstrahlung model with best-fitting spectral parameters reported in Appendix \ref{Appendix:Tables}, Table \ref{Tab:Xrayjoint} and shown in Fig.\ \ref{Fig:NH} and Fig.\ \ref{Fig:Temperature}.
     The empty black boxes identify the spectral location of an excess of emission with respect to this model.  Following \cite{Margutti2017}, we associate the excess of emission around $6.7$ keV (black boxes) to a K$\alpha$ line transition in H-like or He-like Fe atoms. 
     \emph{Right Panels}: Zoom-in on the Fe-line region, here modeled with a local continuum plus Gaussian line model. The best-fitting parameters are reported in Appendix \ref{Appendix:Tables}, Table \ref{Tab:Iron} and the observed flux evolution of the Fe emission is shown in Fig.\ \ref{Fig:IronFlux}. Data have been rebinned for graphical purposes, but model fitting has been done on unbinned data.}
    \label{Fig:Bremss1}
\end{figure*}

\begin{figure*}[t!]
    \centering
  \includegraphics[width=0.97\textwidth]{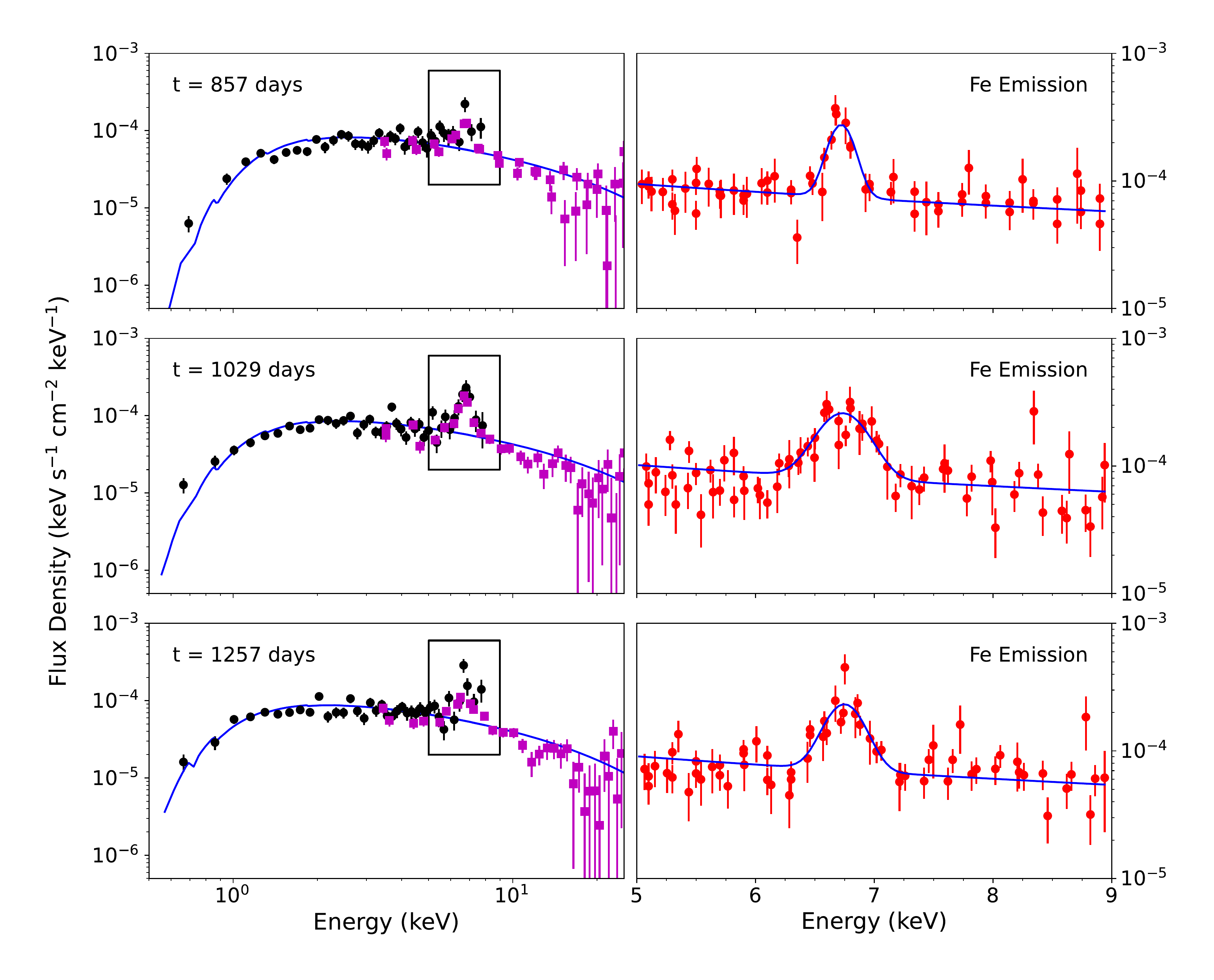}
     \caption{Broad-band X-ray spectral evolution of SN\,2014C as captured by \emph{CXO} and \emph{NuSTAR} observations at $\delta$t = 857, 1029, and 1257 days. These panels follow the same color scheme as in Fig.\ \ref{Fig:Bremss1}.}
    \label{Fig:Bremss2}
\end{figure*}

\begin{figure*}
    \centering
  \includegraphics[width=0.97\textwidth]{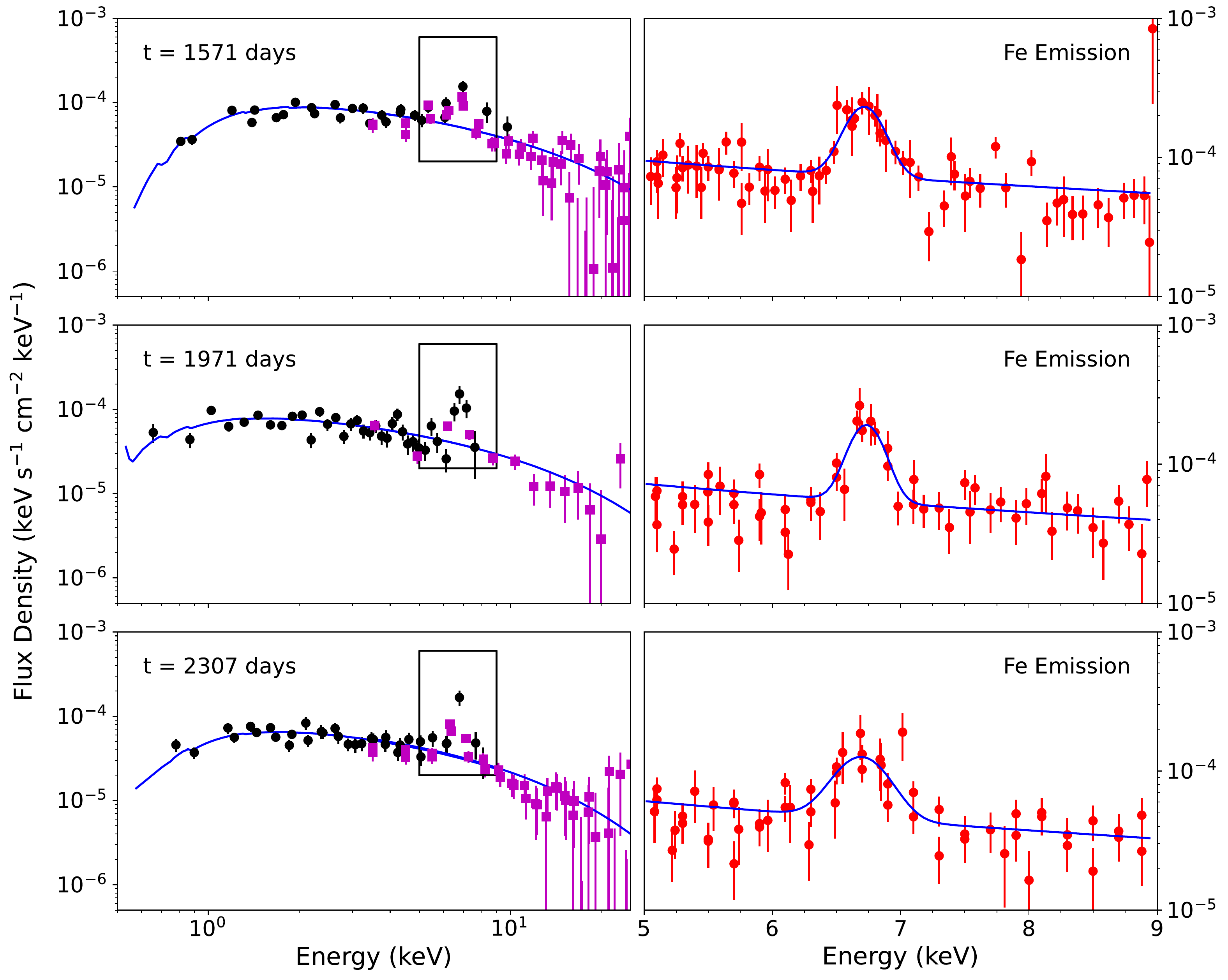}
    \caption{Broad-band X-ray spectral evolution of SN\,2014C as captured by \emph{CXO} and \emph{NuSTAR} observations at $\delta$t = 1571, 1971, and 2307 days. These panels follow the same color scheme as in Fig.\ \ref{Fig:Bremss1}.}
    \label{Fig:Bremss3}
\end{figure*}

\section{Soft and hard X-ray Observations}
\label{Sec:Obs}
We present a homogeneous analysis of our entire broadband X-ray campaign of SN\,2014C for the first $\sim 2307\,\rm{days}$ of the evolution, using the \emph{CXO} and \emph{NuSTAR}. For consistency, we also re-analyze the X-ray data acquired at $\delta t < 500\,\rm{days}$, which we originally published in \cite{Margutti2017}. While the present paper was in an advanced stage of preparation, the paper by \cite{Thomas22} appeared on the arXiv presenting the data from our extensive soft and hard-X-ray campaign of SN\,2014C. In the following we comment on the differences between the two independent analyses. 
This data set offers the unprecedented opportunity to study the evolution of the hard X-ray emission from an extra-galactic SN over a baseline of seven years.

\subsection{Soft X-ray Observations with the CXO}
Prior to the explosion, the field of SN\,2014C was  observed by the \emph{CXO} on 2001 January 27 (PI Zezas, ID 2198) for a total exposure time of $\sim$29.5 ks. No X-ray emission was detected at the location of SN\,2014C with a 3$\sigma$ absorbed flux limit of $< 2.8 \times 10^{-15}$ erg s$^{-1}$ cm$^{-2}$ (0.3-10 keV). We refer the reader to \S 2.4 of \cite{Margutti2017} for more details on \emph{CXO} pre-explosion observations.

We started monitoring SN\,2014C with the \emph{CXO} on 2014 November 3rd ($\delta t = 308$ days since explosion, PI Soderberg, ID 16005). Further \emph{CXO} observations were acquired starting from 2015 January 30th to 2020 April 18th, $\approx$1\,year  to $\approx$6\, years post-explosion, with total exposure time of $\sim$242\,ks (Table \ref{Tab:XrayCXOInfo}, PI  Margutti).

We reduced the \emph{CXO} data with the software package \texttt{CIAO} v. 4.12 by applying standard ACIS filtering criteria and using the latest calibration database (\texttt{CALDB v. 20190813}). We used the \texttt{wavdetect} task within \texttt{CIAO} to perform blind point-source detection. A bright X-ray source is blindly detected with high confidence ($> 40\sigma)$ at the location of SN\,2014C throughout these observations. The corresponding significance of detection and count-rates are reported in Appendix \ref{Appendix:Tables},  Table \ref{Tab:XrayCXOInfo}. We note that the observed 0.5-8 keV count-rate increases until $\delta t\approx1258$ days as a result of the combination of the larger intrinsic luminosity of the source and the smaller intrinsic absorption with time.   

For each observation, we extracted a spectrum from a 1.5\arcsec\, radius region around the source using \texttt{specextract} within \texttt{CIAO}. We extracted the background spectrum from a large source-free region away from chip edges of $\sim$ 20\arcsec. 
We fit the spectrum from each \emph{CXO} observation independently with the exception of IDs 18343+21077 and 21640+23216. Each pair of observation IDs were acquired very close in time ($\Delta t/t<10^{-3}$) and can be thus considered effectively part of the same epoch of observations. For this first round of spectral modeling we used an absorbed power-law model \texttt{tbabs*ztbabs*pow} in \texttt{XSPEC}. We adopted a Galactic neutral hydrogen column density in the direction of SN\,2014C as $NH_{\rm{MW}} = 6.14 \times 10^{20}\,\rm{cm^{-2}}$ \citep{Kalberla05}. We employed Cash statistics to constrain the spectral parameters and their uncertainties. The inferred best fitting parameters as well as absorbed and unabsorbed fluxes are reported in Appendix \ref{Appendix:Tables}, Table \ref{Tab:XrayCXOFit}. Our measurements are consistent with the X-ray analysis presented in \cite{Jin19}.

\subsection{Hard X-ray Observations with NuSTAR}
\label{SubSec:NuSTAR}
We acquired hard X-ray (3--79 keV) observations of SN\,2014C with \emph{NuSTAR} starting on 2015 January 29 ($\delta t$  $\sim$ 400 days) through 2020 April 30 ($\delta t$  $\sim$ 2300 days, PI Margutti), for a total  exposure of $\sim$ 347 ks. Since the emission from the background largely dominates the spectrum above $\sim$ 40 keV, in the following we focus our analysis on the 3--40 keV spectral window. We extracted and cleaned the event files with \texttt{nupipeline} and \texttt{nuproducts} within  \texttt{NuSTARDAS} (v.1.9.5) using standard filtering criteria and the latest files available in the \emph{NuSTAR} calibration database (\texttt{CALDB v.20190607}). Specifically, for the data screening and filtering step we follow the updated prescriptions of \texttt{NuSTARDAS} (v.1.9.5).

With these parameters, we find that a source of hard X-ray emission is blindly detected at the location of SN\,2014C with high significance ($>15\sigma$ c.l.) throughout the entire duration of our monitoring campaign. Importantly, these observations establish SN\,2014C as the first SN with well-monitored bright hard X-ray emission over its first $\sim$ 7 years of evolution post explosion.

For each observation, we extracted a source spectrum from a region of 1$\arcmin$\, radius around the position of SN\,2014C. For the background spectrum we used a source-free annulus region centered at the source location with an inner and outer radius of $1.1 \arcmin$\, and 3$\arcmin$\,, respectively. 
\emph{NuSTAR} observations complement the \emph{CXO} observations at higher X-ray energies. In \S \ref{SubSec:Joint} we exploit the high angular resolution of coordinated \emph{CXO} observations to model and account for the partial contamination of the low-energy \emph{NuSTAR} data by unrelated sources in the host galaxy of SN\,2014C, which results from the larger \emph{NuSTAR} Point Spread Function (PSF, with Full Width at Half Maximum (FWHM) of $\sim$18\arcsec).

\subsection{Joint spectral analysis of \emph{CXO} and NuSTAR data}
\label{SubSec:Joint}
In this section we perform broadband X-ray spectral modeling in the 0.3--40 keV energy range using observations from both \emph{CXO} and \emph{NuSTAR}.  

We performed a joint spectral fit of each of the epochs for which we have coordinated  \emph{CXO} and \emph{NuSTAR} observations (Appendix \ref{Appendix:Tables} Table \ref{Tab:Xrayjoint}). Following  \cite{Margutti2017}, we use an absorbed thermal bremsstrahlung spectral model (\texttt{tbabs*ztbabs*bremss}) for the SN emission.\footnote{Note that the Fe line emission and the underlying continuum do not necessarily originate from the same emitting region, and we thus avoid using the \texttt{vapec} model of collisionally-ionized diffuse gas adopted by \citealt{Thomas22}.} While the fine angular resolution of the \emph{CXO} allows for the \emph{CXO} spectrum to be entirely dominated by the SN emission, the more extended \emph{NuSTAR} PSF includes important contributions  from the host galaxy emission that is unrelated to the SN. We account for the presence of contamination in the \emph{NuSTAR} spectrum by adding an absorbed power-law (\texttt{tbabs*pow}) spectral component to model the \emph{NuSTAR} data.

We initially tie the \texttt{bremss} model parameters of \emph{CXO} and \emph{NuSTAR} observations and fit for the intrinsic neutral hydrogen column density ($NH_{\rm{int}}$), plasma temperature ($T$) and the emission measure ($EM$) associated to the \texttt{bremss} model and the photon index ($\Gamma$) of the power-law component. The best-fitting power-law model obtained in this way quantifies the contribution of the contamination by the host-galaxy emission to the \emph{NuSTAR} spectrum. We then freeze the power-law model parameters to the best-fitting values obtained in the previous step, untie the hard and soft X-ray $EM$s and perform a final joint-fit to constrain the spectral parameters associated with the bremsstrahlung emission. Even with the bremsstrahlung normalization constants untied between \emph{CXO} and \emph{NuSTAR}, they were consistently well within 10\% of each other. In this way we account for potential \emph{CXO}-\emph{NuSTAR} intercalibration uncertanties. 

\begin{figure}
    \centering
  \includegraphics[width=0.49\textwidth]{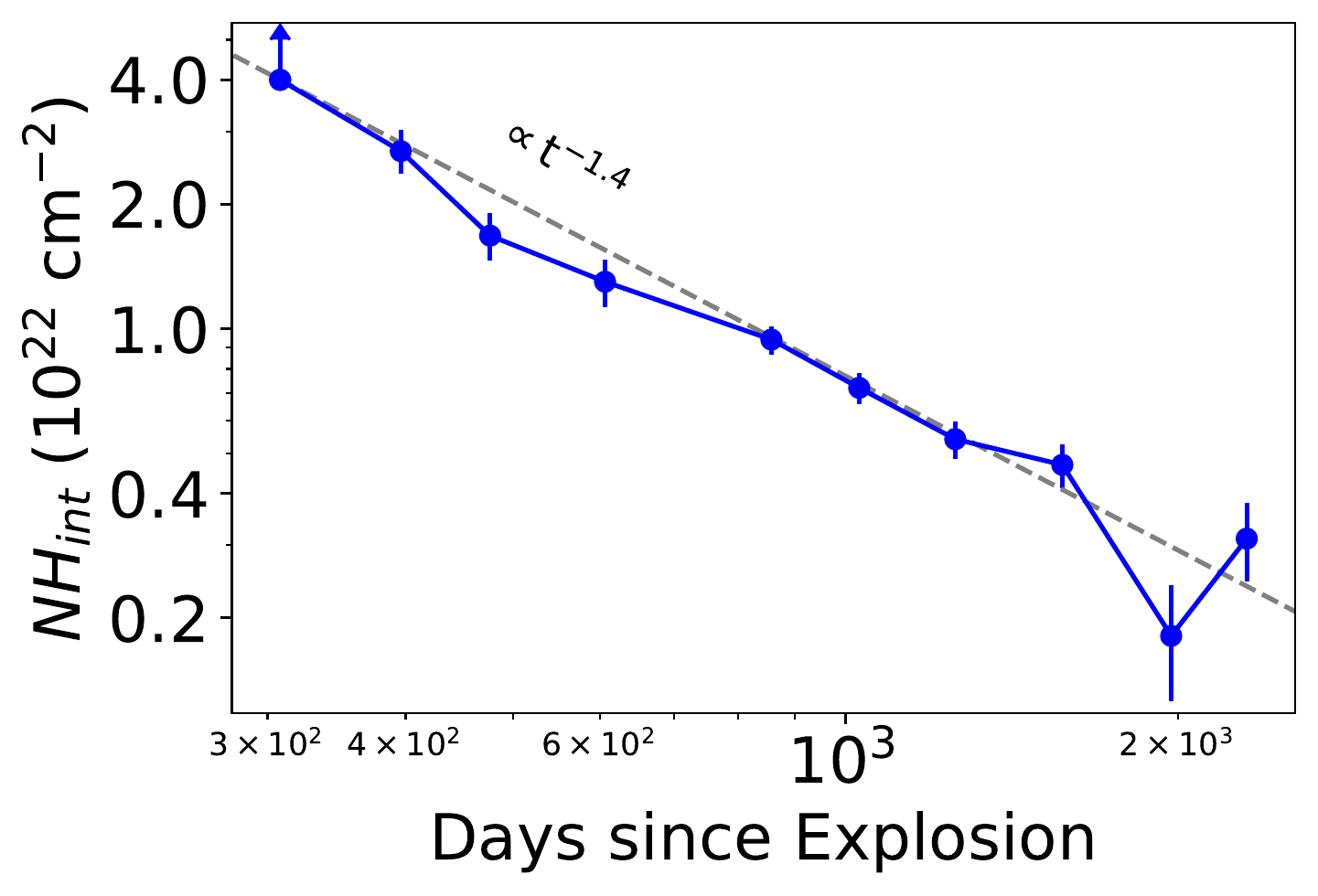}
    \caption{Evolution of intrinsic neutral hydrogen column density with time as revealed by our broad-band \emph{CXO}+\emph{NuSTAR} spectral modeling. The first epoch lacks \emph{NuSTAR} coverage, which leads to a lower limit on the $NH_{\rm{int}}\gtrsim 4\times 10^{22}\,\rm{cm^{-2}}$ \citep{Margutti2017}. A grey dashed line marks a $t^{-1.4}$ evolution to guide the eye.}
    \label{Fig:NH}
\end{figure}

\begin{figure}[t!]
    \centering
  \includegraphics[width=0.49\textwidth]{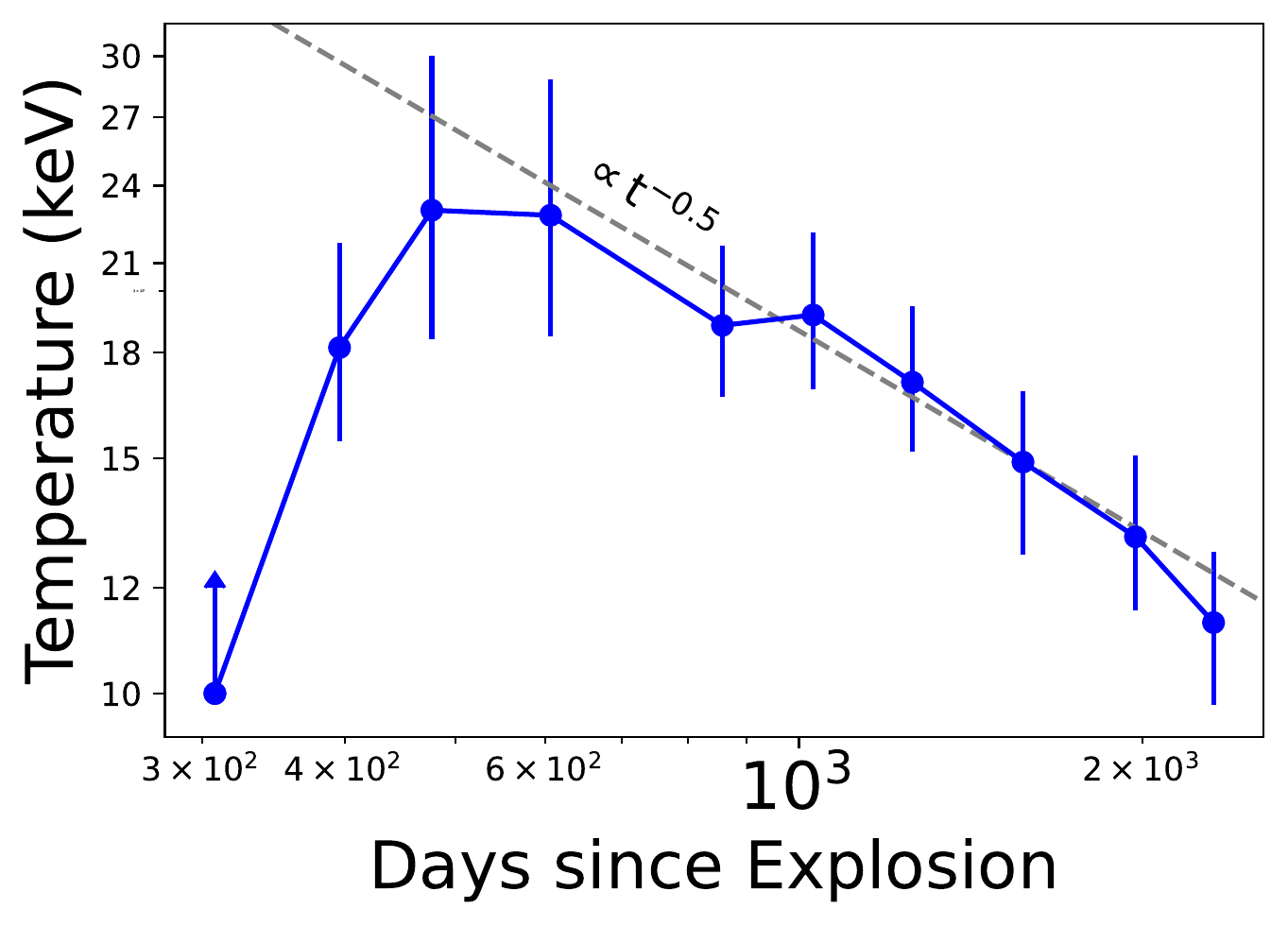}
    \caption{Temperature evolution with time as revealed by our broad-band X-ray modeling. From the first \emph{CXO} observation \cite{Margutti2017} inferred $T>10$ keV, which we plot here. The dashed grey line marks a $t^{-0.5}$ power-law decay to guide the eye.}
    \label{Fig:Temperature}
\end{figure}

\begin{figure}
    \centering
  \includegraphics[width=0.49\textwidth]{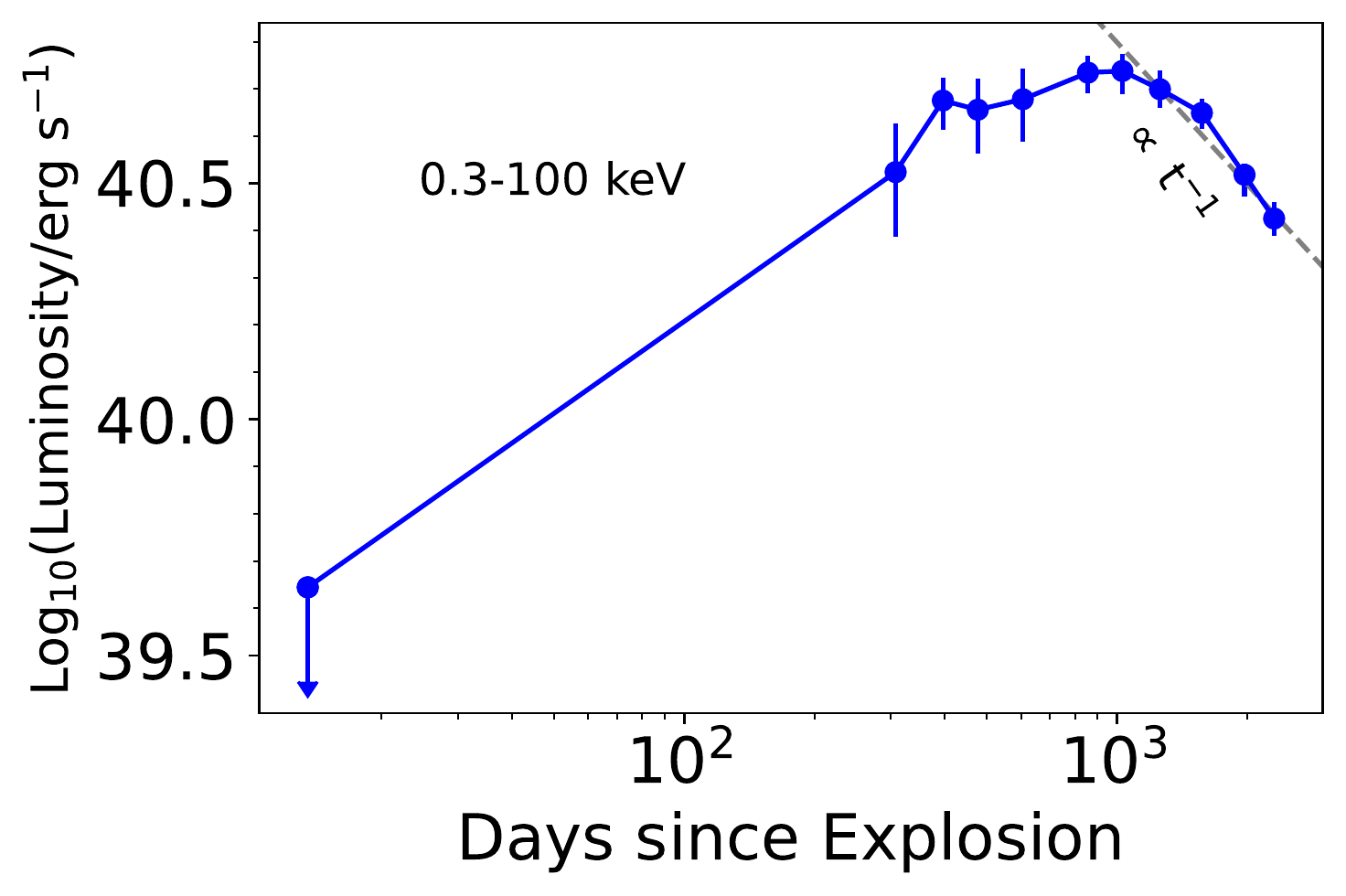}
    \caption{Evolution of 0.3--100 keV X-ray luminosity with  time as revealed by our broad-band spectral modeling. Very early \emph{Swift}-XRT observations acquired at $\delta t$ = 7--20 days led to $L_x<4.41 \times 10^{39}\,\rm{erg\,s^{-1}}$ \citep{Margutti2017}. A grey dashed line marks a $\propto t^{-1}$ decay to guide the eye. }
    \label{Fig:Lum2}
\end{figure}

Table \ref{Tab:Xrayjoint} in Appendix \ref{Appendix:Tables} reports the inferred best-fitting parameters values and the derived absorbed and unabsorbed fluxes in the 0.3--100 keV range, plasma temperature, and intrinsic neutral hydrogen column density from the broadband X-ray spectral fitting. The best-fitting spectra of SN\,2014C are portrayed in Figures \ref{Fig:Bremss1}, \ref{Fig:Bremss2}, and \ref{Fig:Bremss3}. The evolution of the $NH_{\rm{int}}$, the plasma temperature $T$ and the 0.3--100 keV X-ray luminosity are shown in Figures \ref{Fig:NH}, \ref{Fig:Temperature}, and \ref{Fig:Lum2}, respectively. Our findings in the first 500 days of evolution are consistent with the results presented by \cite{Margutti2017}, and broadly similar to \cite{Thomas22}. However,  \cite{Thomas22} did not account for the contamination of the \emph{NuSTAR} PSF by unrelated sources. We find that this leads to a systematic underestimate of the plasma temperature $T$.  We can reproduce the $T\approx$10 keV values reported by \cite{Thomas22} by setting to zero the normalization of the power-law spectral component that quantifies the contamination by other X-ray sources and by including the \emph{NuSTAR} data above 50 keV in the fit, which are background dominated.

From our analysis we find that $NH_{\rm{int}}(t)$ shows a monotonic decline with
$NH_{\rm{int}}(t)\propto t^{-1.4}$ from $\sim$ 2.7 $\times 10^{22}$ cm$^{-2}$ at $\delta$t $\sim$ 396 days to $\sim$ 3 $\times 10^{21}$ cm$^{-2}$ at $\delta$t $\sim$ 2307 days, which suggests that the $NH_{\rm{int}}$ is completely dominated by material outside of the Milky Way and local to the SN explosions at all times (Figure \ref{Fig:NH}). 
The temperature peaks at $T\sim $ 23 keV at $\delta t\approx 500-600$ days, to later decay as $\propto t^{-0.5}$ until the end of our monitoring at $\delta t\approx$ 2300 days (Figure \ref{Fig:Temperature}). The resulting 0.3--100 keV luminosity inferred from the unabsorbed fluxes reaches $L_x\sim 5.5\times 10^{40}\,\rm{erg s^{-1}}$ at  $\sim 1000$ days. The post-peak decline is currently well described by a $\propto t^{-1}$ decay  (Figure \ref{Fig:Lum2}). The physical implications of these observational findings are discussed in detail in the following section.

Finally, we note the presence of an excess of emission with respect to the thermal bremsstrahlung model in the energy interval $\sim$6.5--7.1 keV in our broadband X-ray modeling (right-side panels of Figures \ref{Fig:Bremss1}, \ref{Fig:Bremss2}, and \ref{Fig:Bremss3}). We fit the emission in the energy range 5-9 keV with a Gaussian model in addition to the inferred best-fit bremsstrahlung and power-law models. As noted in \cite{Margutti2017}, we associate this emission with H- and He-like Fe atom transitions, in particular with the resulting K$\alpha$ emission line. Table \ref{Tab:Iron} reports the best-fitting values and the inferred unabsorbed flux in the 6.5--7.1 keV energy range of the Gaussian component. The best-fitting models are shown in the right-side panels of Figures \ref{Fig:Bremss1}, \ref{Fig:Bremss2}, and \ref{Fig:Bremss3}. Unfortunately, the data  lack the spectral resolution to further constrain the potentially complex line emission from highly ionized iron that might be expected \citep{Mewe85, Mewe86,Liedahl95}.

\begin{figure}
    \centering
  \includegraphics[width=0.49\textwidth]{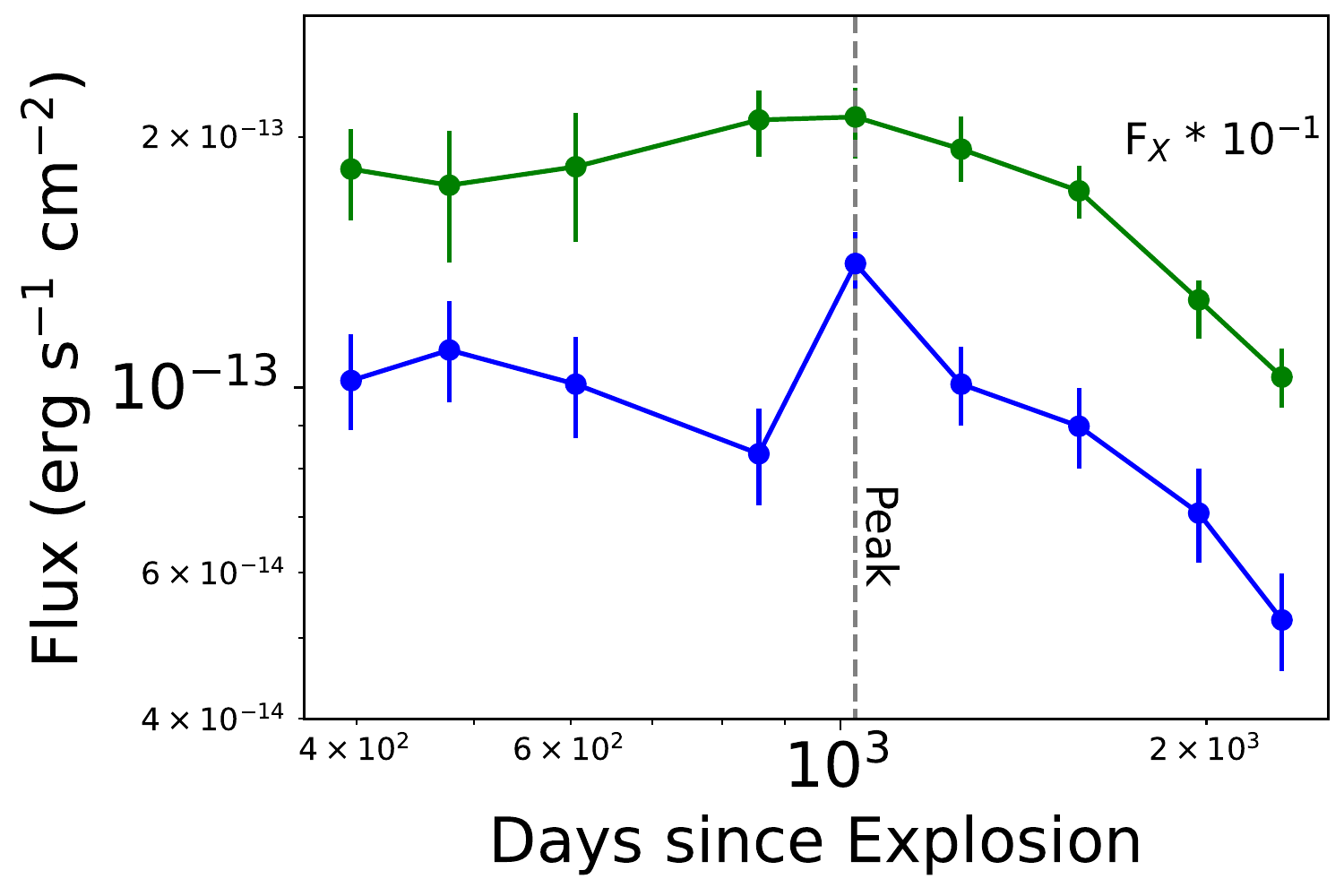}
    \caption{Evolution of the Fe line emission (blue) and total emission (green, rescaled by 10$^{-1}$) with time. The Fe line emission is consistent with no change until $\delta t\approx 850$ days. The line emission starts to decay at $t> 1000$ days  along with the X-ray continuum of Figure \ref{Fig:Lum2}. The equivalent width of the Fe line emission remains roughly constant with time, suggesting a physical connection between the Fe line emission and the continuum emission.   
    }
    \label{Fig:IronFlux}
\end{figure}

\section{Inferences on the Shock and Environment Properties}
\label{Sec:Inf}
In this section we derive inferences on the shock and environment physical parameters using the constraints on the temperature $T(t)$, intrinsic neutral hydrogen absorption $NH_{\rm{int}}(t)$, and Emission Measure $EM(t)$:
\footnote{ 
Within \emph{XSPEC} the normalization of the bremsstrahlung spectrum $C$ is defined as with all units in cgs: 
 \begin{equation}\label{Eq:bremssnorm}  C = 
 \frac{3.02 \times 10^{-15}}{4 \pi \left(\frac{d_L}{\rm{cm}}\right)^{2}} 
 \int{
 \frac{n_{i}}{\rm{cm}^{-3}} 
 \frac{n_{e}}{\rm{cm}^{-3}} 
 \frac{dV}{\rm{cm}^{-3}}} 
 \end{equation} 
 where $d_L$ is the luminosity distance, and the integral is the $EM$. 
}

\begin{equation}\label{Eq:EM}
EM \equiv \int{n_{I} n_{e} dV}\approx n_{I} n_{e} V_{\rm FS}= \frac{\rho_{\rm FS}^2}{\mu_e \mu_I m_p^2} V_{\rm FS}
\end{equation}

where $n_{I}$ and $n_{e}$ (assumed constant over the emitting volume) are the ion and electron number densities in the shocked region, respectively, and the integral is over the emitting volume $V_{\rm FS}$ (i.e., the shocked region).  Following \cite{Margutti2017}, we identify the emitting region with the shocked CSM, i.e., the region that has been shocked by the forward shock, based on the similar velocity inferred from the H$\alpha$ line and the one inferred from the X-ray modeling. We do not repeat here the argument and we refer the reader to \cite{Margutti2017} for details.  $\rho_{\rm FS}$ is the shocked CSM matter density, while  $\mu_e$ and $\mu_I$ are the electron mean molecular weight and the ion mean molecular weight, respectively, and they reflect the chemical composition and ionization state of the CSM. For the Solar-like composition and full ionization that we assume for the CSM, $\mu_e\approx1.25$ and $\mu_I\approx1.15$.

\subsection{General Considerations}
\label{SubSec:Considerations}
We start with three considerations. First, the temporal decay of the temperature of emission at $\delta t \gtrsim 500$ days (Fig.\ \ref{Fig:Temperature}) is a likely indication that the shock front has broken out from the densest part of the  shell and is moving through additional, albeit much less dense, material. 
From \cite{Fransson96}, while the shock is strongly interacting with the surrounding medium, the post-shock temperature  is related to the velocity of the shockwave $v$ by the equation
\begin{equation}
    T \approx 2.27 \times 10^{9} \mu v_4^2 \,\rm{K}
\label{Eq:TemptoVel}
\end{equation}
where $T$ is the temperature of the shocked region, $\mu$ is the mean molecular weight of the shocked medium ($1/\mu\equiv 1/\mu_e +1/\mu_I$), and $v_4\equiv v/(10^4 \rm{km\; s^{-1}})$ is the shock velocity. At peak we measure $T\approx 23$ keV (Figure \ref{Fig:Temperature}), which implies $v\approx$ 4400 km s$^{-1}$ for a CSM with completely ionized, Solar-like composition (i.e., $\mu_{\rm FS}= 0.61$).

Second, further supporting evidence for the shockwave emerging from a dense shell is the decrease in $NH_{\rm{int}}$ with time (Figure \ref{Fig:NH}). $NH_{\rm{int}}$ traces the amount of \emph{neutral} material between the emitting region and the observer, therefore providing a lower limit on the total amount of material as we expect a fraction of the material to be ionized and hence transparent to X-rays. $NH_{\rm{int}}$ has two main components: a local contribution from the immediate environment of SN\,2014C (i.e., the dense shell), and a component from material along the line of sight in the SN host galaxy, NGC 7331. The rapid and dramatic decline of $NH_{\rm{int}}(t)$ indicates that it is dominated by material local to SN\,2014C (i.e., the unrelated host-galaxy component is subdominant). 
Importantly, the decrease in $NH_{\rm{int}}(t)$ is indicative that there is progressively less material in front of the shockwave, which can be interpreted as evidence that the shockwave is moving through the CSM shell. 

Third, the post-peak broad-band X-ray luminosity evolution suggests a decline less steep than  the $\propto t^{-2}$ that is expected in the case of purely adiabatic expansion (e.g., \citealt{Margalit22}).  
These inferences are consistent with the
the shockwave moving through lower-density material after breaking out of the thick CSM shell. An intriguing possibility is that the shock at $\delta t>1000$ days is interacting with mass lost via winds that belonged to the progenitor's evolutionary phase that preceded the shell ejection. Alternatively, we are sampling the CSM structure developed from the interaction of winds by the stellar progenitor in two different evolutionary phases. Both options will be
further discussed in \S \ref{Sec:Disc}.

Finally, the presence of  strong and persistent iron lines in the spectra (right side of Figures \ref{Fig:Bremss1}, \ref{Fig:Bremss2}, \ref{Fig:Bremss3}) indicates that the environment is likely to be clumpy, as was inferred for other strongly interacting SNe, e.g., SNe 1996cr \citep{Dwarkadas10,Dewey11}; 2006jd \citep{Chandra12b}; 2009ip \citep{Margutti14}. The luminosity of iron emission can be a result of either a metallicity of $\sim$ 5\,Z$_\odot$ 
or a clumpy medium \citep{Milisavljevic15,Margutti2017}, since 
the shock is more efficiently decelerated within the higher density clumps, which leads to a lower emission $T$ (e.g., Equation \ref{Eq:EM}) that allows more prominent iron lines, as seen in SN 1993J \citep{Fransson96}. 
Based on their numerical modeling of the broad-band spectra of SN\,2014C, \cite{Vargas21} find that in order to generate the observed overall continuum of absorbed thermal bremsstrahlung, their model requires a metallicity of approximately $\sim0.5\,$Z$_\odot$, further supporting the idea of a clumpy medium.

\begin{figure*}
    \centering
  \includegraphics[width=0.85\textwidth]{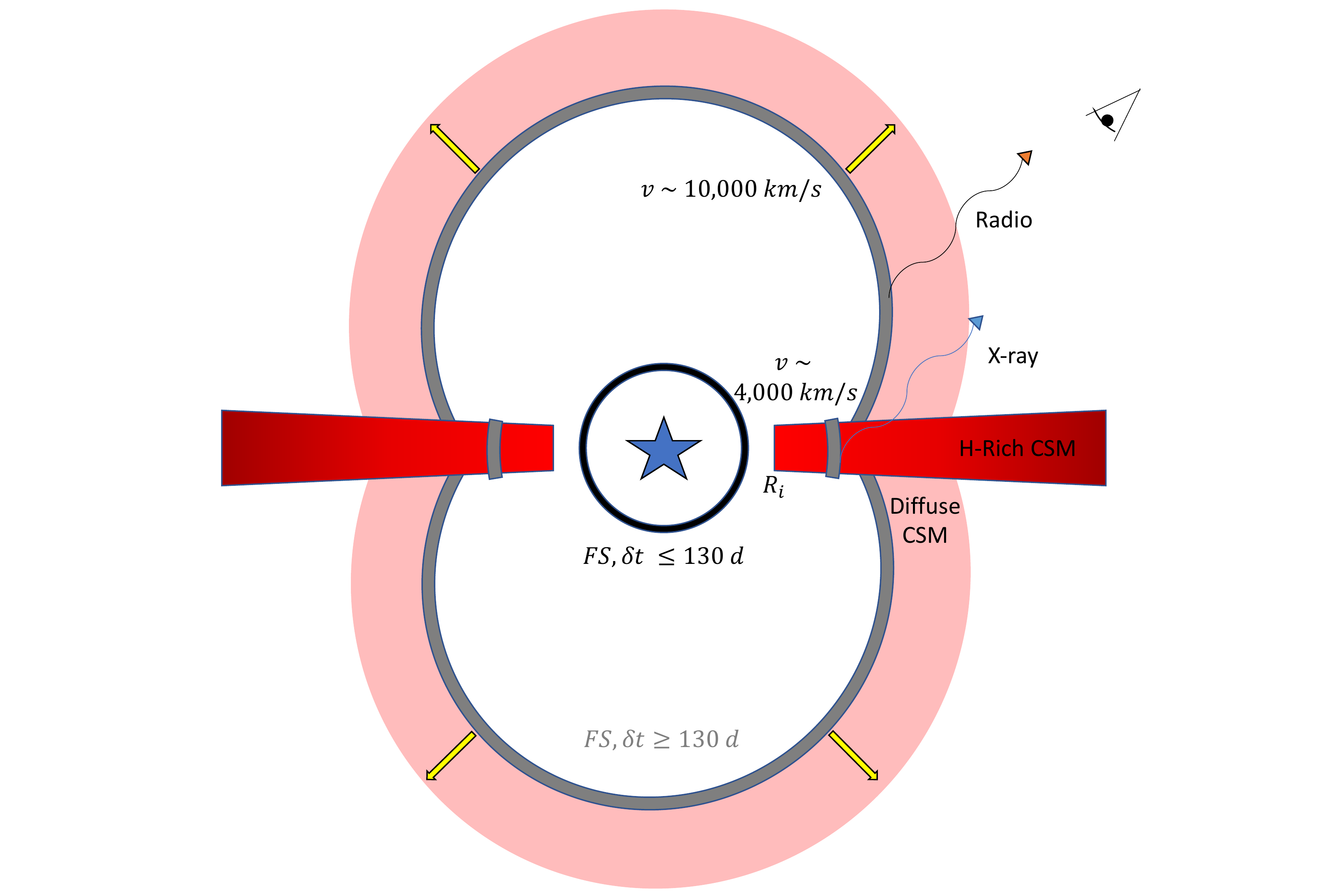}
    \caption{Side view cartoon depiction of the geometry around SN 2014C. The CSM is contained within the equatorial plane, beginning at a radius $R_i \sim 2.2\times10^{16}$ cm, which the FS achieved within 130 days post explosion. Prior to 130 days, the FS was mostly spherical. Upon striking the equatorial CSM, that portion of the FS slowed down from $\sim 10^4$ km s$^{-1}$ to $\sim 4\times10^3$ km s$^{-1}$. This interaction of FS and CSM generates the observed X-rays, while the FS that was only slightly decelerated by a significantly lower density medium at $\sim10^4$ km s$^{-1}$ generates the observed radio emission. 
    As the FS is only impeded in the equatorial plane, the radio emission creates an hourglass shape, which is consistent with the spherical symmetry found by \cite{Bietenholz21}. A similar geometry has been invoked by \cite{Milisavljevic15} and, more recently, by \cite{Thomas22} for the specific case of SN\,2014C and for other SNe as well (see e.g., \citealt{Andrews18,Smith15,Brennan21a}) .
    }
    \label{Fig:Cartoon}
\end{figure*}

\subsection{Constraints on the shock dynamics and geometry of the emitting regions}
\label{SubSec:Constrain}

The forward shock (FS) radius $R(t)$ evolution is constrained by both Eq.\ \ref{Eq:TemptoVel} (from X-ray data) and by Very Long Baseline Interferometry (VLBI) radio observations \citep{Bietenholz21}, which provides a direct measurement of the size of the radio emitting region. The two inferences are expected to agree with each other \emph{if} the X-ray and radio emitting region are the same. From \cite{Bietenholz21}, the best-fitting model of  the radio-emitting forward shock radius evolution with time reads: 
\begin{equation}
    R_{\rm FS}(t) = (6.27 \pm 0.22)\Big (\frac{t}{\rm{yr}}\Big )^{0.77 \pm 0.03} \Big(\frac{d_A}{15.1\, \rm{Mpc}}\Big ) 10^{16} \rm{cm}
    \label{Eq:BietRadii}
\end{equation}
\noindent
where $d_A$ is the angular diameter distance, which we use 14.7 Mpc \citep{Freedman01}. The time derivative of Equation \ref{Eq:BietRadii} implies a FS velocity $v_{\rm FS}\gtrsim$ 10000 km s$^{-1}$, at the same time we derive a  $v_{\rm FS}\approx 4000\,\rm{km s^{-1}}$ from the X-rays, inferred from Eq.\ \ref{Eq:TemptoVel} (for the observed $T\sim20$ keV).\footnote{We note that the $v_{\rm FS}$ derived from VLBI imaging of SN\,2014C is also significantly larger than the value reported by \cite{Anderson16}. We ascribe this difference to the assumption of synchrotron radiation in the self-absorbed regime in the modeling of 15 GHz data of SN\,2014C by \cite{Anderson16}, which might not be realistic. In the following we adopt the direct constraints on the size of the radio emitting region obtained with VLBI techniques.  } 
This discrepancy suggests that the radio and X-ray observations are tracing two different emitting regions. Broadly speaking, we expect the radio synchrotron emission to trace the fastest moving material, while X-rays are powered by bremsstrahlung radiation and are therefore expected to trace the densest material (provided that the photons are not absorbed locally by the material).

Additionally, any valid model would need to reconcile two independent and important observational constraints: (i) from VLBI observations, \cite{Bietenholz21, Bietenholz17} found the emission consistent with a projection of a spherical shell into the sky plane, implying that the fastest moving material is consistent with (albeit can still deviate from) a spherical source; and (ii) a broad H$\alpha$ spectral component with width of few $1000\,\rm{km\,s^{-1}}$ emerged in the spectra of SN\,2014C at 127 days since explosion, at the same time of the X-ray and radio re-brightening \citep{Milisavljevic15,Mauerhan18,Thomas22}. Since the ejecta of SN\,2014C are H-poor (SN\,2014C was originally classified as a type-Ib SN), following the reasoning of \cite{Chugai06} for SN\,2001em, \cite{Margutti2017} associated the width of the broad H$\alpha$ component with shocked CSM material, which implies a FS velocity of few $1000\,\rm{km\,s^{-1}}$, comparable to the inferred velocity of the X-ray emitting material. This line of reasoning supports the association of the X-ray emitting material with H-rich, dense shocked  CSM. Instead, the radio emitting region imaged by VLBI is located at a larger radius and the radio emitting material is expanding at significantly larger velocity in the environment (i.e., the radio shock traced by VLBI was not as heavily decelerated as the X-ray emitting material).

The inferences from the X-ray, optical, and radio VLBI observations of SN\,2014C can be reconciled in a scenario where the CSM is highly asymmetric, e.g.,  in the shape of an equatorial ``disk'' as shown in the cartoon in Fig.\ \ref{Fig:Cartoon}. Within the inner disk radius $R_{i}$, the expansion of the FS is spherically symmetric and the FS has the same expansion velocity in all directions. At distances $>R_{i}$, the FS dynamics are impacted by the presence of the dense equatorial disk of CSM. As a consequence,  material ejected along the polar directions will interact with CSM material of significantly lower-density than  that of the denser CSM on the equatorial plane. We associate the faster, quasi-spherical FS component with the VLBI emitting region (Fig.\ \ref{Fig:Cartoon}), and the significantly decelerated equatorial FS with the X-ray and H$\alpha$ emitting material. 
Departures from spherical symmetry for the CSM around SN\,2014C were initially suggested by \cite{Milisavljevic15}. Additionally, a very similar conclusion has been reached by \cite{Thomas22} also from the discrepancy between the radio and X-ray velocities. 

\begin{figure}
    \centering
  \includegraphics[width=0.49\textwidth]{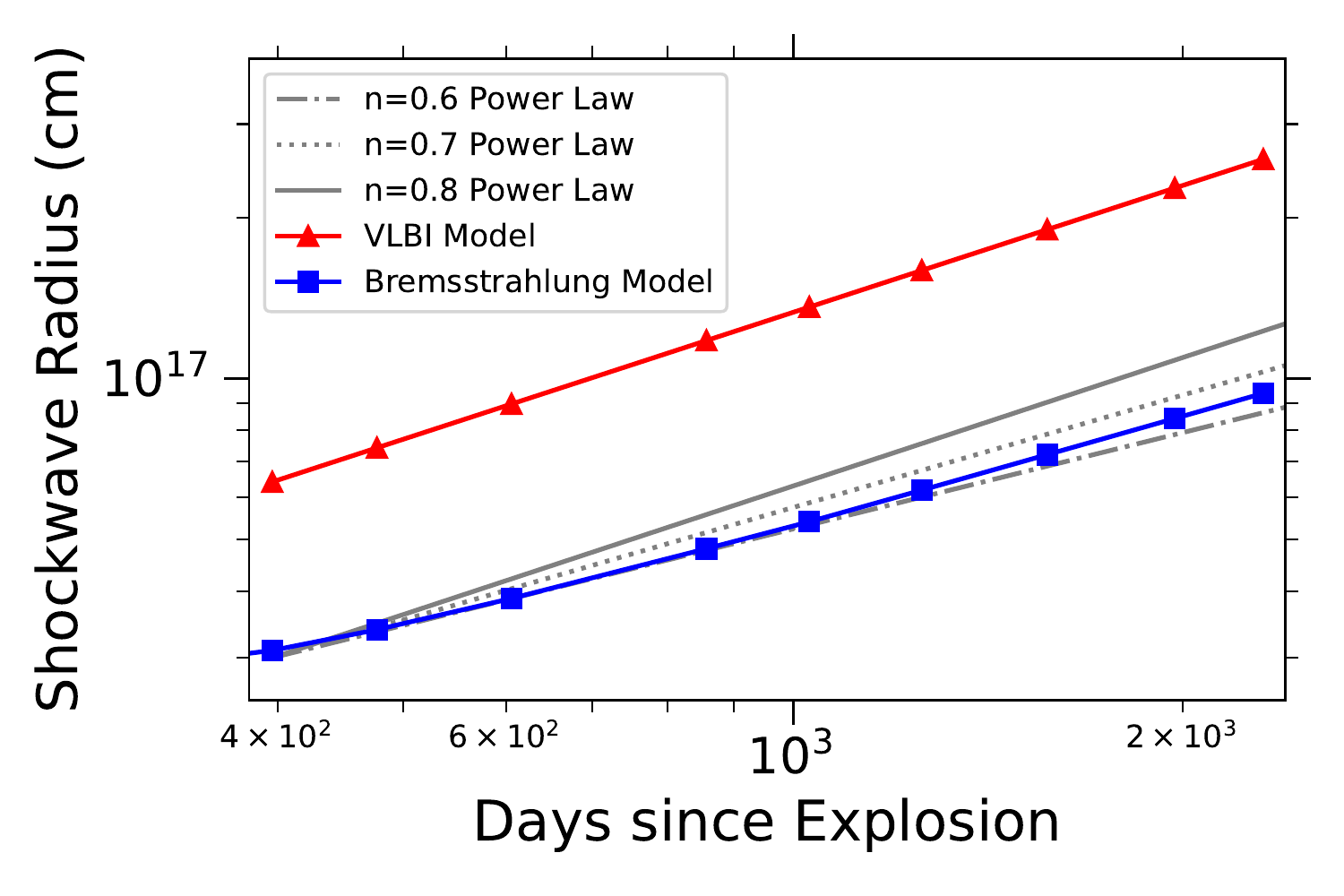}
    \caption{Inferred temporal evolution of the radius of the shockwave that dominates the radio emission detected by VLBI (red triangles, Eq.\ \ref{Eq:BietRadii} from \citealt{Bietenholz21}), and the inferred radius of the X-ray emitting material (blue squares, \S\ref{SubSec:Constrain}) calculated using the shock velocity inferred from Eq.\ \ref{Eq:TemptoVel}. The thick, dotted, and dot-dashed grey lines are power-law evolution of the radius $R\propto t^{n}$ with $n=0.8$, $n=0.7$ and  $n=0.6$, respectively, to guide the eye. }
    \label{Fig:RadiiEvo}
\end{figure}

Within this ``disk+quasi-spherical polar outflow'' model, prior to the interaction with the CSM equatorial disk at $R_i$, the FS front was spherical (thick black line in Fig.\ \ref{Fig:Cartoon}) and later was  not strongly decelerated as it did not encounter the thick disk. Thus, we  extrapolate the position of the shock using the inferences from VLBI imaging (i.e., Eq.\ \ref{Eq:BietRadii}) to the onset time of strong interaction ($\delta t \sim 100$ days, as constrained by optical spectra from \citealt{Milisavljevic15}) and then we use the shock velocities inferred from the X-ray spectral modeling (Eq.\ \ref{Eq:TemptoVel}) to estimate the shock radius evolution with time within the thick disk (Fig.\ \ref{Fig:RadiiEvo}, blue squares). 
In this scenario, the CSM inner radius is $R_i\approx 2.2\times10^{16}$ cm and the X-ray observations presented in this paper probe the CSM out to a radius $\sim10^{17}$ cm at $\delta t = 2307$ days (Fig.\ \ref{Fig:RadiiEvo}).

\subsection{Properties of the CSM probed by X-ray Observations}
\label{SubSec:CSMConstrain}

We estimate the total mass and density profile of the CSM probed by X-ray observations (i.e., the ``disk'' in Fig.\ \ref{Fig:Cartoon}) using two independent methods: the evolution of the $NH_{\rm{int}} (t)$ and $EM(t)$. We expect both methods to lead to the same order of magnitude estimate of the CSM mass. However, the $NH_{\rm{int}}(t)$ provides a direct measurement of the neutral material (i.e., material that absorbs X-ray radiation through photo-electric effects) along the line of sight, while the $EM(t)$ depends on the volume and density of the emitting material.

\begin{figure*}
    \centering
  \includegraphics[width=0.88\textwidth]{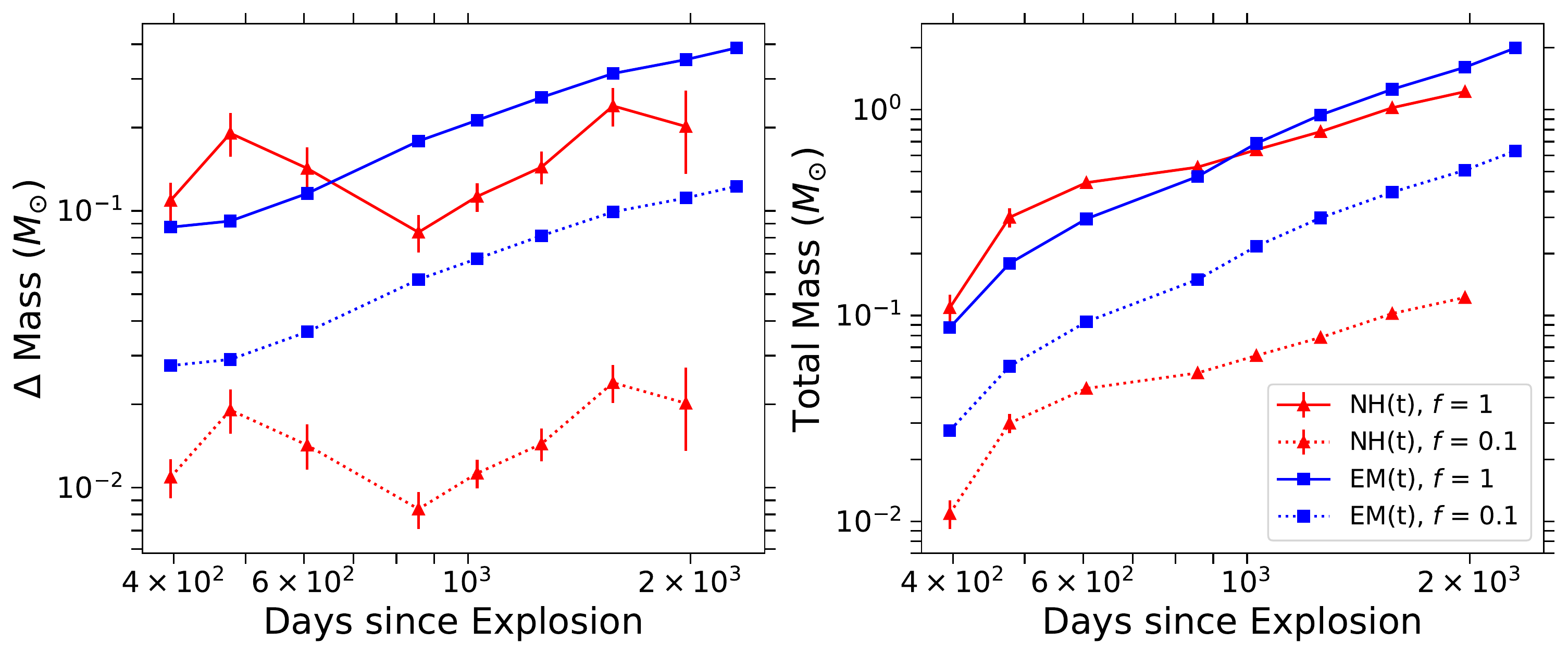}
    \caption{Incremental mass added between observations (left) and total mass (right) based on $NH_{\rm{int}}(t)$ (red triangles) and $EM(t)$ (blue squares). The $NH_{\rm{int}}(t)$ masses are calculated via Equation \ref{Eq:nHtomass} using radii from X-ray emission, while the $EM(t)$ masses are calculated using Equations \ref{Eq:FSVol} and \ref{Eq:FSden} under an assumption that shell thickness is 0.1$R$. Importantly, each measurement of accumulated mass from $NH_{\rm{int}}(t)$ requires a following observation, hence there is one fewer data point compared to the $EM(t)$. The non-solid lines represent a filling factor of 0.1 applied to each method, as estimated by line emission in \cite{Milisavljevic15}.  We note that the plotted error bars reflect statistical uncertainties only (statistical uncertainty of $EM(t)$ is smaller than the squares).}
    \label{Fig:nHMass}
\end{figure*}
\begin{figure*}
    \centering
  \includegraphics[width=0.97\textwidth]{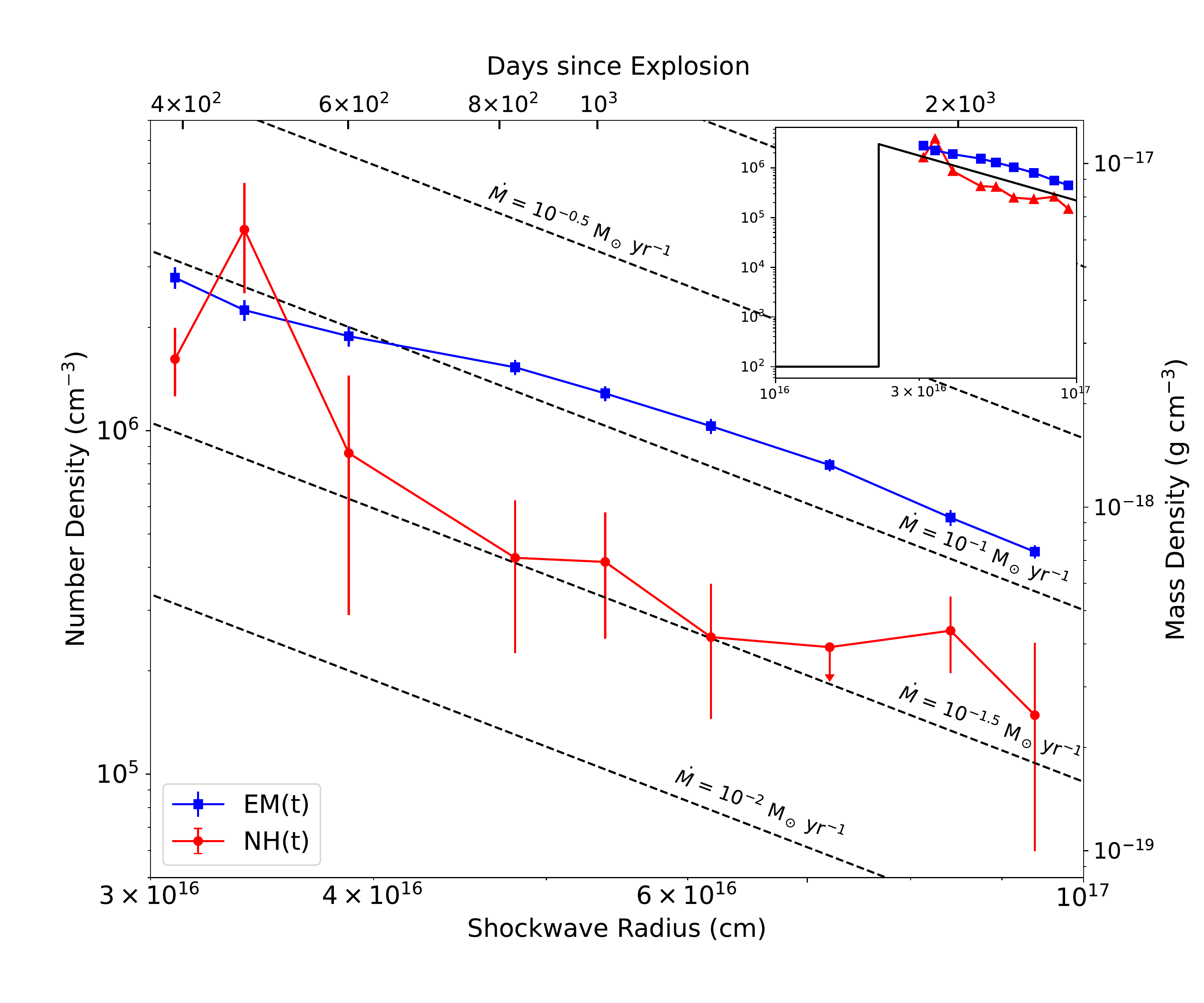}
    \caption{Density profile of the CSM assuming solar chemical composition, as derived from $NH_{\rm{int}}(t)$ (red triangles) and $EM(t)$ (blue squares) following the assumptions stated in \S \ref{SubSec:CSMConstrain}. We assume \textit{f} = 1, $\Delta R$ = $0.1R(t)$, $\mu_e$ = 1.25, and $\mu_I$ = 1.15 in these calculations. Black dashed lines:  wind-density profiles for a range of mass-loss rates $\dot M$, i.e., ($\rho_{\rm CSM}=\dot M/(4\pi v_w R^{2})$) for  $\log({\dot M/\rm{M_{\odot} yr^{-1}}}) = [-2,-1.5, -1, -0.5, 0]\, $ with an assumed wind speed of $v_w=$1000 km s$^{-1}$. Due to a small change in $NH_{\rm{int}}$ between observations for $\sim 7 \times 10^{16}$ cm,  the point is an upper limit as the error bars are approximately equal to the density measurement. Fitting with a power law  the density profile is best fit with $\rho_{\rm CSM} \propto$ $R^{-2.42 \pm 0.17}$ and $R^{-1.50 \pm 0.01}$, derived from $NH(t)$ and $EM(t)$, respectively. Both profiles are inconsistent than a wind profile, albeit with an extreme mass-loss rate. We note that the plotted error bars reflect statistical uncertainties only.
    \label{Fig:RadiivNumDen}}
\end{figure*}

First, we compute the CSM mass constraints inferred from  $EM(t)$. Following \cite{Margutti2017} and our discussion in \S\ref{SubSec:Constrain} we identify the emitting region as the shocked CSM, which has volume:

\begin{equation}
\label{Eq:FSVol}
    V_{\rm FS} \approx 4 \pi R^{2} \Delta R \textit{f}
\end{equation}
\noindent
where $\Delta R$ is the thickness of the FS shocked CSM shell and \textit{f} is a filling factor that quantifies the deviation of the shocked material from spherical symmetry.
Combining Equation \ref{Eq:FSVol} with  Equation \ref{Eq:EM}, the density of the shocked CSM is:

\begin{equation}
\label{Eq:FSden}
    \rho_{\rm FS}(t) = \sqrt{\frac{ EM(t)\mu_{e}^{\rm FS}\mu_{I}^{\rm FS} }{V_{\rm FS}(t)}} m_{p} 
\end{equation}
\noindent
The density of the CSM upstream is $\rho_{\rm CSM} \sim \frac{1}{4} \times \rho_{\rm FS}$ for a shock compression factor $R\sim 4$ appropriate of strong shocks and monoatomic ideal gas.
Additionally, we assume that only $\sim$ half of the radiation can reach the observer, and so multiply $EM(t)$ by 2. 
Assuming that at any time $t$ the X-ray emission is dominated by newly shocked material at radius $R(t)$ in a shell of shocked material of thickness $\Delta R$,\footnote{We expect this assumption to break down at late times, i.e., once the shock emerges from the thicker part of the H-rich CSM ``disk''.} the resulting pre-shock CSM density is thus given by: 
\begin{equation}
\rho_{\rm CSM}(t) = \sqrt{\frac{EM(t)\mu_{e}^{\rm FS}\mu_{I}^{\rm FS}}{32\pi R(t)^2 \Delta R\,f}}m_p
\end{equation}
We show our results in terms of shell mass, total shocked mass,  mass density and particle number density in  Figures \ref{Fig:nHMass} and \ref{Fig:RadiivNumDen} (blue squares).

Second, we derive constraints on the CSM mass from  $NH_{\rm{int}}(t)$.
$NH_{\rm{int}}$ is the column density of the equivalent amount of neutral hydrogen between an observer and the object, and is defined as
\begin{equation}
    NH_{\rm{int}} = \int_{R_E}^{R_{obs}} n_{H}(r) dr
    \label{Eq:NHFormula}
\end{equation}
\noindent
where $n_{H}$ is the number density of hydrogen, $R_{obs}$ is the distance of the observer from the object, and $R_E$ is the radius at which radiation is produced. The observed difference of $NH_{\rm{int}}(t)$ between two observations carried out at $t_1$ and $t_2$, where  $R_E(t_1)\equiv R_1$ and $R_E(t_2)\equiv R_2$ is:
\begin{equation}
    \label{Eq:DeltanH}
    \Delta NH_{\rm{int}} = \int_{R_1}^{R_2} n_H(r) dr\approx n_H(R_1)\times (R_2-R_1)
\end{equation}
where we assumed that  $n_{H}$ is approximately constant between $R_1$ and $R_2$. 

It follows that the total mass the shock has traveled through between two observations at $t_1$ and $t_2$ is: 

\begin{equation}
   \Delta M_{\rm CSM}(t) = \frac{ \,\,m_{p}}{X_H}  \Big( \frac{\Delta NH_{\rm{int}}(t)}{R_2-R_1} \Big) V_{R_1,R_2}(t)
    \label{Eq:nHtomass}
\end{equation}
where $X_H$ is the fraction by mass of hydrogen, $V_{R_1,R_2}(t)$ is the spherical volume between $R_1$ and $R_2$ (and therefore $V_{R_1,R_2}(t) \propto \textit{f}$\,) and $X_H=0.7381$ for solar abundances \citep{Asplund09}. 
We show $\Delta M_{\rm CSM}(t) $ and the mass sampled by the shockwave as it expands in the CSM in Figure \ref{Fig:nHMass}, red symbols.
Finally, we find that the total amount of \emph{neutral} mass sampled by the shockwave between $\delta t$ = 396 days to $\delta t$ = 2307 days is $M_{\rm CSM}\approx (1.2 \pm 0.03)\textit{f}\,\, M_\odot$. 
 
Similarly, we estimate the mass of the CSM using the density profile from $EM(t)$ (which  is sensitive to ionized material emitting free-free radiation) as  $M_{\rm CSM}\approx (2.0  \pm 0.04)\sqrt{f}\, M_\odot$. These two estimates agree within a factor $<2$.  We provide the statistical uncertainties from standard error propagation only. However, it is clear that a major source of uncertainty is of systematic nature and related to the geometry of the CSM, which we quantify with the filling factor $f$, and to the thickness of the shocked region (represented by the $\Delta R$ parameter).  With these caveats in mind, we find that the best-fitting CSM ``disk'' density profile scales as $\rho_{\rm CSM}(R)\propto R^{-1.50 \pm 0.01}$ and  $\rho_{\rm CSM}(R)\propto R^{-2.42 \pm 0.17}$  for the $EM(t)$ and the $NH_{\rm{int}}(t)$ methods, respectively (Fig.\ \ref{Fig:RadiivNumDen}). For $f=1$ and a wind velocity $v_w=1000\,\rm{km\,s^{-1}}$, the inferred $\rho_{\rm CSM}(r)$ profiles correspond to very large mass-loss rates in the range $\dot M\approx (0.03-0.1)\,\rm{M_{\odot}yr^{-1}}$ (Fig.\ \ref{Fig:RadiivNumDen}).

\section{Discussion}
\label{Sec:Disc}
SN\,2014C was initially spectroscopically identified as a type Ib SN and transitioned to a strongly interacting type IIn over approximately one year \citep{Milisavljevic15}. From our modeling of the evolution of the broad-band X-ray emission from SN\,2014C we infer the presence of dense H-rich asymmetric CSM starting at $R_i\approx 2\times 10^{16}$ cm and with mass in the range  $M_{\rm CSM}\approx(1.2f-2.0\sqrt f)M_{\odot}$, which is consistent with previous estimates by \cite{Margutti2017,Vargas21,Harris20} (Fig.\ \ref{Fig:nHMass}). Intriguingly, \cite{Harris20} also note that radio observations would indicate a significantly smaller CSM mass than the X-rays, further supporting a disassociation between the radio and X-ray data. In this section we start by putting the CSM shell parameters of SN\,2014C into the broader context of other core-collapse SNe that showed clear signs of shock interaction with a dense medium at some point during their evolution (\S\ref{SubSec:observationalcontext}).  Second, we explore in \S\ref{SubSec:masslosstheory} the physical mechanism(s) that might be behind the mass-loss phenomenology currently observed in core-collapse SNe.

\begin{figure*}
    \centering
  \includegraphics[width=0.97\textwidth]{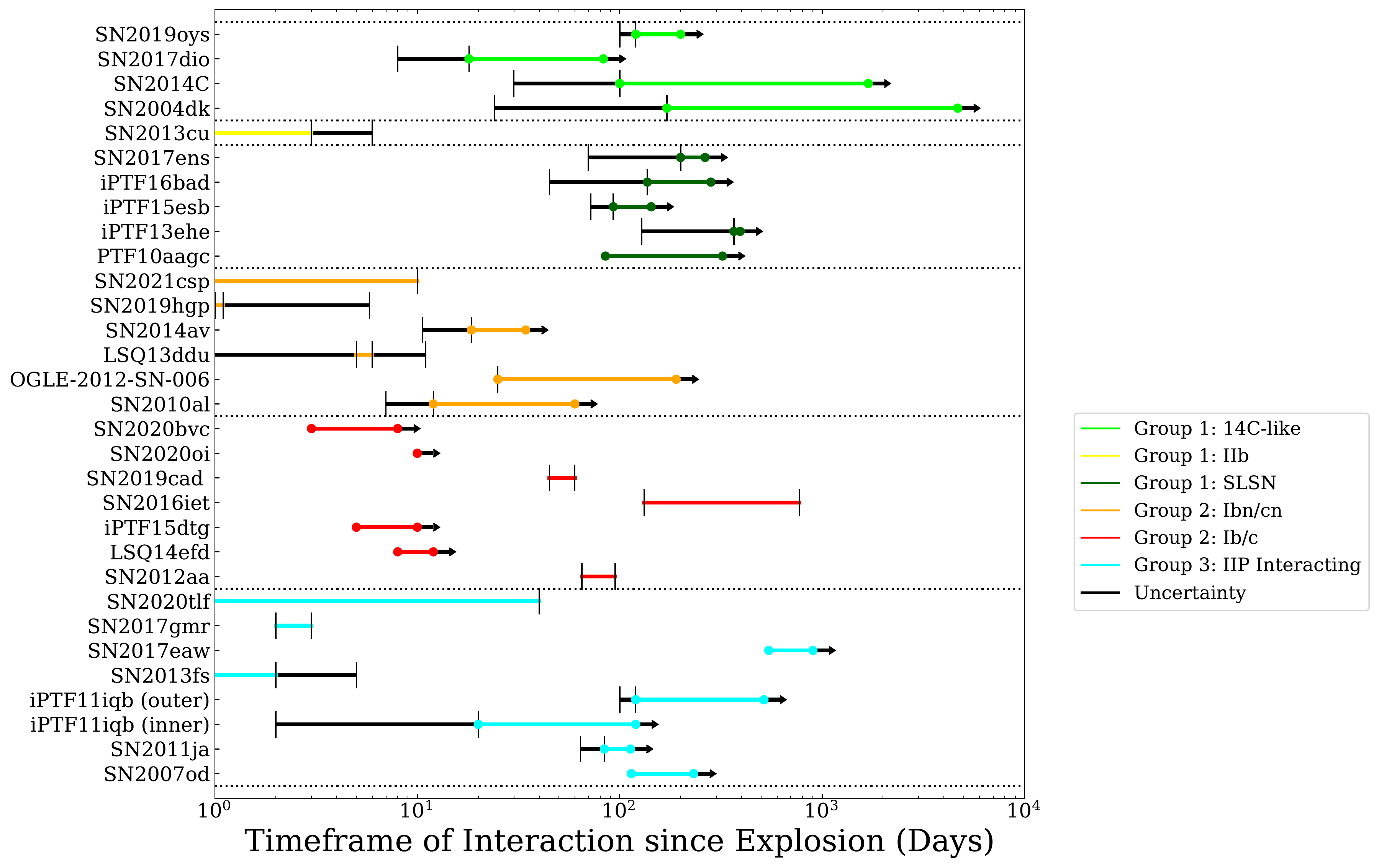}
    \caption{A comparison of the timeframes of interaction using the same color scheme as Figures \ref{Fig:CSMRad} and \ref{Fig:Burn} (the legend also reflects the order that categories appear). Most points are lower limits as interaction was still ongoing as of the latest observation when the transient faded below the threshold of detection. Uncertainties in interaction onset are dominated by the time between consecutive observations.  Additionally, we exclude type IIn SNe (normal and SLSN) from this plot as their interaction begins immediately after explosion and continues to the latest observation. Circles represent when the data include a lower limit.
    \label{Fig:InteractComp}}
    
\end{figure*}

\begin{figure*}
    \centering
  \includegraphics[width=0.97\textwidth]{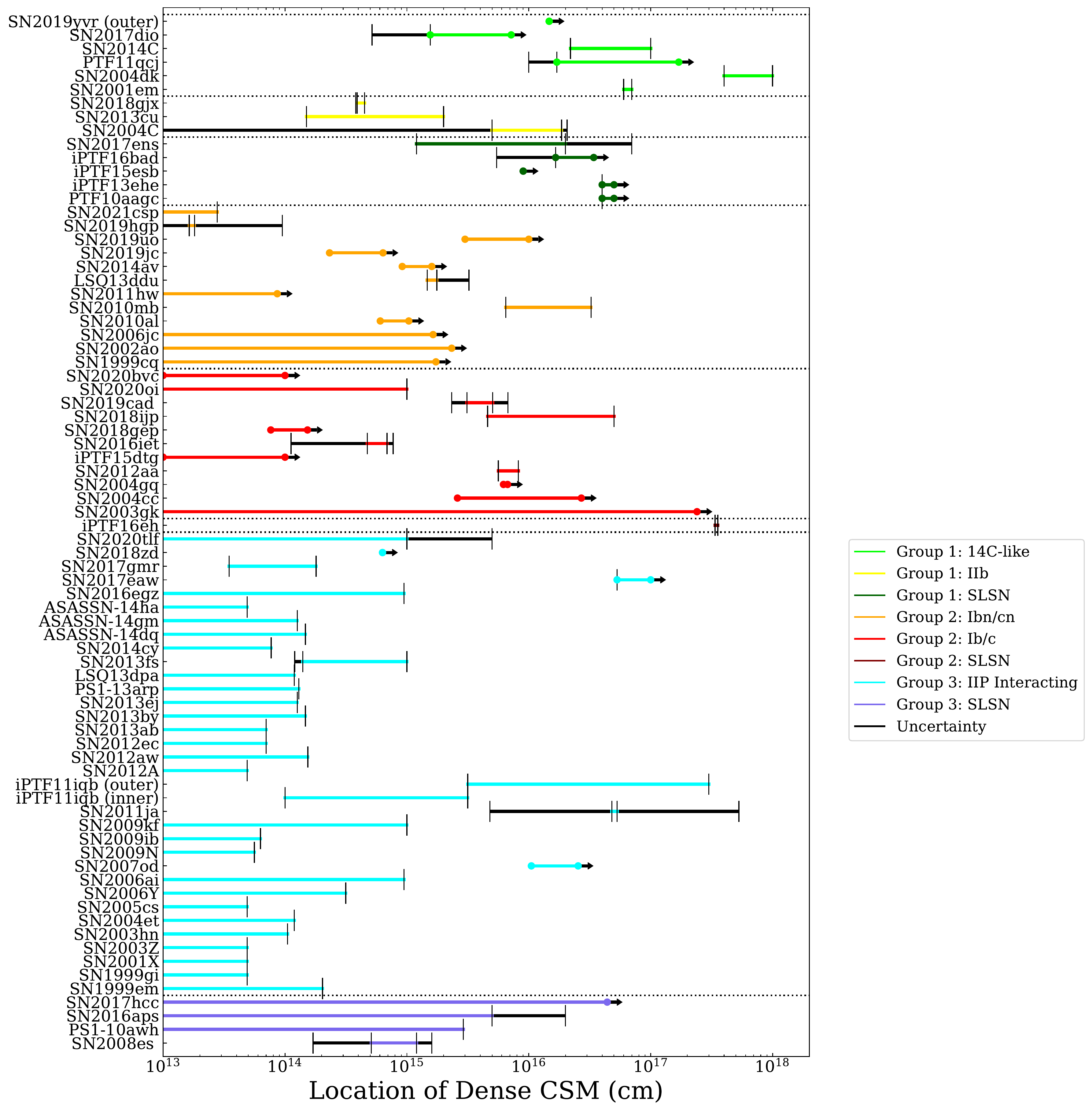}
    \caption{Radial location (or extension) of the regions of high-density material for different types of SNe that showed evidence for CSM interaction at some point during their evolution. Type IIn SNe are excluded from this plot as their CSM extend across  the whole range plotted. This figure uses the same color scheme as Figure \ref{Fig:InteractComp} and \ref{Fig:Burn} (the legend also reflects the order that categories appear).  Circles represent when the data includes a lower limit.
    \label{Fig:CSMRad} }
    
\end{figure*}

\begin{figure*}
    \centering
  \includegraphics[width=0.97\textwidth]{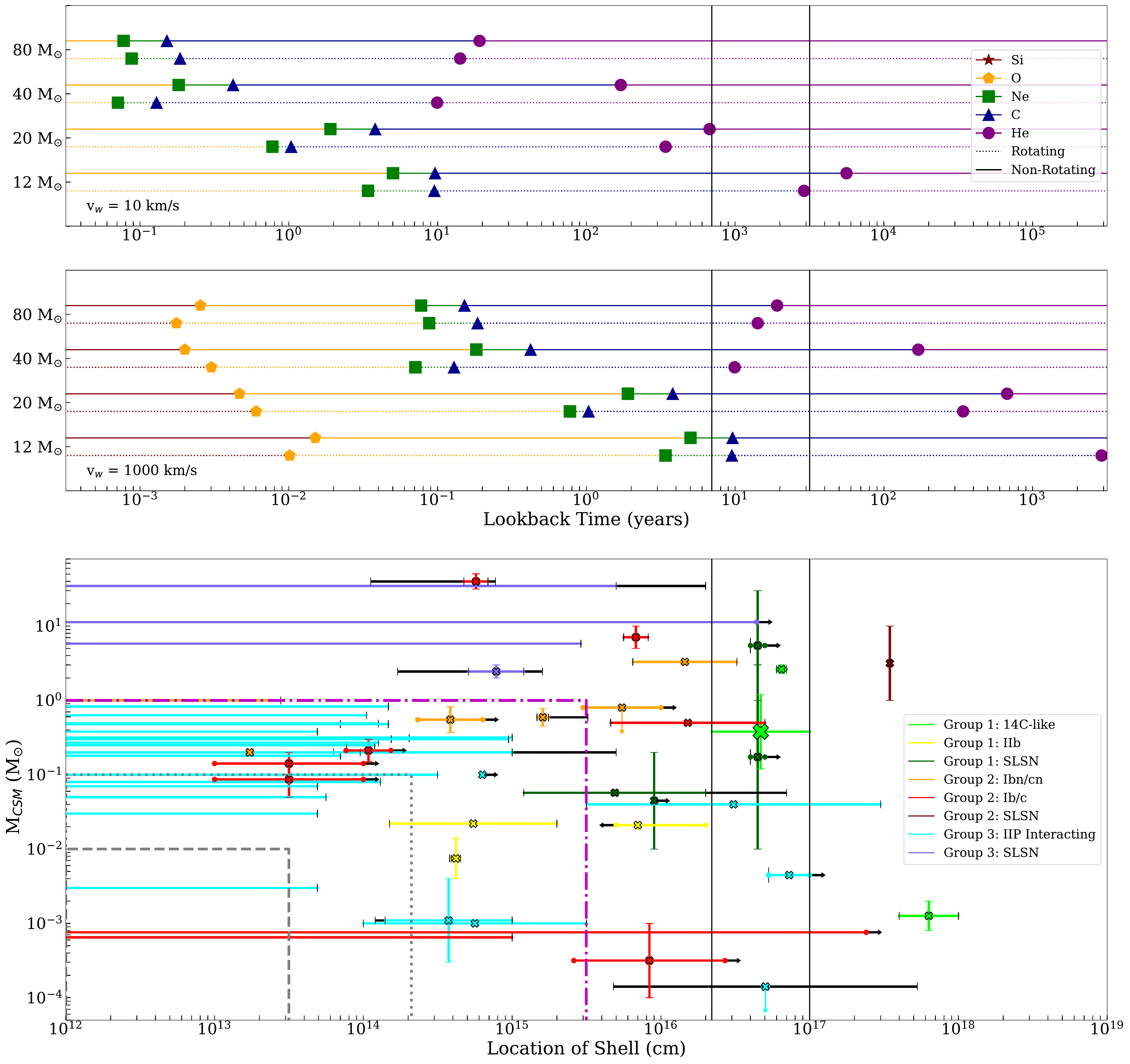}
    \caption{\emph{Upper and Middle Panels:} Core nuclear burning stages of massive stars with $M=12-80\,\rm{M_\odot}$ as a function of time before collapse. Solid and dotted lines represent non-rotating and rotating stars with $\Omega$ = 0.5, as described in \cite{Shiode14}, respectively. 
    \emph{Lower Panel:} amount of mass lost plotted against the location of the thick CSM for a subset of SNe in Fig.\ \ref{Fig:CSMRad} and follows the same color scheme. The x-axis of this plot maps into the x-axes of the middle and upper panels for an assumed wind speed of 10 km s$^{-1}$ (top panel) or 1000 km s$^{-1}$ (middle panel). Vertical black solid lines emphasize the location of the CSM material for SN\,2014C. For the CSM mass of SN\,2014C, we use an upper limit of $f=1$ and a lower limit of $f=0.1$. Additionally, we show the boundaries for which three wave-driven mass-loss episodes occurred in simulations. The magenta dot-dash line represents a \cite{Wu20} model, and the grey dotted and dashed lines represent a H-rich and H-poor model from \cite{Leung21b}. Details of the model are described in \S \ref{SubSec:masslosstheory}. 
    \label{Fig:Burn}}
    
\end{figure*}

\subsection{SN\,2014C in the context of interacting core-collapse SNe}
\label{SubSec:observationalcontext}
We collected from the literature a large sample of 119 core-collapse SNe that showed observational signatures of strong CSM interaction (Table \ref{Tab:4Interactors}). Shock interaction with a dense CSM can manifest through at least one of the following observational features:
 (i) spectral narrow lines (transient or persistent) due to shock ionization of wind material ahead of the shock ; (ii) photometric variability of the SN light-curve (including multiple bumps); (iii) luminous X-ray/radio emission. 
 
 The core-collapse SN interactors  fall into three major physical categories (Table \ref{Tab:4Interactors}):
\begin{itemize}

\item \emph{Group 1: SNe with H-poor ejecta interacting with dense, H-rich CSM}.  Type Ib and type Ic SN\,2014C-like transitional objects and SLSNe-I with late H$\alpha$ emission belong to this group. 
\item \emph{Group 2: SNe with H-poor ejecta interacting with H-poor CSM.} Belonging to this group are type Ibn SNe and the recently identified class of type Icn SNe, as well as type Ib/c SNe that interacted with H-poor CSM without producing narrow spectral lines.  
\item \emph{Group 3: SNe with H-rich ejecta interacting with dense, H-rich CSM}.
Belonging to this group are normal type IIn SNe and SLSNe IIn,  UV-bright type IIP/IIL SNe and interacting type IIP/IIL SNe. 
\end{itemize}

\begin{deluxetable*}{l|c|cc}[h!]
\tabletypesize{\scriptsize}
\setlength{\tabcolsep}{16pt}
\tablecolumns{3}
\tablewidth{14pc}
\tablecaption{Sample of strongly interacting core-collapse SNe. See Appendix \ref{Appendix} for references.  
\label{Tab:4Interactors}}
\tablehead{\colhead{} & H-Rich CSM & H-Poor CSM \\[0.0 cm]}
\startdata
  H-rich Ejecta & \textbf{IIn} & \\
   & 1996cr, 1987F, 1988Z & \\
   & 1994W, 1994aj, 1995G & \\
   & 1995N, 1996L, 1997ab & \\
   & 1997eg, 1998S, 2005gl & \\
   & 2005ip, 2005kj, 2006aa & \\
   & 2006bo, 2006jd, 2006qq & \\
   & 2006tf, 2008fq, 2008iy, & \\
   & 2009ip$^{*}$, 2010jl, 2010mc & \\
   & 2011fh, 2011ht, 2013L & \\
   & 2015da, 2016jbu & \\
   & & \\
   & \textbf{IIP} & ? \\
   & 1999em, 1999gi, 2001X & \\
   & 2003Z, 2003hn, 2004et & \\
   & 2005cs, 2006Y, 2006ai & \\
   & 2007od, 2009N, 2009ib & \\
   & 2009kf, 2011ja, iPTF11iqb & \\
   & 2012A, 2012aw, 2012ec & \\
   & 2013ab, 2013by, 2013ej & \\
   & 2013fs, PS1-13arp, LSQ13dpa  & \\
   & 2014cy, ASASSN-14dq, ASASSN-14gm & \\
   & ASASSN-14ha, iPTF14hls, 2016egz & \\
   & 2017eaw, 2017gmr, 2018zd & \\
   & 2020faa, 2020pni, 2020tlf & \\
   & & \\
   & \textbf{Type II SLSNe} & \\
   & 2006gy, 2008es, 2016aps & \\
   & 2017hcc & \\
   & & \\
   & \textbf{Type II} & \\
   & 2017ahn & \\
   & & \\
   \hline
   & & \\
  H-Poor Ejecta & \textbf{14C-like Events}  & \textbf{Ibn}  \\
  & 2001em, 2004dk, PTF11qcj & 1999cq, 2002ao, 2006jc  \\
  & 2014C, 2017dio, 2017ens  &  2010al, 2011hw, OGLE-2012-SN-006   \\
  & 2019oys, 2019yvr & LSQ13ddu, 2014av, ASASSN-14ms\\
  & & 2019uo\\
  & \textbf{IIb} & \\
  & 1993J, 2004C, 2011dh &  \textbf{Icn} \\
  &  2013cu, 2013df, 2018gjx  & 2010mb, 2019jc, 2019hgp \\
   & ASASSN-15no$^{+}$ & 2021ckj, 2021csp \\
   &  &  \\
   & \textbf{Type I SLSNe} & \textbf{Type I SLSNe} \\
   & PTF10aagc, iPTF13ehe, iPTF15esb & iPTF16eh  \\
   & iPTF16bad, 2017ens &  \\
   & & \textbf{Type Ib/c} \\
   &  & 2003gk, 2004cc, 2004gq \\
   & &  2012aa, LSQ14efd, iPTF15dtg \\
   & &   2016iet, 2018gep, 2018ijp \\
   & &  2019cad, 2019ehk, 2020oi \\
   & &  2020bvc   \\
   & & \\
  \hline
\enddata
$^{*}$ SN\,2009ip refers to the 2012 explosion. $^{+}$ ASASSN-15no interacted with H-poor material. ASASSN-15no was spectroscopically classified as a type Ic at early times \citep{Class:ASASSN-15no:Early}, then later reclassified as type II \citep{Benetti18}.

\end{deluxetable*}

The physical parameters of interest for our study are: (i) the onset time and the end time of the interaction $t_{start}$ and $t_{end}$, respectively; (ii) the location of the dense CSM parameterized by an inner and outer radius of the CSM, $R_{\rm in}$ and $R_{\rm out}$, respectively; the total CSM mass $M_{\rm CSM}$; (iii) the wind velocity $v_w$ (i.e., the ejection velocity of what later becomes CSM); (iv) and the implied mass-loss rate, $\dot{M}$, assuming a wind profile. When possible, we calculated (or retrieved) the values of these parameters using the information provided in the papers  referenced in Table \ref{Tab:4Interactors}. Our detailed reasoning for each SN is reported in Appendix \ref{Appendix} for the reader to be able to fully reproduce our results and figures. 

Before commenting on the results from this exercise (displayed in Fig.\ \ref{Fig:InteractComp}, \ref{Fig:CSMRad} and \ref{Fig:Burn}),  we emphasize that the physical parameters have been estimated with a variety of observational ``tracers''. This is particularly true for the CSM mass for which various methods are known to lead to different estimates. Additionally, our collected sample of interacting core-collapse SNe represents the status quo of the field, but it is hardly complete nor homogeneous. The goal is to illustrate the variety of CSM properties that have been claimed in the literature so far. 

With  the caveats above in mind, we find that our heterogeneous sample of core-collapse SNe showed CSM interaction over a large dynamic range of times since explosion (from time of first light until several $1000$ days, Figure \ref{Fig:InteractComp}), which maps into a large dynamic range of distances from the explosion site $\sim 3\times 10^{13}-10^{18}$ cm (Figure \ref{Fig:CSMRad}, \ref{Fig:Burn}). The range of inferred CSM masses is also broad, with $M_{\rm CSM}$ spanning $ \sim 10^{-4}\,\rm{M_{\odot}}$ up to 10s $M_{\odot}$ of material (Figure \ref{Fig:Burn}). The large dynamic ranges of these key parameters likely reflect the diversity of stellar progenitors (and hence ejection velocity and composition of the CSM material) as well as a variety of envelope mass ejection mechanism that we discuss in more detail in \S\ref{SubSec:masslosstheory}.

Below we comment on some specific types of SN interactors that are directly relevant to SN\,2014C. While we show interacting type IIP SNe for completeness in Figures \ref{Fig:InteractComp}, \ref{Fig:CSMRad}, and \ref{Fig:Burn}, their red supergiant stellar progenitors \citep{Smartt15} likely imply mass ejection mechanisms different from those at play in the more compact H-poor evolved massive stars. 

\subsubsection{SN2014C-like type Ib/c SNe}

The same finding of a large dynamic range of CSM distances and masses holds even restricting the sample to core-collapse SNe from H-poor progenitors (like SN\,2014C) that also showed evidence for interaction with CSM of different chemical composition and geometries.  The transition from type Ib/c to type IIn SN due to H-rich CSM interaction has been reported in at least eleven SNe so far, i.e., SNe 2001em \citep{Class:2001em2}, 2004dk \citep{Mauerhan18}, iPTF11qcj \citep{Corsi14}, iPTF13ehe \citep{Yan17}, 2014C \citep{Milisavljevic15}, iPTF15esb \citep{Yan17}, iPTF16bad \citep{Yan17}, 2017dio \citep{Kuncarayakti18}, 2017ens \citep{Chen18}, 2018gjx \citealt{Prentice20}), and 2019yvr \citep{Kilpatrick21} while interaction with an H-poor medium has been invoked for SNe 2012aa \citep{Roy16}, 2019ou \citep{Strotjohann21}, 2004cc \citep{Wellons12}, 2004gq \citep{Wellons12}, iPTF15dtg \citep{Jin21}, and 2020bvc \citep{Jin21}. We note that for SNe 2004gq, iPTF15dtg, and 2020bvc the composition of the CSM was not spectroscopically verified. Most of these SNe show clear signs of interaction between $10-100$ days after explosion (Fig.\ \ref{Fig:InteractComp}), corresponding to a typical CSM  radius $\sim 10^{16}$ - 10$^{17}$ cm (Fig.\ \ref{Fig:CSMRad}), but otherwise display largely different CSM properties.
For example, SN\,2004dk is a SN2014C-like event for which 
thick H-rich CSM was inferred to be present starting as close as $\sim 5 \times 10^{15}$ cm and with densities  $\sim 2 \times 10^4$ cm$^{-3}$ (i.e., $\sim 100$ times less dense than in SN\,2014C, \citealt{Mauerhan18}).

In addition,  while here we focus on the small  sample of interacting type Ib/c SNe, it is worth noting that the vast majority of type Ib/c SNe  do \emph{not} show evidence for interaction with a thick CSM medium for years of radio monitoring after explosion (e.g., \citealt{Margutti2017}), which implies that most type Ib/c stellar progenitors do not experience substantial mass-loss for $\sim 10^3-10^5$ years prior to explosion depending on the  assumed $v_w$. A large sample study of radio Ib/c SNe by \cite{Bietenholz20a} infers a typical mass-loss rate $\dot M=10^{-5.4 \pm 1.2}$ M$_\odot$ yr$^{-1}$, which is significantly smaller than in SN2014C-like events, which require   $\dot M\approx (0.03-0.1)\,\rm{M_{\odot}yr^{-1}}$ (Fig.\ \ref{Fig:RadiivNumDen}).
The diversity of CSM properties close to the explosion's site can be an indication of very different mass-loss mechanisms at play or of the wildly different timescales over which the same mass-loss mechanism can operate in H-poor SN stellar progenitors (e.g., \citealt{Margutti2017}).

\subsubsection{Type Ibn and Icn SNe}
Known types of H-poor SNe interacting with an H-poor medium include type Ibn and type Icn SNe. For type Ibn SNe the CSM is He-rich (e.g., \citealt{Pastorello08}), while for the recently discovered type Icn SNe the CSM is both H- and He-poor (and show instead prominent narrow emission of oxygen and carbon from the CSM, e.g., 
SNe 2021csp, 2021ckj, 2019hgp, and possibly 2010mb, \citealt{Pellegrino22,Fraser21,Gal-Yam21Astronote,Perley22}). These SNe represent the case of stripped progenitors interacting with material that was lost from inner  layers, i.e.,  after H-envelope removal. Interestingly, the inferred location of this H-poor material and its mass also span large ranges: up to few $M_{\odot}$ of material (e.g., $M_{\rm CSM} \sim$1 M$_\odot$ and $M_{\rm CSM} \sim$3.3 M$_\odot$  for SNe 2021csp and 2010mb, respectively, \citealt{Fraser21,Ben-Ami14}), extending from the stellar surface to $>10^{16}$ cm.

\subsubsection{SLSNe}
One potential explanation for the large luminosities of SLSNe (e.g., \citealt{Quimby12,Chomiuk11}; see \citealt{Gal-Yam19} for a recent review) is efficient kinetic energy conversion into radiation due to shock interaction with a thick medium. Within this scenario, extreme mass-loss histories  have been invoked to power the persistent, luminous displays that are typical of SLSNe. For example for the type IIn SLSNe 2006gy \citep{Ofek07,Smith08,Fox15} and 2016aps \citep{Nicholl20,Suzuki21} and the H-stripped SLSNe 2016iet \citep{Gomez19} and iPTF16eh \citep{Lunnan18} several $M_{\odot}$ of CSM material have been inferred (Fig.\ \ref{Fig:Burn}). These dense regions of CSM are typically located at distances far from the progenitor at $\sim$10$^{16}$ cm and often extending far beyond to $\sim$10$^{17}$ and even to $\sim 10^{18}$ cm, implying mass ejections that occurred on timescales of hundreds to tens of thousands of years prior explosion.

Within interacting SLSNe, the most relevant comparisons to the phenomenology of SN\,2014C are SLSNe-I for which H$\alpha$ emerged at late times, presumably from the SN shock interaction with H-rich material, as seen in iPTF13ehe, iPTF15esb, and iPTF16bad \citep{Yan17} as well as SN 2017ens \citep{Chen18}. The typical time of interaction onset is $\sim$ 100s of days, with uncertainty driven by the time between spectra (See Figure \ref{Fig:InteractComp}). Intriguingly, the estimated CSM mass for these events ranges from 0.05 - 3 M$_\odot$, which is broadly similar to our calculations of SN 2014C's shell and of a similar timescale.

\subsubsection{Type IIn}
Finally, we comment on type IIn SNe \citep{Schlegel90}, as they might represent extreme cases of 14C-like events where the mass ejection occurred just before stellar core collapse, as was suggested, for example,  for SN\,1996cr \citep{Dwarkadas10}. The link between LBV progenitors and type IIn (e.g., SN 2005gl \citealt{Gal-Yam07}, SN 2009ip \citep{Mauerhan13,Margutti14,Smith16}, and SN 2010mc \citealt{Ofek13}) can explain the rarity and associated mass ejections.

In this scenario there might be a continuum of properties between at least some ordinary type IIn SNe, 14C-like SNe and ordinary type Ib/c SNe, as was suggested by \cite{Margutti2017}. Type IIn SNe are typically associated with large inferred mass-loss rates $\dot M \sim (0.0001 - 0.3)$ M$_\odot$ yr$^{-1}$ \citep{Kiewe12}, not dissimilar to our findings for SN\,2014C. However, differently from SN\,2014C, interaction with an H-rich CSM is detected from the very beginning (indicating the presence of thick CSM all the way to the progenitor surface). 
Additionally, there are many  type IIn SNe that fade below the detection level before the H$\alpha$ emission subsides, providing a lower limit to the radial extent of the H-rich thick CSM, and some for which the H$\alpha$ is observed to vanish before the SN itself fades \citep[e.g.,][]{Taddia13}. Lower limits on the radial extent of the CSM range from $\sim10^{15}$ cm to more than $\sim3\times10^{17}$ cm, indicating an extensive time range over which the stellar progenitors were experiencing substantial mass loss.

\subsection{Mass-Loss Mechanisms}
\label{SubSec:masslosstheory}
We discuss in this section a variety of mass-loss mechanisms and their applicability to the specific case of SN\,2014C, as well as to the broader range of observational constraints on mass loss from evolved massive stars of the previous section. Specifically, we discuss in \S\ref{SubSubSec:windwind} the interaction of the fast Wolf-Rayet (WR) winds and slower red supergiant (RSG) winds from the previous evolutionary phase; in \S\ref{SubSubSec:binary} binary interaction and mass transfer through Roche lobe overflow and common envelope; in  \S\ref{SubSubSec:nuclear} the complete ejection of the H-envelope triggered by nuclear burning instabilities or gravity waves.

\subsubsection{RSG Wind - WR Wind Interaction}
\label{SubSubSec:windwind}
A first possibility is that the progenitor of SN\,2014C experienced typical line-driven wind mass loss during a RSG phase. For RSGs, $v_w$ of a few $10\,\rm{km\,s^{-1}}$, typical mass-loss rates are
$\dot M\approx 10^{-6}-10^{-5}\,\rm{M_{\odot}\,yr^{-1}}$ reaching $\dot M\approx 10^{-3}\,\rm{M_{\odot} yr^{-1}}$ in more extreme cases \citep[e.g.,][]{deJager88,Marshall04,vanLoon05}. 
The RSG phase was then followed by an anomalously short WR phase, during which winds have typical $v_w$ of a few $1000\,\rm{km\,s^{-1}}$ with $\dot M\sim 10^{-5}\,\rm{M_{\odot} yr^{-1}}$ \citep[e.g.,][]{Crowther07}.  The stellar progenitor then exploded as a type Ib SN in the bubble generated by its own WR winds against its previous phase of RSG winds. The ejecta traveled freely for approximately 130 days until they impacted and began to strongly interact with the hydrogen shell producing the transition from type Ib to type IIn.

The interaction between the lighter, faster WR winds with the slower, thicker RSG winds is known to lead to the formation of ``shell-like'' overdensity structures in the CSM around stars \citep{Dyson89}. However, the documented cases of WR-RSG wind-wind interaction are associated with  ``bubbles'' at typical  distances of $\sim 10^{19}$ cm \citep{Marston97},  significantly farther than the few $10^{16}$ cm distances inferred for SN\,2014C. In this context, the proximity of the H-rich CSM shell in SN\,2014C implies an extremely short WR phase with duration of $\sim $3.5 $(\frac{v}{2000\, \rm{km\; s}^{-1}})^{-1}$ yrs,  conflicting with the $\sim 10^5$ yr duration of the WR phase expected in the case of isolated massive stars, similar to what we concluded in \cite{Margutti2017}. We thus explore alternative scenarios. 
This conclusion extends to all H-poor SNe interacting with H-rich shells at $\le 10^{16}$ cm, for which an even shorter WR phase would have to be invoked (Fig.\ \ref{Fig:Burn})   
In the broader context of mass loss in evolved massive stars, the sizes of typical wind-driven bubbles are potentially consistent only with the farthest out CSM shells observed around the SN interactors of our sample.

\subsubsection{Binary Interaction}
\label{SubSubSec:binary}
A second possibility is that the progenitor of SN\,2014C underwent stripping due to a binary companion. It is well understood that the hydrogen rich envelope can be lost through binary interaction \citep[e.g.,][]{Podsiadlowski92}, and the majority ($\sim$ 69\%) of young massive stars live in interacting binary systems \citep{Sana12}. Modeling  archival pre-explosion Subaru and HST images at the location of SN\,2014C, \cite{Milisavljevic15} found evidence for a massive star cluster with favored age of $\sim 30$ Myr, and a turn-off mass of the stellar population between 3.5 and 9.5 M$_\odot$, thus indicating a massive progenitor in the low-mass bin of stars that experience core collapse. Consistent with this finding, \cite{Sun20} inferred for SN\,2014C a progenitor mass of 11 M$_\odot$ and suggest a binary model where the primary star experienced envelope stripping by the companion, likely through either Case C or Case BC mass transfer then followed by an eruption event or strong metallicity-dependent winds (we refer the reader to \citealt{Sun20} for further details). We also note that binary interaction is likely to naturally lead to an asymmetric CSM distribution in the environment (e.g., a disk or torus, \citealt{Ivanova13}), which is consistent with our inferences of \S \ref{SubSec:Constrain} and Figure \ref{Fig:Cartoon}.

While binary interaction is a viable option for SN\,2014C, from a population perspective we expect 14C-like events due to binary interaction to be quite rare in the last $\le 1000$ yr of a massive star life \citep{Margutti2017}.\footnote{Unless the last phases of stellar evolution are coupled with substantial envelope inflation, e.g., as a consequence of wave heating in H-poor stars \citep{Wu22} or nuclear burning instabilities \citep{Smith14b}.} Despite a large fraction of young massive stars which are in interacting binary systems \citep{Sana12}, the interaction is not necessarily synchronized with stellar death of the primary star. If binary interaction occurs too early and strips the progenitor of its hydrogen envelope, the CSM produced will be too far away to generate detectable signs of interaction by the time of explosion. 

In the broader context of mass-loss events in our sample of SN interactors, binary interaction may be the cause of the wide diversity in CSM distances due to its lack of direct synchronization with stellar death.\footnote{Note that the underdensity of events beyond $\sim10^{17}$ cm is likely the result of observational selection effects. Material at those distances is difficult to observe as the associated electromagnetic emission will be faint by the time the shock reaches those distances. }
However, we note that the abundance of type IIP SNe with thick CSM at such close proximity to their progenitor site (Fig.\ \ref{Fig:Burn}) and the recent detection of a pre-SN outburst \citep{Jacobson-Galan22}  suggests a separate mass-loss mechanism that is timed with core-collapse. 

\subsubsection{Nuclear Burning Instabilities and Gravity Waves}
\label{SubSubSec:nuclear}
Eruptive mass loss in the final stages of evolution in massive stars in massive stars can be consistent with nuclear burning instabilities and gravity waves, especially during O and Ne burning, in the few months to years prior to core collapse \citep{Arnett11b,Quataert12,Shiode13,Shiode13b,Smith14b,Woosley15,Morozova20}. 
Specifically, energetic gravity waves are driven by vigorous convection during the last $\sim$ 10 years of a massive star's life and have the potential to deposit energy into the envelope and unbind material as originally proposed by \citealt{Quataert12}. Numerous simulations have attempted to resolve the last years of progenitors ranging from 10-80 M$_\odot$ and have become increasingly sophisticated with time (e.g., \citealt{Wu20,Leung21b,Wu22}).

For lower mass stars, which is likely to be the case for the progenitor of SN\,2014C as detailed by \cite{Milisavljevic15,Sun20}, \cite{Wu20} identified an 11 M$_\odot$ RSG model capable of ejecting $\sim 1$ M$_\odot$ of CSM approximately 10 years prior to explosion at a velocity of 100 km s$^{-1}$ (magenta dot-dash outline in bottom panel of Figure \ref{Fig:Burn}). Similarly, \cite{Leung21b} found a H-poor model that was able to eject $\sim 0.01$ M$_\odot$ out to $\sim3\times10^{13}$ cm (gray dashed line in bottom panel of Figure \ref{Fig:Burn}). An H-rich progenitor counterpart in the same study was able to eject $\sim 0.1$ M$_\odot$ out to $\sim2\times10^{14}$ cm (gray dotted line in bottom panel of Figure \ref{Fig:Burn}).

Both \cite{Leung21b} and \cite{Wu20} show that there is strong scatter in their simulations among the amount of energy (and therefore mass ejected) and the timing of ejection, indicating that the simulations are highly sensitive to input parameters.
However, updated simulations from \cite{Wu22} suggest that wave-driven mass-loss might be significantly weaker than
previously estimated, and might instead be unable to 
remove more than $\approx 10^{-6}\,\rm{M_\odot}$ of material. Furthermore, for the specific case of H-stripped progenitors, \cite{Wu22} find that instead of unbinding material, the energy deposited by gravity waves leads to an envelope inflation up to a factor $\approx 10$ in radius, 
which could trigger binary interaction specifically within the last decade of a star's life. H-rich progenitors, such as RSGs, experience only minor variations in surface properties and are not expected to initiate binary interaction.

It is clear from Figure \ref{Fig:Burn} that gravity waves are insufficient at explaining CSM beyond $\sim$ few$\times10^{15}$ cm, even with the more optimistic models from \cite{Wu20} and \cite{Leung21b}. Even before the updated models of \cite{Wu22}, it was clear that gravity waves were unable to transfer a sufficient energy to unbind the CSM mass estimated for type IIP events by roughly an order of magnitude for many cases, indicating a large discrepancy between observation and current theory. 

There are other mass-loss mechanisms that could be responsible for this discrepancy such as silicon deflagration \citep{Woosley15} or  nuclear burning instabilities \citep{Arnett11b}. 
These mechanisms are synchronized with stellar death and can potentially lead to unbinding some of the envelope material.

\section{Conclusion}
\label{Sec:Conc}
Observations of SNe in the recent years have demonstrated our inadequate understanding of how mass loss proceeds in evolved massive stars in the centuries to years before core collapse (e.g., \citealt{Smith14}). Here  we presented the analysis of data from our broadband hard and soft X-ray campaign of SN\,2014C with the  \emph{CXO} and \emph{NuSTAR} extending from $\delta t = 396$  days to $\delta t = 2307$ days. While the number of SNe with evidence for interaction with a thick CSM is rapidly growing, SN\,2014C is still \emph{the only} event for which we were able to detect broad-band soft and hard X-ray emission over a timescale of several years. 
We interpreted this unique set of observations in the context of an absorbed thermal bremsstrahlung radiation model, and we constrained key physical parameters of the emission and their temporal evolution such as emission temperature $T(t)$, intrinsic absorption $NH_{\rm{int}}(t)$, and luminosity output $L_x(t)$. 
We used the evolution of these parameters with respect to time to infer important parameters of the system, particularly the density profile of the CSM at sub-pc scales around the progenitor.

Similar to what we found in \cite{Margutti2017}, we find that SN\,2014C exploded in a low-density cavity extending to  a radius of $\sim 2\times10^{16}$ cm. Beyond this radius we find evidence for a high-density CSM with number density $\sim 10^6$ cm$^{-3}$ and neutral density profile not too dissimilar from a wind-like medium, $\rho_{\rm CSM}\propto R^{-2.42 \pm 0.17}$  (Fig.\ \ref{Fig:CSMRad}). This CSM is hydrogen rich \citep{Milisavljevic15,Mauerhan18}. Furthermore, we find that the combination of the results from the radio monitoring of \cite{Bietenholz20a} and \cite{Bietenholz21} and our broad-band X-ray monitoring can be explained in the context of a highly asymmetric CSM with a ``disk-like'' geometry (Figure \ref{Fig:Cartoon}). Quantifying the deviation from spherical symmetry with a geometrical filling factor $f$, we revise the estimate of the CSM mass using two methods, leading to $M_{\rm CSM}\sim 2.0 (\sqrt{\textit{f}}\,\,) $ M$_\odot$ based on $EM(t)$ or $M_{\rm CSM}\sim 1.2 (\textit{f}\,) \rm{M}_\odot$ based on $NH_{\rm{int}}(t)$. These analytical estimates are in broad quantitative agreement with the inferences by \cite{Vargas21} from numerical modeling and simulations of SN\,2014C. 

The presence, mass, location, and chemical composition of the CSM are not consistent with traditional line-driven winds, and require the exploration of alternative models of time-\emph{dependent} mass-loss mechanisms in evolved massive stars. In this paper we considered three models:
WR-RSG wind-wind interaction, interaction with a binary companion, or a H-envelope ejection model triggered by nuclear burning instabilities or gravity waves. While the wind-wind interaction model has observational counterparts in our Galaxy and does not require eruptive mass loss, the exceedingly short WR phase implied makes it highly unlikely.
Current models of wave-driven mass loss or mass loss related to nuclear burning instabilities tend to predict small mass ejections at timescales very close to core collapse ($\lesssim $yrs) that translate in CSM distances that are significantly smaller than in SN\,2014C (Figure \ref{Fig:Burn}, lower panel). Additionally, for the small mass of the stellar progenitor of SN\,2014C ($\sim 3.5-11 M_{\odot}$, \citealt{Milisavljevic15,Sun20}), instabilities would need to extend to the C-burning phase (Figure \ref{Fig:Burn}, upper and middle panels).

We conclude that mass loss associated with the interaction with a binary companion is the most likely explanation for the phenomenology of SN\,2014C. However, the large dynamic range of CSM masses and timescales of the mass-loss episodes before explosion inferred for our large sample of SN interactors is likely a signature of a variety of mass-loss mechanisms at play. For example, while binary interaction may be able to  account for the extremes of the distributions of CSM masses and timescales of the observed values, it does not offer a natural explanation for the extremely compact ($r\lesssim 10^{14}$ cm), thick CSM environments around some type IIP SNe (Figure \ref{Fig:Burn}, lower panel). For these cases of very compact CSM around the explosion site, mass-loss mechanisms related with the nuclear burning history of the primary star offer a more natural solution to the observed timing with core collapse.

Deep observations by sky surveys to look for pre-SN emission combined with rapid follow-up in the X-ray and radio band can reveal important information about the underlying mass-loss mechanisms within massive, evolved stars. Better constraints on CSM mass, location, and pre-explosion ejection time will subsequently provide an avenue to dispel uncertainty surrounding these crucial mass-loss mechanisms, such as which mechanisms dominate in particular stars, the timescales at which those mechanisms dominate, and the observational signatures of each mechanism.

\vspace{5mm}
\section*{Acknowledgments}
We thank Cristiano Guidorzi, Brian Grefenstette, and Sayan Chakraborti  for their contributions to the proposals that led to the data presented, without which this paper would not be possible.

We are extremely grateful to the efforts by the entire \emph{CXO} and \emph{NuSTAR} teams without which this unique dataset would simply not exist.

This work  and made use of data from the NuSTAR mission, a project  led  by  the  California  Institute  of  Technology,  managed  by  the  Jet  Propulsion  Laboratory,  and  funded  by  the National  Aeronautics  and  Space  Administration.  This  research  has  made  use  of  the NuSTARData  Analysis  Software (NuSTARDAS)  jointly  developed  by  the  ASI  Science  Data  Center  (ASDC,  Italy)  and  the  California  Institute  of Technology (USA). This project is partially supported by NASA under contracts NNX17AI13G and NNX17AG80G. Support for this work was provided by the National Aeronautics and Space Administration through Chandra Awards Number GO6-17054A, GO9-20060A, issued by the Chandra X-ray Center, which is operated by the Smithsonian Astrophysical Observatory for and on behalf of the National Aeronautics Space Administration under contract NAS8-03060.
This research has made use of data obtained from the Chandra Data Archive and the Chandra Source Catalog, and software provided by the Chandra X-ray Center (CXC) in the application packages CIAO and Sherpa.

D.B acknowledges support from the NASA Illinois Space Grant and the Northwestern Summer Undergraduate Research Grant. 
R.M. acknowledges support by National Science Foundation under Award Nos.~AST-1909796 and AST-1944985. D.~M.\ acknowledges NSF support from grants PHY-1914448 and AST-2037297.
F.V. and F.D.C. acknowledge support
from the UNAM-PAPIIT grant AG100820.
L.D. is grateful for the partial financial support of the IDEAS Fellowship, a research traineeship program funded by the National Science Foundation under grant DGE-1450006.
W.J-G is supported by the National Science Foundation Graduate Research Fellowship Program under Grant No.~DGE-1842165. W.J-G acknowledges support through NASA grants in support of {\it Hubble Space Telescope} program GO-16075 and 16500.

\facilities{NuSTAR,CXO}

\software{astropy, python, pyplot, CIAO v. 4.12, NuSTARDAS v. 1.9.5, XSPEC, CALDB}

\appendix

\section{Inferences on the environments of SNe in the comparison sample}
\label{Appendix}

Here we describe our methodology in collecting and estimating information on the mass loss of stars from their SNe. We organize them chronologically and group them by the paper they were announced in.

\subsection{SNe 1987F, 1988Z, 1994W, 1994aj, 1995G, 1995N, 1996L, 1997ab, 1997eg, 1998S, 2005gl, 2005ip, 2006tf, and 2008iy}
SNe of type-IIn with mass-loss rates inferred from a combination of optical spectroscopy and light-curve modeling. We calculate the amount of mass between $10^{11}$ cm and $6\times 10^{16}$ cm to draw a direct comparison to SN 2014C by assuming a wind profile density with wind velocity and mass-loss rate from Table 9 of \cite{Kiewe12}. Onset of interaction is assumed to be 0 days because the earliest spectra of each SNe showed type IIn features. The lookback time for each SN was calculated by dividing the assumed $6\times10^{16}$ cm by the wind velocity.

Original classification spectra are as follows: 1987F \citep{Class:1987F}, 1988Z \citep{Class:1988Z}, 1994W \citep{Class:1994W}, 1994aj \citep{Class:1994aj}, 1995G \citep{Class:1995G}, 1995N \citep{Class:1995N}, 1996L \citep{Class:1996L}, 1997ab \citep{Class:1997ab}, 1997eg \citep{Class:1997eg}, 1998S \citep{Class:1998S}, 2005gl \citep{Class:2005gl}, 2005ip \citep{Class:2005ip}, 2006tf  \citep{Class:2006tf}, and 2008iy \citep{Class:2008iy}.

While we adopt the type IIn SN classification of \cite{Kiewe12}, we note that some of these classifications are contested. For example,  \cite{Pastorello18} point out odd properties such as missing nucleosynthesis elements that would be expected if SN 2005gl were a true SN as well as some similarities to SN 2009ip (see details below), and SN 2006tf could nearly be classified as a SLSN \citep{Smith08b}.

\subsection{SNe 1993J and 2011dh}

SNe 1993J and 2011dh are type IIb SNe \citep{Class:1993J,Class:2011dh} that showed strong radio and X-ray emission \citep{Fransson96,Soderberg12}. Based on the modeling of these observations, \cite{Kundu19} infer a mass-loss rate of $4\times10^{-5}$ and $4\times10^{-6}$ M$_\odot$ yr$^{-1}$ for SNe 1993J and 2011dh, respectively, that lasted for the 6500 and 3000 years prior to explosion assuming a 10 km s$^{-1}$ wind velocity. We infer a CSM mass of 0.26 and 0.012 M$_\odot$, respectively, by multiplying the mass-loss rate by the number of years prior to explosion the mass loss began. Additionally, \cite{Kundu19} assume that the CSM of 1993J extends out to $2\times10^{17}$ cm.

\subsection{SN 1996cr}

SN 1996cr is a type IIn SN explosion \citep{Class:1996cr} that showed a late-time increase in radio and X-ray luminosity \citep{Dwarkadas10}. Despite exploding in 1996, the first optical spectroscopic observations occurred 11 years post-explosion and so early time typing is unknown. \cite{Dwarkadas10} estimate a dense CSM of density $\sim8.2\times10^{-22}$ g/cm$^{3}$ from the progenitor that extends to $1.5\times10^{17}$ cm. They predict a mass-loss episode occurred within the 10$^4$ years before stellar death and generated a 0.64 M$_\odot$ CSM. Additionally, they calculate a lower limit of mass-loss rate for 1000 km s$^{-1}$ winds to be $\sim 3.3\times 10^{-9}$ M$_\odot$ yr$^{-1}$ and an upper limit of $\sim 2\times10^{-5}$ M$_\odot$ yr$^{-1}$.

\subsection{SNe 1999cq, 2002ao, 2006jc}

SNe 1999cq, 2002ao, and 2006jc are type Ib SNe \citep{Class:1999cq,Class:2002ao,Class:2006jc} observed in the optical band photometrically and spectroscopically. They each interacted with an H-poor but He-rich environment. \cite{Foley07} present optical spectra of each, focusing on SN\,2006jc, and show  spectra with evidence for interaction at 20, 27, and 19 days, respectively. \cite{Foley07} measure a wind speed of 500 km s$^{-1}$ for SN\,2006jc and estimate that the CSM had been ejected $\sim 2$ years prior to explosion. At an assumed ejecta velocity of 10000 km s$^{-1}$, considering the SNe must have been interacting at the time of the spectra presented in \cite{Foley07}, the CSM must have resided within  $\sim2\times10^{15}$ cm for each SN.

\subsection{SNe 1999em, 1999gi, 2001X, 2003Z, 2003hn, 2004et, 2005cs, 2009N, 2009ib, 2012A, 2012aw, 2012ec, 2013ab, 2013by, 2013ej, LSQ13dpa, 2014cy, ASASSN-14dq, ASASSN-14gm, and ASASSN-14ha}

SNe of type-IIP/L with CSM mass, radial extension, and density parameter $K$ inferred from a combination of optical and UV light-curve modeling (see Table 2 in \citealt{Morozova18}). We calculate the mass-loss rate by assuming a wind density profile and multiplying $K$ by $4\pi v_w$, where $v_w$ is their assumed wind speed (10 km s$^{-1}$). The onset time of interaction is assumed to the explosion time, as the UV excesses that are signatures of interaction are detected from the start of UV observations.
Lookback time is calculated by dividing the radial extension by the wind speed.

Additionally, we note that SN 2005cs is considered a peculiar 'sub-luminous' type IIP \citep{Pastorello06} while SN 2009N is considered an intermediate between 'sub-luminous' and typical \citep{Takats14}.

\subsection{SN 2001em}
SN 2001em is a type Ib SN \citep{Class:2001em1,Shivver19} that later evolved into a IIn stellar explosion \citep{Class:2001em2}   and it was  observed in the radio, X-ray, and optical \citep{Chugai06,Chandra20}. \cite{Chugai06} estimate a $\sim 3$ M$_\odot$ CSM at contained between $\sim6-7\times10^{16}$ cm. They estimate that a mass-loss rate of 2-10 $\times 10^{-3}$ M$_\odot$ yr$^{-1}$ over the course of 1000-2000 years prior to explosion could generate such a CSM. 

\subsection{SN 2003gk}
SN 2003gk is a type Ib SN \citep{Class:2003gk} observed in the radio with VLA \citep{Bietenholz14}. We estimate the mass and mass-loss rate using the density parameter $A$ as defined by \cite{Bietenholz14}. The mass-loss rate is defined as $A 4\pi v_w$, and the density profile is defined as $\rho_{\rm CSM} (R) = A r^{-2}$ for a wind. We integrate this density profile out to  $2.4\times10^{17}$ cm, the location of the shockwave at 8 years post explosion. Using an assumed minimum and maximum wind velocity of 10 and 1000 km s$^{-1}$, we calculate a mass-loss rate of 10$^{-7}$-10$^{-5}$ M$_\odot$ yr$^{-1}$ and a CSM mass of $7.6\times10^{-4}$ M$_\odot$.

\subsection{SN 2004C}
SN 2004C is a type IIb SN observed at radio frequencies. \cite{DeMarchi22} estimate an extended CSM at $\sim5\times10^{15}$ to $10^{16}$ cm consisting of at least 0.021 M$_\odot$ of density $1.5\times10^{-18}$ g cm$^{-3}$. They estimate that this corresponds to a mass-loss rate of at most $\sim5\times 10^{-3}$ M$_\odot$ yr$^{-1}$. Assuming a wind speed of 1000 km s$^{-1}$, dividing the radius by the wind velocity gives that the mass-loss episode must have occurred approximately 6.5 years prior to explosion.

\subsection{SN 2004cc}
SN 2004cc is a type Ic SN \citep{Class:2004cc} observed in the radio band. \cite{Wellons12} estimate that between 10 and 100 years prior to explosion, the progenitor ejected between $10^{-4}$ and $10^{-3}$ M$_\odot$ of material at a rate of $1.3\times10^{-4}$ M$_\odot$ yr$^{-1}$, under an assumed 1000 km s$^{-1}$ wind velocity. While the outer radius of the shell is unconstrained, \cite{Wellons12} calculate that the shell inner radius is between $2.6\times10^{15}$ and $2.7\times10^{16}$ cm.

\subsection{SN 2004dk}
SN 2004dk is a type Ib SN \citep{Class:2004dk1} that later showed H$\alpha$ emission from optical spectra and evolved into a type IIn \citep{Mauerhan18}, and showed strong H$\alpha$ emission out to 13 years post discovery. \citep{Mauerhan18} estimate an inner CSM radius of $\sim4\times10^{17}$ cm, corresponding to $\delta$ t = 1660 days at an assumed 0.1$c$ shock velocity, which was the beginning of radio rebrightening. As interaction is ongoing as of the latest spectra at $\delta$ t = 4684 days, they place a lower limit on the outer radius to be 10$^{18}$ cm. Additionally, \cite{Mauerhan18} assume a wind speed of 400 km s$^{-1}$ based on spectral feature measurements at late time, which would indicate the mass-loss episode occurred roughly 320 years prior to explosion. They also specify that an increased wind speed of 1000 km s$^{-1}$ would instead correlate to approximately 125 years prior to explosion. The mass-loss rate calculated from radio data indicates $6.3\times10^{-6}$ M$_\odot$ yr$^{-1}$. Using these measurements, we assume a wind density profile and estimate a CSM mass of 0.002 M$_\odot$ by integrating the density over the inner and outer radii.

\subsection{SN 2004gq}
SN 2004gq is a type Ib SN \citep{Class:2004gq} observed in the radio. \cite{Wellons12} estimate that between 10 and 100 years prior to explosion, the progenitor ejected material at a rate of $\sim9\times10^{-6}$ M$_\odot$ yr$^{-1}$, under an assumed 1000 km s$^{-1}$ wind velocity. While the outer radius of the shell is unconstrained, \cite{Wellons12} calculate that the shell inner radius is between $6.2\times10^{15}$ and $6.7\times10^{15}$ cm. 

\subsection{SNe 2005kj, 2006aa, 2006bo, 2006jd, 2006qq, 2008fq}

These SNe are of spectroscopic type IIn; optical photometric and spectroscopic data were acquired for each object. Table 21 of \cite{Taddia13} provides the wind speed, mass-loss rate, inner and outer radii of CSM, and the time before explosion. In the cases that both the mass-loss rate and time before explosion are estimated, we infer the mass of the CSM by multiplying their given values.

\subsection{SN 2006Y, 2006ai, 2016egz}
These SNe are of spectroscopic type IIP \citep{Class:2006Y/2006ai,Class:2016egz} with short-lived plateaus ($\sim10s$ of days) partially powered by CSM interaction. \cite{Hiramatsu20b} use optical/NIR light curves and spectroscopy to inform simulations and fit models to find best fit parameters of the CSM. All simulations use an assumed RSG wind velocity of 10 km s$^{-1}$, and find that all three SNe experienced a wind-driven mass-loss rate of 0.01 M$_\odot$ yr$^{-1}$. They find that SN 2006Y is best fit by a 0.01 M$_\odot$ CSM, SN\,2006ai with 0.03 M$_\odot$, and SN\,2016egz with 0.03 M$_\odot$. From this, they estimate SN\,2006Y experienced mass-loss for 10 years prior to explosion, while SN\,2006ai and SN\,2016egz for 30 years prior to explosion. We calculate by multiplying the time frame of mass loss by the wind speed that the CSM extends to $\sim3\times10^{14}$ cm for SN\,2006Y and $\sim9\times10^{14}$ cm for SN\,2006ai and SN\,2016egz.

\subsection{SN 2006gy}

SN 2006gy is a type IIn SLSN \citep{Class:2006gy}. \cite{Moriya13} fit the light curve of SN 2006gy using a variety of models with varying density profiles, FS velocity, and CSM location. \cite{Moriya13} report that wind-like profiles fail to fit the light curve, while a much steeper profile, $\rho_{\rm CSM} \propto R^{-5}$, and a constant density profile fit the light-curve better. All results are reported in Table 1 of \cite{Moriya13}. The average mass-loss rate of all models is 0.1 M$_\odot$ yr$^{-1}$, while the average CSM mass is 15 M$_\odot$. The average inner radius of the CSM is $\sim3\times10^{15}$ cm, and the average outer CSM radius is $\sim1.6\times10^{16}$ cm. Using the assumed wind velocity of 100 km s$^{-1}$, the CSM would have been ejected $\sim50$ years prior to explosion.

\subsection{SN 2007od}

SN 2007od is a type IIP  stellar explosion \citep{Class:2007od} that showed atypically steep declines after the plateau phase ($\delta t \approx 100$ days) in brightness and 'intermediate width' (1500 km s$^{-1}$) H$\alpha$ features in optical photometry and spectroscopy \citep{Andrews10}. \cite{Andrews10} show models that estimate the CSM inner radius is between 700 - 1700 AU (1-2.5$\times10^{16}$ cm) that had been ejected between 300 and 800 years prior to explosion from a 1500 km s$^{-1}$ wind velocity. Additionally, \cite{Andrews10} report that the multi-peaked nature of the H$\alpha$ emission is consistent with a ring or torus-shaped CSM. No estimate of mass or mass-loss rate is given.

\subsection{SN 2008es}
SN 2008es is a SLSN type II \citep{Class:2008es} that at early times showed no narrow features in its optical spectra. \cite{Bhirombhakdi19} estimate from optical and infrared photometry that its CSM mass is between 2-3 M$_\odot$ located between $\sim2\times10^{14}$ to $\sim2\times10^{15}$ cm. They also calculate that this would have required a mass-loss rate between 0.1-1 M$_\odot$ yr$^{-1}$ under their assumption of 100 km s$^{-1}$ wind velocity. This additionally gives a lookback time of 0.5-1.6 years prior to explosion. \cite{Bhirombhakdi19} also report that similar IR/optical measurements could be obtained by a magnetar spin-down model instead of a CSM powered luminosity.

\subsection{SN 2009ip Outburst A and B in 2012}
SN 2009ip is a type IIn SN for which observations exist across the electromagnetic spectrum (radio, optical, X-rays, gamma rays) \citep{Margutti14}. \cite{Margutti14} estimate that a dense CSM shell was located at $\sim5\times10^{14}$ cm from the progenitor and extended to $\sim4\times10^{16}$ cm.

\subsection{SN 2009kf}
SN 2009kf is a type IIP SN \citep{Class:2009kf} observed in the optical and UV bands. Through light-curve modeling, \cite{Moriya15} estimate that hydrogen-rich material extended to $\sim10^{15}$ cm and was the result of  mass-loss rate of 0.01 M$_\odot$ yr$^{-1}$ (under their assumption of 10 km s$^{-1}$ winds). We estimate by dividing the radius by the wind velocity that the progenitor would have to be losing mass for approximately 32 years prior to explosion, which would have produced $\sim 0.3$ M$_\odot$ of CSM.

\subsection{SNe 2010al and 2011hw}
SNe 2010al and 2011hw are both type Ibn SNe \citep{Pastorello15a,Class:2011hw} that were monitored in the UV, IR, and optical bands, both spectroscopically and photometrically. \cite{Pastorello15a} identify that SN\,2010al required re-classification based on a $\delta t \approx 12$ days spectrum to Ibn, and that SN\,2010al continued to show signs of interaction until the epoch of the last spectrum presented at $\delta t \approx 60$ days with a derived wind velocity of at least 1000 km s$^{-1}$. SN 2011hw, however, was not H-free and \cite{Pastorello15a} suggests it could be a transitional SN between Ibn and IIn, and present spectra from $\delta t \approx 1$ day to $\delta t \approx 72$ days with signs of interaction. These estimates, assuming a 10000 km s$^{-1}$ ejecta velocity, would place a constraint on the CSM inner radius within $\lesssim6\times10^{14}$ and $8\times10^{13}$ cm for SNe 2010al and 2011hw, respectively.

\subsection{SN 2010jl}

SN 2010jl is a particularly luminous type IIn SN \citep{Class:2010jl} that showed signs of an $\gtrsim$ 3 M$_\odot$ CSM in the IR, optical, UV, and X-rays, resulting from an extensive period of $\gtrsim 0.1$ M$_\odot$ yr$^{-1}$ mass loss \citep{Fransson14}. These estimates are lower limits due to uncertainties in the shock velocity. \cite{Fransson14} infer a wind velocity of $\sim 100$ km s$^{-1}$ from the optical spectra, which they argue rules out an RSG progenitor. Strong interaction continues until the last optical spectra at $\sim 1100$ days, placing a lower limit of the CSM outer radius to be $2\times10^{16}$ cm. Using the measured wind speed, this indicates the mass ejection had been occurring for at least $\sim60$ years prior to explosion. \cite{Fransson14} also discusses an IR light echo caused by dust at a distance of $6\times10^{17}$ cm, which if part of a previous mass ejection would indicate a lookback time of $\sim 1900$ years.

\subsection{SN 2010mb}
SN 2010mb is a type Ic stellar explosion \citep{Class:2010mb} that interacted with hydrogen-poor material observed in the optical and could possibly be reclassified as a type Icn, with lines of width $\sim800$ km s${-1}$. \cite{Ben-Ami14} observed SN\,2010mb in the optical and UV bands, obtaining both photometry and spectroscopy data. From their modeling, they estimate the CSM mass of approximately 3.3 M$_\odot$,  ejected 2.2 years prior to explosion. We estimate the radius of the CSM shell by multiplying their ejecta velocity (5000 km s$^{-1}$) by the time until interaction onset (150 days) to get $\sim6.5\times10^{15}$ cm. Similarly, we estimate the outer radius by using the time at which interaction ceases (750 days). \cite{Ben-Ami14} measure the wind speed to be 800 km s$^{-1}$, which we then use to calculate the mass-loss rate of $\sim$0.25 M$_\odot$ yr$^{-1}$ by setting the CSM mass equal to the integral of the wind density profile and integrating over the shell.

\subsection{SN 2010mc}
SN\,2010mc is a type IIn SN \citep{Class:2010mc} for which an outburst was observed 40 days prior to explosion. \cite{Ofek13} used optical spectra to model the surrounding CSM. They estimate the CSM was located between $7\times10^{14}$ cm and $10^{16}$ cm and contained 0.01 M$_\odot$. 

\subsection{PTF10aagc}

PTF10aagc is a SLSN that was originally a type I but later showed signs of strong H$\alpha$ emission \citep{Yan15}. \cite{Yan15} derive constraints on the surrounding CSM that caused the transition using optical spectroscopy and photometry and find that based on the width of the narrow H$\alpha$, the CSM had a velocity between 230 and 400 km s$^{-1}$. Additionally, \cite{Yan15} estimate an ejecta velocity of 13000 km s$^{-1}$. Because the H$\alpha$ emission appeared in a $\delta t \approx 322$ days, \cite{Yan15} estimate a CSM radius of $\sim4\times10^{16}$ cm and assume a CSM shell width of 10\%. In order to travel that distance at 230-400 km s$^{-1}$, \cite{Yan15} calculate the ejection must have occurred at least 40 years prior to explosion. \cite{Yan15} also report that due to a lack of absorption, they can place an upper limit of the CSM mass to be $\lesssim 30 M_\odot$.

\subsection{SN 2011fh}

SN 2011fh is a type IIn SN \citep{Class:2011fh} that showed a high degree of similarity to SN 2009ip both spectroscopically and photometrically \citep{Pessi21}, displaying multiple eruptive phases in the years following the initial ``explosion'' in 2011 in the NIR and optical. Prominent H$\alpha$ lines remain strong from the first spectrum at 3 days post explosion out to 1359 days post explosion. \cite{Pessi21} assume a CSM velocity of 100 km s$^{-1}$, which gives a mass-loss rate of $4\times10^{-2} M_\odot$ yr$^{-1}$.

\subsection{SN 2011ht}
SN 2011ht is a type IIn SN \citep{Class:2011ht}, modeled by \cite{Roming12} using optical and UV photometry and spectroscopy. They calculate that the CSM resides between $5\times10^{14}$ and $10^{15}$ cm and contains between 0.01 and 1 M$_\odot$. Using their assumed wind velocity of 600 km s$^{-1}$, they state this implies  a mass-loss rate of $3-5\times10^{-4}$ M$_\odot$ yr$^{-1}$ and the mass loss occurred $\sim$1 year prior to explosion. 

\subsection{SN 2011ja}

SN\,2011ja is a type IIP SN \citep{Class:2011ja} that showed late ($\sim$ 60-80 days post explosion) signs of H$\alpha$ and  H$\beta$ in optical spectra. \cite{Andrews16} interpret this as interaction with a disc-like CSM tilted at a 45$\degree$  angle from the viewing direction. \cite{Andrews16} estimate a mass-loss rate of 0.02-1$\times10^{-5}$ M$_\odot$ yr$^{-1}$ for their measured wind velocity of 180 km s$^{-1}$ and an inner CSM radius of $\sim(4-50)\times10^{15}$. Given the wind velocity, we infer a lookback time of $\sim8-14$ years prior to explosion, which in turn combined with the mass-loss rate would give a CSM mass of $\sim0.01-1\times10^{-4}$ M$_\odot$.

\subsection{iPTF11iqb}
iPTF11iqb is a type IIn SN that later evolved into a type IIL/P that showed evidence for a two-component  medium \citep{Smith15}. \cite{Smith15} suggest a CSM shaped by two different mass-loss rates: a mass-loss rate of 0.4$\times10^{-5}$ M$_\odot$ yr$^{-1}$ for  $(0.1-3)\times10^{15}$ cm, and a mass-loss rate of  1.5 $\times10^{-5}$ M$_\odot$ yr$^{-1}$ at $(3-300)\times10^{15}$. These distances correspond to upper limits of 8 years and 1000 years prior to explosion respectively, using their assumption of 100 km s$^{-1}$ winds. Finally, \cite{Smith15} calculate a  0.001 and 0.04 M$_\odot$ CSM mass for the inner and outer component, respectively.

\subsection{PTF11qcj}
PTF11qcj is a type Ic SN \citep{Class:PTF11qcj} with radio, IR, and optical light-curve coverage in addition to optical spectroscopy \citep{Corsi14}. \cite{Corsi14} calculate that the CSM inner radius is $\sim 10^{16}$ cm. They estimate the CSM had been ejected at a speed of 1000 km s$^{-1}$ with a mass-loss rate of $\sim10^{-4}$ M$_\odot$ yr$^{-1}$ approximately 2.5 years prior to explosion. While they do not estimate the outer radius, they do give a lower limit of $\sim2\times10^{17}$ cm.

\subsection{OGLE-2012-SN-006}
OGLE-2012-SN-006 is a type Ibn SN observed in the IR and optical bands both spectroscopically and photometrically. \cite{Pastorello15b}. \cite{Pastorello15b} show the light curve was significantly shallower than that expected of typical radioactive $^{56}$Co--$^{56}$Fe decay starting at 25 days post explosion and spectra showed signatures of H-poor interaction through $\sim190$ days, and measure a wind speed of 250 km s$^{-1}$. Assuming an ejecta velocity of 10000 km s$^{-1}$, this would imply an inner CSM radius of at least $\sim5\times10^{13}$ cm.

\subsection{SN 2012aa}
SN\,2012aa is a highly luminous type Ic SN \citep{Class:2012aa} that was shown to experience interaction with hydrogen-free material \citep{Roy16}. While SN\,2012aa did not reach the luminosity of a typical SLSN, it still reached magnitude -20 in the optical band, placing it at an intermediate luminosity between typical SN type Ic and SLSN. SN 2012aa went on to show a second bump in its optical light curve, which \cite{Roy16} explain as interaction with a dense CSM. They show that interaction begins approximately 65 days post explosion and ends 95 days post explosion. Using their assumed ejecta velocity of $10^4$ km s$^{-1}$, this translates to a shell located between $\sim5-8\times10^{16}$ cm. They estimate the CSM contains between 5-10 M$_\odot$, which we use to estimate a mass-loss rate between $\sim0.006-0.01$ M$_\odot$ yr$^{-1}$ using their assumed wind velocity of 1000 km s$^{-1}$. Similarly, we also use the same assumed wind velocity to estimate the lookback time as the radius divided by the velocity to get $\sim2-3$ years prior to explosion.

\subsection{SN 2013L}
SN 2013L is a type IIn SN \citep{Class:2013L} that showed prominent hydrogen spectral features and was observed across the UV, optical, and IR \citep{Taddia20}. \cite{Taddia20} estimate an enhanced mass-loss rate of 0.017-0.15 M$_\odot$ yr$^{-1}$ in the 25-40 years prior to explosion. Additionally, they calculate that this corresponds to a mass of 3.8-6.3 M$_\odot$ with a measured wind velocity of 120-240 km s$^{-1}$ extending from the progenitor surface to $\sim0.9-3\times10^{16}$ cm.

\subsection{SN 2013cu}
SN 2013cu is a type IIb \citep{Class:2013cu} that was discovered very early and was rapidly classified 15.5 hours post explosion. The SN already showed signs of interaction with a wind ($v_w \sim 100$ km s$^{-1}$) that was rich in both H and He \citep{Groh14}. \cite{Groh14} fit the optical spectrum and estimates a mass loss rate of $3\times10^{-3} $M$_\odot$ yr$^{-1}$. Their modeling also estimates the CSM inner radius $R_{\rm in} = 1.5\times10^{14}$ cm and extending to $2\times10^{15}$ cm, for a total mass of 0.022 M$_\odot$. Using the wind speed from \cite{Groh14} and their modeled radius, this would indicate a mass-loss period ranging from $\sim6-60$ years prior to explosion. \cite{Groh14} predicts that these properties could be the result of a Yellow Hyper Giant or Luminous Blue Variable star progenitor.

\subsection{SN 2013df}

SN 2013df is a type IIb SN \citep{Class:2013df}, showing signs of interaction in its spectrum within the first 10 days. \cite{Kamble16} observed SN 2013df in the radio and X-rays and estimate a mass-loss rate of $\sim0.7-1.4\times10^{-4}$ M$_\odot$ yr$^{-1}$ for an assumed wind velocity of 10 km s$^{-1}$.

\subsection{PS1-13arp}
PS1-13arp is a type IIP SN \citep{Class:PS1-13arp2} with an early UV excess of emission in addition to its optical light curve \citep{Class:PS1-13arp2,Haynie20}. \cite{Haynie20} model the light curve in the context of a shock breaking through a dense CSM located at $\sim1.3\times10^{14}$ cm and containing 0.08 M$_\odot$. Using that radius and their assumed wind speed of 10 km s$^{-1}$, it would require the material to have been ejected approximately 4 years prior to the explosion.

\subsection{LSQ13ddu}

LSQ13ddu is a type Ibn SN \citep{Smartt14} that  exhibited a very short interaction phase that produced prominent helium lines in the optical spectra. The phase of strong interaction started sometime between explosion and the first spectrum at 5 days and ended between 6 and 11 days post explosion \citep{Clark20}. Using the ejecta velocity from \cite{Clark20} of 34600 km s$^{-1}$ and the constraints on interaction times, this would place the CSM between $\sim1.5-3\times10^{15}$ cm. Additionally, \cite{Clark20} estimate the mass of the CSM to be 0.59 M$_\odot$.

\subsection{SN 2013fs}

SN\,2013fs is a type IIP SN \citep{Yaron17} that showed signs of interaction within hours of discovery. \cite{Yaron17} model the optical light curve and present optical spectra to show that the flash ionization spectra observed was likely due to a dense CSM extending from $\sim1.2-10\times10^{14}$ cm. Their models derive a CSM wind speed between 15-100 km s$^{-1}$, which indicate the mass-loss was occurring on a timescale of $\sim$1 year prior to explosion. By calculating the time it would have taken for the CSM to reach the edge the maximum distance of 10$^{15}$ cm, we find the progenitor star would have been shedding mass for $\sim3-20$ years prior up until explosion.  Considering the estimated mass-loss rate from \cite{Yaron17} of $0.3-4\times10^{-3}$ M$_\odot$ yr$^{-1}$, this would translate to a CSM mass between $\sim9\times10^{-4}$M$_\odot$ - $8\times10^{-2}$ M$_\odot$.

\subsection{iPTF13ehe, iPTF15esb, and iPTF16bad}
SLSNe of type I that later showed H$\alpha$ emission in their optical spectra \citep{Yan17}. \cite{Yan17} estimate their CSMs begin between $\sim4-9\times10^{16}$ cm and contain between 0.05-3 M$_\odot$. We estimate the lookback time by dividing the radius by assumed wind velocities of 100 and 1000 km s$^{-1}$ to give $\sim1-30$ years prior to explosion.

\subsection{SN 2014av}
SN 2014av is a type Ibn SN \citep{Pastorello16} observed in the optical photometrically and spectroscopically. \cite{Pastorello16} find that the spectrum taken at $\delta t \approx 18$ days post explosion show features of a Ibn, while the spectra at $\delta t \approx 10$ days is largely featureless. Assuming an ejecta velocity of 10000 km s$^{-1}$, this would place a CSM radius between 9-20$\times10^{14}$ cm.

\subsection{LSQ14efd, iPTF15dtg, 2020bvc}
Type Ic SNe that interacted with hydrogen-poor material. \citep{Jin21} used optical lightcurves to inform simulations and constrain parameters of possible CSM that could result in the observed lightcurves. Their simulations produced 0.1-0.2 M$_\odot$ of hydrogen poor material beginning at either 10$^{13}$ or 10$^{14}$ cm, which had been ejected $\sim 2.4$ months prior to explosion. Based on their assumed wind velocity of 200 km s$^{-1}$, they report this would correspond to a mass-loss rate of 0.6-13 M$_\odot$ yr$^{-1}$.

\subsection{ASASSN-14ms}

ASASSN-14ms is a highly luminous type Ibn \citep{Class:ASASSN-14ms} that begin interacting by the time the first optical spectrum (which was acquired at 7 days) and continued through the last spectrum at 44 days post explosion \citep{Vallely18}. \cite{Vallely18} model the optical photometry and extract a CSM mass of 0.51 M$_\odot$.

\subsection{iPTF14hls}
iPTF14hls is a type IIP SN \citep{Class:iPTF14hls} that showed extremely long-lasting optical emission and photometric variability \cite{Arcavi17}. A potential interpretation of this phenomenology is the SN shock interaction with a complex medium. Within this scenario, \cite{Arcavi17} estimate the CSM contains tens of solar masses due to an ejection a few years prior to explosion.

\subsection{SN 2015da}
SN 2015da is a type IIn SN \citep{Class:2015da} that was observed to have IR echoes due to a surrounding CSM shell \citep{Tartaglia20a}. \cite{Tartaglia20a} estimate, using optical light curve modeling and spectra, the shell contains between $\sim5-10$ M$_{\odot}$ due to enhanced mass-loss rate of 0.6-0.7  M$_\odot$ yr$^{-1}$.

\subsection{ASASSN-15no}

ASASSN-15no is a type Ic SN \citep{Class:ASASSN-15no:Early}, but later spectra from \cite{Benetti18} revealed the development of broad H lines in optical spectra, which led to a re-classification as type II SN. Combined with optical photometric data, \cite{Benetti18} interpret its spectral evolution and photometric data as evidence for a two component structured CSM, composed of two shells: a H-rich CSM inner shell and a H-poor outer shell. \cite{Benetti18} estimate the inner CSM to be a shell at $\sim2\times10^{14}$ cm of 1-2 M$_\odot$ while the outer H-poor shell has an inner radius  $\sim3\times10^{15}$ cm and extends to $\sim8\times10^{15}$ cm. Using an assumed wind velocity of 10-100 km s$^{-1}$, \cite{Benetti18} calculate lookback times of 0.7-7 and 9-90 years for the inner and outer CSM, respectively.

\subsection{iPTF16eh}
iPTF16eh is a type I SLSN based on UV observations and optical spectroscopy \citep{Lunnan18}. \cite{Lunnan18} modeled the CSM structure based on simulations and estimate the dense hydrogen-poor shell has inner radius of $\sim3\times10^{17}$ cm that extends to $\sim3.5\times10^{17}$ cm and was ejected $\sim32$ years prior to explosion. They calculate the shell contains 1-10 M$_\odot$ with a measured wind speed of 3300 km s$^{-1}$.

\subsection{SN 2016aps}

SN 2016aps is a type IIn SLSN \citep{Class:2016aps} with optical photometry and spectroscopy data. The extreme energy output by this SLSN is thought to be a result of massive ejecta colliding with an extremely massive CSM. There are two separate models seeking to fit the optical light curve from \cite{Nicholl20} and \cite{Suzuki21}, which estimate a CSM mass of $\sim150$ M$_\odot$ (minimum of 40 M$_\odot$) and 8 M$_\odot$, respectively. \cite{Suzuki21}, using a radiation hydrodynamics simulation code designed for SLSNe, find that an outer CSM radius of 10$^{16}$ cm best fits the data. This would indicate that for a wind range of 10 to 1000 km s$^{-1}$, the beginning of mass loss would have to be between $\sim$3-300 years prior to explosion. We emphasize the calculations from \cite{Suzuki21} due to the use of specific numerical models.

\subsection{SN 2016iet}
SN 2016iet shows two optical peaks separated by approximately 100 days \citep{Gomez19}. \cite{Gomez19} estimate based on optical light-curve modeling and spectroscopy that the CSM was extremely massive at $\sim38$ M$_\odot$ located between $1-7\times10^{14}$ cm, which corresponds to a mass-loss rate of 7 M$_\odot$ yr$^{-1}$ assuming a 100 km s$^{-1}$ velocity. They calculate that this would require ejection between 2-7 years prior to explosion, but also note that if the assumed velocity is increased to 1000 km s$^{-1}$ this would place ejection within 70-260 days prior to explosion.

\subsection{AT 2016jbu}

AT 2016jbu is thought to be an SN-impostor event akin to 2009ip before the 2012  \citep{Class:2016jbu}, that appeared spectroscopically similar to a type IIn SN. In two papers, \cite{Brennan21a} and \cite{Brennan21b} propose that the event likely occurred as a result of a highly asymmetric outburst of 0.05-0.14 M$_\odot$ yr$^{-1}$ with an assumed wind velocity of 250 and 750 km s$^{-1}$ based on the overall luminosity and narrow H$\alpha$ lines. 

\subsection{SN 2017ahn}

SN 2017ahn is a type II SN \citep{Class:2017ahn} that displayed signs of interaction in its spectra within 6 days of explosion. \cite{Tartaglia21} use the spectra and optical light-curve modeling to predict a mass-loss rate of 2.7-4$\times10^{-3}$ M$_\odot$ that generated a CSM of inner radius $\sim2\times10^{13}$ cm. Using the assumed wind speed from \cite{Tartaglia21} of 150 km s$^{-1}$, we infer a lookback time of $\sim$ 18 days.

\subsection{SN 2017dio}
SN 2017dio is a type Ic SN \citep{Class:2017dio1} that later evolved into a type IIn, showing narrow H and He lines in its optical spectra \citep{Kuncarayakti18}. \cite{Kuncarayakti18} estimate a mass-loss rate of $\sim0.02$ M$_\odot$ yr$^{-1}$ that peaked in the few decades prior to explosion. They estimate a 10$^4$ km s$^{-1}$ ejecta speed, which based on the lack of narrow features in the $\delta$t = 6 days spectrum and the presence of narrow features indicates  the shockwave would have traveled between $\sim 0.5-1\times10^{15}$ cm before encountering the CSM. While there is no estimate for the outer radius, we can at least  place a lower limit as the last spectra taken presented was 83 days post explosion, corresponding to at least $\sim7\times10^{15}$ cm. The CSM has a measured speed of 500 km s$^{-1}$, which integrating over a wind  density profile from the inner radius to the lower limit produces a minimum CSM mass of $\sim0.089$ M$_\odot$.

\subsection{SN 2017eaw}
SN 2017eaw is a type IIP \citep{Class:2017eaw} that showed extensive signs of interaction at 900 days post explosion from a transition of a narrow H$\alpha$ feature to a box-shaped H$\alpha$ feature \citep{Weil20}. \cite{Weil20} calculate from optical spectroscopy and light curve modeling that 2017eaw had a mass-loss rate of $3\times10^{-6}$ M$_\odot$ yr$^{-1}$ at least 1700 years prior to explosion. They also estimate the CSM shell begins at $\sim5.3\times10^{16}$ cm from their assumed 10 km s$^{-1}$ wind velocity of red supergiants, and while additional observations have not been taken to determine the extent of the CSM, if we assume the shell extends to a similar distance as SN 2014C ($\sim10^{17}$ cm), we estimate a mass of approximately 0.004 M$_\odot$.

\subsection{SN 2017ens}
SN 2017ens is a superluminous type Ic-BL that later evolved into a type IIn \citep{Class:2017ens} as revealed by optical and UV observations in addition to spectroscopy \citep{Chen18}. \cite{Chen18} estimate that the explosion interacted with a dense shell located between $\sim10^{15}$ and $7\times10^{16}$ cm. They were also able to measure a wind velocity of between 50-60 km s$^{-1}$ and therefore calculate that the mass-loss rate was $\sim5\times10^{-4}$ M$_\odot$ yr$^{-1}$. Assuming a wind density profile, we estimate the amount of mass contained within the shell by integrating the wind density profile over the extent of the shell to get $\sim0.06$ M$_\odot$. Additionally, we estimate the lookback time by diving the radius by the wind speed to get a lookback time of $\sim7$ years.

\subsection{SN 2017gmr}

SN 2017gmr is a type IIP \citep{Class:2017gmr} that did not show narrow H features in optical spectra at early times ($\sim$ 2 days post explosion), but did have an bump in the U and B filters that \cite{Andrews19}  suggest to be asymmetric CSM interaction. Interaction began at least 2 days post explosion and is ongoing as of the last spectrum presented in \cite{Andrews19} at 180 days post explosion. From the combined UV, optical, and IR light curve, \cite{Andrews19} estimate a progenitor radius (and therefore lower limit on the CSM radius) of $\sim3\times10^{13}$ cm that extends to an outer radius of 1.8$\times10^{14}$ cm. From this radius and a measured wind velocity of 55 km s$^{-1}$, \cite{Andrews19} predict a lookback time of ``years to decades.''

\subsection{SN 2017hcc}
SN 2017hcc is a type IIn SLSN \citep{Class:2017hcc} for which a highly asymmetric CSM structure was invoked in order to explain the irregular H emission and absorption features \citep{Smith20}. \cite{Smith20} measure a 50 km s$^{-1}$ wind velocity, and an H$\alpha$ component out to at least 848 days post explosion. Using the measured broad line H$\alpha$ emission that \cite{Smith20} associate to the ejecta of 6000 km s$^{-1}$, that would place a lower limit on the outer radius of 4.4$\times10^{16}$ cm. \cite{Smith20} estimate that the enhanced mass-loss episode was occurring for 6-12 years prior to explosion at a rate of 1.4 M$_\odot$ yr$^{-1}$, translating to 8-16 M$_\odot$ of CSM.

\subsection{SN 2018zd}
SN\,2018zd is a type IIP SN that shows hydrogen lines at 4.9 days post explosion and eventually developed a plateau in its optical light curve and spectra \citep{Hiramatsu20}. \cite{Hiramatsu20} estimate that approximately 10 years prior to explosion, SN 2018zd ejected 0.1 M$_\odot$, at a rate of 0.01 M$_\odot$ yr$^{-1}$. Using their assumed wind velocity of 20 km s$^{-1}$, we calculate that the shell would extend to $\sim6\times10^{14}$ cm by multiplying the lookback time by the wind velocity.

\subsection{SN 2018gep}

SN 2018gep is a Ic-BL stellar explosion \citep{Class:2018gep} with an exceptionally short rise time of 0.5-3 days \citep{Ho19}, placing it amongst the Fast Blue Optical Transients (FBOTs). \cite{Leung21} estimate this was due to a $\sim0.3$ M$_\odot$ CSM approximately beginning $\sim0.7-1.5\times10^{15}$ cm from the progenitor based on simulations of the optical light curve.

\subsection{SN 2018gjx}
SN 2018gjx  went through three distinct phases: (I) a hot blue spectrum with signatures of ionized CSM, (II) signatures of a type IIb, then (III) interaction with a helium-rich CSM that led to the classification as a type Ibn SN \citep{Prentice20}. \cite{Prentice20} calculate that the CSM is composed of 0.004-0.014 M$_\odot$ of material located in a thin shell between $\sim3.8\times10^{14}$ and $\sim4.5\times10^{14}$ cm. They assume a wind speed between 150-500 km s$^{-1}$, which we use to estimate a lookback time by diving the radius by the wind velocity to get $\sim0.2-0.8$ years prior to explosion. Similarly, using the assumed wind velocity, \cite{Prentice20} calculate a mass-loss rate of $(5-510)\times10^{-4}$ M$_\odot$ yr$^{-1}$.

\subsection{SN 2018ijp}
SN 2018ijp is a type Ic-BL stellar explosion \citep{Class:2018ijp} that was discovered by ZTF.  Follow-up data include optical photometry and spectroscopy \cite{Tartaglia20b}. Modeling these data,  \cite{Tartaglia20b} estimate a dense CSM located between $\sim4.6\times10^{15}$ and $5\times10^{16}$ cm. The CSM contains 0.5 M$_\odot$ and corresponds to a mass-loss rate of 0.2 M$_\odot$ yr$^{-1}$ originating 10-100 years prior to explosion.

\subsection{SNe 2019jc, 2021ckj}

SNe 2019jc and 2021ckj are both type Icn stellar explosion \citep{Pellegrino22} with optical photometry and spectroscopy. \cite{Pellegrino22} model the light curve of SN\,2019jc as being powered by shock interaction with a wind-density profile $(\rho_{\rm CSM}(R) \propto R^{-2})$, and find the CSM contains $0.58$ M$_\odot$ of H- and He-poor material at an inner radius of $\sim4.04\times10^{14}$ cm. \cite{Pellegrino22} also measure a wind speed between 500 and 1000 km s$^{-1}$, which would implies a lookback time of $\sim26$ days. SN\,2021ckj has sparser data: \cite{Pellegrino22} present a spectrum at $\sim 13$ days with evidence for interaction. Assuming a minimum ejecta speed of 7000 km s$^{-1}$ (as for SN\,2019jc, \citealt{Pellegrino22}), this ejecta speed implies a rough CSM distance of $\sim 7.7\times10^{14}$ cm.

\subsection{SN 2019uo}
SN 2019uo is a type Ibn SN \citep{Class:2019uo}, for which the CSM properties were inferred by \cite{Strotjohann21} through optical light-curve modeling. SN\,2019uo was observed to have an outburst 320 days prior to explosion, which ejected at most 0.8 M$_\odot$. It is unknown how far the CSM extends, but \cite{Strotjohann21} estimate inner shell radius to be between $3\times10^{15}$ and $10^{16}$ cm, with a measured wind velocity of 880 km s$^{-1}$.

\subsection{SN\,2019cad}

SN\,2019cad is a type Ic stellar explosion \citep{Class:2019cad} that had a second peak in its optical lightcurve approximately 45-60 days post explosion and showed prominent C and Si lines in its spectra \citep{Gutierrez21}. Using the ejecta velocity derived in \cite{Gutierrez21} of 14000 km s$^{-1}$ and the time and duration of the second peak, we infer that a CSM shell could exist between $\sim2-6\times10^{15}$ cm.

\subsection{SN 2019ehk}

SN 2019ehk is a  Calcium-rich transient \citep{Class:2019ehk} that showed signs of shock interaction with a H-rich CSM in its optical spectra $\sim1.5$ days post explosion.
The CSM mass inferred from optical and X-ray observations is  $\sim$0.007 M$_\odot$ for a wind velocity of $\sim500$ km s$^{-1}$  \citep{Jacobson-Galan20}. The spectral features associated with shock-interaction   vanished by $\delta t \approx 2.4$ days, 
indicating an end to interaction with the dense CSM \citep{Jacobson-Galan20}. This phase experienced a mass-loss rate of $\sim 0.01$ M$_\odot$ yr$^{-1}$ for $\sim0.6$ years prior to explosion, using the 500 km s$^{-1}$ wind velocity from \cite{Jacobson-Galan20}. Additionally, \cite{Jacobson-Galan20} estimate an upper limit for the mass-loss rate of $10^{-5}$ M$_\odot$ yr$^{-1}$ beyond the thick CSM shell at $10^{16}-10^{17}$ cm based on radio observations. 
Since the nature of SN\,2019ehk as a core-collapse event is  disputed, and a connection with white-dwarf explosions is not excluded, we leave SN\,2019ehk out of Figures \ref{Fig:InteractComp}, \ref{Fig:CSMRad}, and \ref{Fig:Burn}. Similarly, we leave SN\,2021gno, another Calcium-rich SN \citep{Jacobson-Galan22b} with detected X-ray emission out of the sample.

\subsection{SN 2019hgp}

SN 2019hgp is a type Icn SN \citep{Class:2019hgp} that within a day of explosion began interacting with a 1900 km s$^{-1}$ wind rich in C and O \citep{Gal-Yam22}. \cite{Gal-Yam22} estimate that a CSM mass of 0.2 M$_\odot$ CSM, resulting from a mass-loss rate of 0.004 M$_\odot$ yr$^{-1}$. Using their modeled wind speed and the onset of interaction of $\sim$ 1 day and the end of interaction at $\sim 6$ days as shown by spectra in \cite{Gal-Yam22}, this would place the CSM between $\sim10^{13}-10^{14}$ cm. Again using the 1900 km s$^{-1}$ wind speed, this would correlate to a mass-loss episode within a month before explosion.

\subsection{SN 2019oys}

SN 2019oys is a type Ib SN that transitioned to a type IIn, much like SN 2014C \citep{Class:2019oys1,Sollerman20}. This transition occurred $\sim100$ days post explosion and spectra still showed broad and strong H$\alpha$ to at least 200 days post explosion.

\subsection{SN 2019yvr}

SN 2019yvr is a type Ib \citep{Class:2019yvr1}, but later interacted with a H-rich CSM. \cite{Kilpatrick21} find that the optical photometry and spectroscopic data are best fit by a two-part ejection model, one that occurred 50-100 years prior to explosion and removed most of the H-envelope, followed by a minor outburst within the last 2.6 years that removed the remainder of the H-envelope. Using stellar models combined with spectroscopy, \cite{Kilpatrick21} estimate that the final eruption removed a 0.01-0.03 M$_\odot$ H-envelope from the star. The previous outburst does not have an estimated mass, but \cite{Kilpatrick21} calculate a mass-loss rate of 1.3$\times10^{-4}$ M$_\odot$ yr$^{-1}$ for an assumed wind speed of 100 km s$^{-1}$. Based on the onset time of interaction, \cite{Kilpatrick21} also estimate an inner CSM radius of $\sim1000$ AU, or $\sim1.5\times10^{15}$ cm.

\subsection{SN 2020oi}

SN 2020oi is a type Ic SN \citep{Class:2020oi} that showed deviations from steady-state wind mass loss in the radio. SN 2020oi has been observed extensively in the optical and the UV by \cite{Maeda21} and \cite{Gagliano22}. \cite{Maeda21} estimate that the CSM has strong fluctuations compared to the expected power-law distribution within 10$^{15}$ cm with a mass-loss rate of $\sim0.3-1\times10^{-3}$ M$_\odot$ yr$^{-1}$ and then returns to a smooth power-law beyond 10$^{16}$ cm. According to \cite{Maeda21}, this places the erratic mass loss within 1 year of explosion based on the assumption of 1000 km s$^{-1}$ velocities. We in turn thus infer a CSM mass of at most 0.001 M$_\odot$.

\subsection{SN 2020faa}

SN 2020faa is a type II SN \citep{Class:2020faa} for which optical photometry and spectroscopy has been acquired. H$\alpha$ appeared in its spectrum 12 days post explosion, but was absent at least 6 days post explosion \citep{Yang21}. Based on calculations from \cite{Yang21} of the blackbody radius at early times, we estimate that the CSM inner radius is $\sim10^{15}$ cm.

\subsection{SN 2020pni}
SN 2020pni is a type II SN \citep{Class:2020pni}. Modeling of optical-UV photometry and optical spectroscopy by \cite{Terreran22} led to the inference of a CSM shell extending to $\sim1.3\times10^{15}$ cm and containing between 0.04 and 0.12 M$_\odot$. \cite{Terreran22} assume a wind velocity of 200 km s$^{-1}$, which corresponds to a mass-loss rate of 0.02-0.08 M$_\odot$ yr$^{-1}$ for approximately 2 years prior to explosion.

\subsection{SN 2020tlf}
SN 2020tlf is a type IIP/L stellar explosion\citep{Class:2020tlf}  with detected pre-explosion eruptions, as well as early spectroscopic signatures of shock interaction and luminous UV emission  \cite{Jacobson-Galan22}. By modeling these observations \cite{Jacobson-Galan22} infer 
the presence of dense CSM with an extent  to 10$^{15}$ cm produced by a mass-loss rate of $\sim10^{-2}$ M$_\odot$ yr$^{-1}$ for an assumed wind speed of 50 km s$^{-1}$. 
\cite{Chugai22} provide further constraints, estimating a CSM mass of 0.2 M$_\odot$ that had been ejected within 6 years prior to explosion.

\subsection{SN 2021csp}

SN 2021csp is a type Icn explosion \citep{Class:2021csp} that showed prominent CIII lines of velocity 1800 km s$^{-1}$ \citep{Fraser21}. \cite{Fraser21} estimate the CSM to extend to $\sim400\,R_\odot$ ($2\times10^{13}$ cm) and contain $\sim1$ M$_\odot$ of H- and He-poor material. We infer from the wind velocity that this material was ejected only $\sim2$ days prior to explosion.

\section{Tables}
\label{Appendix:Tables}

\begin{deluxetable*}{lccccccccc}[h!]
\tabletypesize{\scriptsize}
\setlength{\tabcolsep}{10pt}
\tablecolumns{7} 
\tablewidth{30pc}
\tablecaption{Chandra X-ray observations of SN\,2014C
\label{Tab:XrayCXOInfo}}
\tablehead{\colhead{Chandra ID} & \colhead{Date} & \colhead{Date} & \colhead{Exposure} & \colhead{Phase} & \colhead{Count-Rate}& \colhead{Source Significance} &\\[0.0 cm]
 &  (MJD) & (UT) & (ks) & (days) & 0.5-8 keV ($\rm{c\,s^{-1}}$) & $\sigma$ & } 
\startdata
16005 & 56964.33 & 11/3/14 & 12.4 & 308.33 & $1.1 (\pm 0.09) \times 10^{-2}$ & 46.4  \\ 
\hline
17569 & 57052.69 & 1/30/15 & 13.3 & 396.69 & $1.9 (\pm 0.12)\times 10^{-2}$ & 93.2 \\ 
\hline
17570 & 57133.01 & 4/21/15 & 12.5 & 477.01 & $1.5 (\pm 0.13) \times  10^{-2}$ & 106.9  \\
\hline
17571 & 57262.43 & 8/28/15 & 12.1 & 606.43 & $2.8 (\pm 0.15) \times  10^{-2}$ & 117.9 \\
\hline
18340 & 57513.25 & 5/5/16 & 29.9 & 857.25 & $4.4 (\pm 0.12) \times  10^{-2}$ &  296.0 \\
\hline
18341 & 57685.59 & 10/24/16 & 32.2 & 1029.59 & $4.7 (\pm 0.12) \times  10^{-2}$ & 115.7 \\
\hline
18342 & 57913.94 & 6/9/17 & 29.8 & 1257.94 & $5.0 (\pm 0.13) \times  10^{-2}$ & 340.1 \\
\hline
18343 & 58230.77 & 4/22/18 & 12.0 & 1574.77 & $4.1 (\pm 0.18) \times  10^{-2}$ & 154.1 \\
\hline
21077 & 58224.68 & 4/16/18 & 22.3 & 1568.68 & $4.7 (\pm 0.15) \times  10^{-2}$ & 225.3  \\
\hline
21639 & 58627.05 & 5/24/19 & 32.1 & 1971.05 & $4.0 (\pm 0.11) \times 10^{-2}$ & 279.4 \\
\hline
21640 & 58955.57 & 4/16/20 & 20.1 & 2300.57 & 3.2 $(\pm 0.13) \times 10^{-2}$ & 171.9 \\
\hline
23216 & 58957.76 & 4/18/20 & 13.1 & 2301.76 & 3.2 $(\pm 0.16) \times 10^{-2}$ & 143.2 \\
\hline
\enddata
\end{deluxetable*}

\begin{deluxetable*}{lccccccccc}[h!]
\tabletypesize{\scriptsize}

\setlength{\tabcolsep}{10pt}
\tablecolumns{7} 
\tablewidth{24pc}

\tablecaption{Best-fitting parameters from our spectral analysis of Chandra X-ray observation of SN\,2014C 
\label{Tab:XrayCXOFit}}
\tablehead{\colhead{Chandra ID} & \colhead{Phase} & \colhead{$\Gamma$}&  \colhead{$NH_{\rm{int}}$}& \colhead{Flux}& \colhead{Unabsorbed Flux} &\\[0.0 cm]
 & (Days) & & ($10^{22} \rm{ cm^{-2}}$) & 0.3-10 keV $(\rm{erg\,s^{-1}\,cm^{-2}})$ & 0.3-10 keV $(\rm{erg\,s^{-1}\,cm^{-2}})$ }

\startdata
16005 & 308 & $-0.52^{+0.34}_{- 0.30}$ & $0.29^{+0.63}_{-0.29}$ & $7.6^{+0.46}_{- 2.2} \times 10^{-13}$ & $7.7^{+0.46}_{- 2.2} \times 10^{-13}$ \\ 
\hline
17569 & 397  & $0.16 \pm 0.25$ & $0.49^{+0.43}_{-0.41}$ & $1.1^{+ 0.064}_{- 0.18} \times  10^{-12}$ & $ 1.1^{+ 0.064}_{- 0.18} \times 10^{-12}$ \\ 
\hline
17570 & 477 & $-0.0063^{+ 0.18}_{- 0.14}$ & $0.057^{+0.20}_{-0.057} $ & $1.2^{+ 0.096}_{- 0.15} \times 10^{-12}$ & $1.2^{+ 0.096}_{- 0.15} \times 10^{-12}$ \\
\hline
17571 & 606 & $0.72 \pm 0.18$ & $0.56^{+0.23}_{-0.21}$ & $9.3^{+ 0.69}_{- 1.0} \times 10^{-13}$ & $1.0^{+ 0.069}_{- 0.1} \times 10^{-12}$\\
\hline
18340 & 857 & $0.83 \pm 0.08$ & $0.40^{+0.095}_{-0.090}$ & $1.2^{+0.059}_{-0.059} \times 10^{-12}$ & $1.3^{+0.059}_{-0.059} \times 10^{-12}$ \\
\hline
18341 & 1030 & $0.93 \pm 0.08$ & $0.31^{+0.083}_{-0.079}$ & $1.2^{+ 0.055}_{- 0.056} \times  10^{-12}$ & $1.3^{+ 0.055}_{- 0.056} \times 10^{-12}$ \\
\hline
18342 & 1258 & $1.0 \pm 0.08$ & $0.24^{+0.078}_{-0.075}$ & $1.1^{+ 0.049}_{- 0.053} \times  10^{-12}$ & $1.3^{+ 0.049}_{- 0.053} \times  10^{-12}$\\
\hline
21077 and & 1571 & $1.1^{+ 0.083}_{- 0.082}$ & $0.20^{+0.079}_{-0.076}$ & $1.1^{+0.046}_{- 0.053} \times  10^{-12}$ & $1.2^{+0.046}_{- 0.053} \times  10^{-12}$ \\
18343 \\
\hline
21639  & 1971 & $1.3^{+ 0.095}_{- 0.093}$ & $0.083^{+0.089}_{-0.083}$ & $8.4^{+0.35}_{- 0.44} \times  10^{-13}$ &
$9.2^{+0.38}_{- 0.48} \times  10^{-13}$\\
\hline
21640 and & 2301 & $1.3^{+0.107}_{-0.105}$ & $0.15^{+0.102}_{-0.098}$ & $7.2^{0.40}_{-0.49} \times 10^{-13}$ & $8.0^{+0.44}_{-0.54} \times 10^{-13}$ \\
23216 \\
\hline
\enddata
\end{deluxetable*}


\begin{deluxetable*}{lccccccc}[h!]
\tabletypesize{\scriptsize}
\setlength{\tabcolsep}{10pt}
\tablecolumns{7} 
\tablewidth{20pc}
\tablecaption{NuSTAR X-ray Observations of SN\,2014C. \label{Tab:XrayNuSTAR}}
\tablehead{\colhead{NuSTAR ID} & \colhead{Date} & \colhead{Date} & \colhead{Exposure Time} & \colhead{Phase} & \colhead{Count-Rate} & \colhead{Source Significance}  \\[0.0 cm]
 & (MJD) & (UT) & (ks) & (days) & 3-79 keV (c s$^{-1}$) & $\sigma$ }
\startdata
80001085002 & 57051.86 & 1/29/15 & 32.5  & 396.86 & 1.4 $(\pm 0.07) \times 10^{-2}$ & 18.9 \\
\hline
40102014001 & 57122.43 & 4/10/15  & 22.4 & 466.43 & 1.2 $(\pm 0.08) \times 10^{-2} $ & 15.1 \\
\hline
40102014003 & 57263.10 & 8/29/15  & 30.2 & 607.10 & 1.3 $(\pm 0.08) \times 10^{-2}$ & 17.2\\
\hline
40202013002 & 57511.78 & 5/3/16 & 43.0 & 856.78 & 1.6 $(\pm 0.07) \times 10^{-2}$ & 24.0 \\
\hline
40202013004 & 57693.46 & 11/1/16 & 40.9 & 1037.46 & 1.7 $(\pm 0.07) \times 10^{-2}$ & 23.9 \\
\hline
40302002002 & 57920.23 & 6/16/17 & 42.3 & 126.23 & 1.3 $(\pm 0.06) \times 10^{-2}$ & 21.3 \\
\hline
40302002004 & 58242.67  & 5/4/18 & 40.2 & 1587.67 & 1.4 $(\pm 0.07) \times 10^{-2}$ & 20.9\\
\hline
40502001002 & 58635.67 & 6/1/19 & 41.5 & 1980.67 & 9.1 $(\pm 0.5) \times 10^{-3}$ & 16.8\\
\hline
40502001004 & 58969.80 & 4/30/20 & 54.2 & 2314.80 & 7.8 $(\pm 0.4) \times 10^{-3}$ & 17.5 \\
\hline
\enddata

\end{deluxetable*}

\text{}

\begin{deluxetable*}{lcccccccc}[h!]
\tabletypesize{\scriptsize}
\rotate
\setlength{\tabcolsep}{16pt}
\tablecolumns{6} 
\tablewidth{30pc}
\tablecaption{Joint Chandra and NuSTAR X-ray spectral analysis of SN\,2014C.
\label{Tab:Xrayjoint}}
\tablehead{\colhead{Chandra ID} & \colhead{NuSTAR ID} & \colhead{Phase} & \colhead{Temperature}& \colhead{$NH_{\rm{int}}$}& \colhead{Flux}& \colhead{Unabsorbed Flux} &\\[0.0 cm]
 &  & (days) & keV & $10^{22} \rm{cm^{-2}}$  & 0.3--100 keV $(\rm{erg\,s^{-1}\,cm^{-2}})$ & 0.3--100 keV $(\rm{erg\,s^{-1}\,cm^{-2}})$} 
\startdata
17569 & 80001085002 & 396 & $18.15^{+3.6}_{-2.7}$ & $2.69^{+0.34}_{-0.32}$  & $1.43^{+0.17}_{-0.19} \times 10^{-12}$ & $1.83^{+0.21}_{-0.24} \times 10^{-12}$ \\
\hline
17570 & 40102014001 & 477 & $23^{+7}_{-4.6}$ & $1.68^{+ 0.23}_{-0.22}$ & $1.48^{+0.24}_{-0.29} \times 10^{-12}$ & $1.75^{+0.29}_{-0.34} \times 10^{-12}$ \\
\hline
17571 & 40102014003 & 606 & $22.8^{+6}_{-4.3}$ & $1.30^{+0.17}_{-0.17}$ & $1.58^{+0.26}_{-0.30} \times 10^{-12}$ & $1.84^{+0.30}_{-0.35} \times 10^{-12}$ \\
\hline
18340 & 40202013002 & 857 & $18.86^{+2.77}_{-2.19}$ & $0.942^{+0.074}_{-0.077}$ & $1.79^{+0.15}_{-0.17} \times 10^{-12}$ & $2.10^{+0.18}_{-0.20} \times 10^{-12}$ \\
\hline
18341 & 40202013004 & 1029 & $19.2^{+2.95}_{-2.30}$ & $0.72^{+0.063}_{-0.062}$ & $1.84^{+0.16}_{-0.20} \times 10^{-12}$ & $2.11^{+0.18}_{-0.23} \times 10^{-12}$ \\
\hline
18342 & 40302002002 & 1257 & $17.1^{+2.38}_{-1.94}$ & $0.541^{+0.057}_{-0.056}$ & $1.69^{+0.16}_{-0.15} \times 10^{-12}$ & $1.93^{+0.18}_{-0.17} \times 10^{-12}$ \\
\hline
21077 and 18343 & 40302002004 & 1571 & $14.9^{+1.9}_{-2.2}$ & $0.469^{+0.057}_{-0.055}$ & 
$1.50^{+0.11}_{-0.11} \times 10^{-12}$ & $1.72^{+0.13}_{-0.13} \times 10^{-12}$ \\
\hline
21639 & 40502001002 & 1971 & $13.1^{+1.97}_{-1.56}$ & $0.181^{+0.059}_{-0.055}$ &
$1.15^{+0.06}_{-0.12} \times 10^{-12}$ &
$1.27^{+0.07}_{-0.13} \times 10^{-12}$\\
\hline
21640 and 23216 & 40502001004 & 2307 & $11.3^{+1.47}_{-1.19}$ & $0.311^{+0.068}_{-0.066} $& $8.92^{+0.72}_{-0.72} \times 10^{-13} $& $1.03^{+0.84}_{-0.84} \times 10^{-12} $\\
\hline
\enddata
\end{deluxetable*}

\begin{deluxetable*}{lcccccccc}
\tabletypesize{\scriptsize}
\setlength{\tabcolsep}{12pt}
\tablecolumns{4}
\tablewidth{18pc}
\tablecaption{Properties of the Iron Emission from combined \emph{CXO} and \emph{NuSTAR}
 \label{Tab:Iron}}
\tablehead{\colhead{Phase} & \colhead{Central Energy} & \colhead{FWHM} & \colhead{Flux}\\[0.0 cm]
(Days) & (keV) & (keV) & 6.5--7.1 keV (erg s$^{-1}$ cm$^{-2}$)}
\startdata
396 & 6.77 $^{+0.044}_{-0.043}$ & 0.477 $^{+0.10}_{-0.10}$  & $1.02 ^{+0.14}_{-0.13} \times 10^{-13}$ \\
\hline
477 & 6.80 $^{+0.048}_{-0.047}$ & 0.531 $^{+0.11}_{-0.089}$ & $1.11 ^{+0.16}_{-0.15} \times 10^{-13}$ \\
\hline
606 & 6.79 $^{+0.041}_{-0.041}$ & 0.404 $^{+0.11}_{-0.10}$ & 1.01 $^{+0.14}_{-0.14} \times 10^{-13}$ \\
\hline
857 & 6.72 $^{+0.027}_{-0.029}$ & 0.241 $^{+0.11}_{-0.10}$ & 8.33 $^{+1.1}_{-1.1} \times 10^{-14}$ \\
\hline
1029 & 6.73 $^{+0.026}_{-0.026}$ & 0.461 $^{+0.065}_{-0.60}$ & 1.41 $^{+0.13}_{-0.12} \times 10^{-13}$ \\
\hline
1257 & 6.75 $^{+0.027}_{-0.028}$ & 0.364 $^{+0.076}_{-0.073}$ & 1.01 $^{+0.11}_{-0.11} \times 10^{-13}$ \\
\hline
1571 & 6.71 $^{+0.027}_{-0.028}$ & 0.332 $^{+0.069}_{-0.065}$ & 8.99 $^{+1.0}_{-0.99} \times 10^{-14}$ \\
\hline
1971 & 6.73 $^{+0.028}_{-0.030}$ & 0.298 $^{+0.076}_{-0.087}$ & 7.07 $^{+0.92}_{-0.90} \times 10^{-14}$ \\
\hline
2307 & 6.69 $^{+0.056}_{-0.040}$ & 0.467 $^{+0.107}_{-0.091}$ & 5.26 $^{+0.720}_{-0.691} \times 10^{-14}$ \\
\hline
\enddata
\end{deluxetable*}

\bibliography{SN2014C}{}

\begin{thebibliography}{}
\expandafter\ifx\csname natexlab\endcsname\relax\def\natexlab#1{#1}\fi
\providecommand{\url}[1]{\href{#1}{#1}}
\providecommand{\dodoi}[1]{doi:~\href{http://doi.org/#1}{\nolinkurl{#1}}}
\providecommand{\doeprint}[1]{\href{http://ascl.net/#1}{\nolinkurl{http://ascl.net/#1}}}
\providecommand{\doarXiv}[1]{\href{https://arxiv.org/abs/#1}{\nolinkurl{https://arxiv.org/abs/#1}}}

\bibitem[{Anderson {et~al.}(2016)Anderson, Horesh, Mooley, Rushton, Fender,
  Staley, Argo, Beswick, Hancock, Pérez-Torres, \& et~al.}]{Anderson16}
Anderson, G.~E., Horesh, A., Mooley, K.~P., {et~al.} 2016, Monthly Notices of
  the Royal Astronomical Society, 466, 3648, \dodoi{10.1093/mnras/stw3310}

\bibitem[{{Andrews} \& {Smith}(2018)}]{Andrews18}
{Andrews}, J.~E., \& {Smith}, N. 2018, \mnras, 477, 74,
  \dodoi{10.1093/mnras/sty584}

\bibitem[{{Andrews} {et~al.}(2010){Andrews}, {Gallagher}, {Clayton},
  {Sugerman}, {Chatelain}, {Clem}, {Welch}, {Barlow}, {Ercolano}, {Fabbri},
  {Wesson}, \& {Meixner}}]{Andrews10}
{Andrews}, J.~E., {Gallagher}, J.~S., {Clayton}, G.~C., {et~al.} 2010, \apj,
  715, 541, \dodoi{10.1088/0004-637X/715/1/541}

\bibitem[{{Andrews} {et~al.}(2016){Andrews}, {Krafton}, {Clayton}, {Montiel},
  {Wesson}, {Sugerman}, {Barlow}, {Matsuura}, \& {Drass}}]{Andrews16}
{Andrews}, J.~E., {Krafton}, K.~M., {Clayton}, G.~C., {et~al.} 2016, \mnras,
  457, 3241, \dodoi{10.1093/mnras/stw164}

\bibitem[{{Andrews} {et~al.}(2019){Andrews}, {Sand}, {Valenti}, {Smith},
  {Dastidar}, {Sahu}, {Misra}, {Singh}, {Hiramatsu}, {Brown}, {Hosseinzadeh},
  {Wyatt}, {Vinko}, {Anupama}, {Arcavi}, {Ashall}, {Benetti}, {Berton},
  {Bostroem}, {Bulla}, {Burke}, {Chen}, {Chomiuk}, {Cikota}, {Congiu}, {Cseh},
  {Davis}, {Elias-Rosa}, {Faran}, {Fraser}, {Galbany}, {Gall}, {Gal-Yam},
  {Gangopadhyay}, {Gromadzki}, {Haislip}, {Howell}, {Hsiao}, {Inserra},
  {Kankare}, {Kuncarayakti}, {Kouprianov}, {Kumar}, {Li}, {Lin}, {Maguire},
  {Mazzali}, {McCully}, {Milne}, {Mo}, {Morrell}, {Nicholl}, {Ochner},
  {Olivares}, {Pastorello}, {Patat}, {Phillips}, {Pignata}, {Prentice},
  {Reguitti}, {Reichart}, {Rodr{\'\i}guez}, {Rui}, {Sanwal}, {S{\'a}rneczky},
  {Shahbandeh}, {Singh}, {Smartt}, {Strader}, {Stritzinger}, {Szak{\'a}ts},
  {Tartaglia}, {Wang}, {Wang}, {Wang}, {Wheeler}, {Xiang}, {Yaron}, {Young}, \&
  {Zhang}}]{Andrews19}
{Andrews}, J.~E., {Sand}, D.~J., {Valenti}, S., {et~al.} 2019, \apj, 885, 43,
  \dodoi{10.3847/1538-4357/ab43e3}

\bibitem[{{Arcavi} {et~al.}(2011){Arcavi}, {Gal-Yam}, {Yaron}, {Sternberg},
  {Rabinak}, {Waxman}, {Kasliwal}, {Quimby}, {Ofek}, {Horesh}, {Kulkarni},
  {Filippenko}, {Silverman}, {Cenko}, {Li}, {Bloom}, {Sullivan}, {Nugent},
  {Poznanski}, {Gorbikov}, {Fulton}, {Howell}, {Bersier}, {Riou},
  {Lamotte-Bailey}, {Griga}, {Cohen}, {Hachinger}, {Polishook}, {Xu},
  {Ben-Ami}, {Manulis}, {Walker}, {Maguire}, {Pan}, {Matheson}, {Mazzali},
  {Pian}, {Fox}, {Gehrels}, {Law}, {James}, {Marchant}, {Smith}, {Mottram},
  {Barnsley}, {Kandrashoff}, \& {Clubb}}]{Class:2011dh}
{Arcavi}, I., {Gal-Yam}, A., {Yaron}, O., {et~al.} 2011, \apjl, 742, L18,
  \dodoi{10.1088/2041-8205/742/2/L18}

\bibitem[{Arcavi {et~al.}(2017)Arcavi, Howell, Kasen, Bildsten, Hosseinzadeh,
  McCully, Wong, Katz, Gal-Yam, Sollerman, \& et~al.}]{Arcavi17}
Arcavi, I., Howell, D.~A., Kasen, D., {et~al.} 2017, Nature, 551, 210–213,
  \dodoi{10.1038/nature24030}

\bibitem[{{Arnett} \& {Meakin}(2011)}]{Arnett11b}
{Arnett}, W.~D., \& {Meakin}, C. 2011, \apj, 733, 78,
  \dodoi{10.1088/0004-637X/733/2/78}

\bibitem[{{Asplund} {et~al.}(2009){Asplund}, {Grevesse}, {Sauval}, \&
  {Scott}}]{Asplund09}
{Asplund}, M., {Grevesse}, N., {Sauval}, A.~J., \& {Scott}, P. 2009, \araa, 47,
  481, \dodoi{10.1146/annurev.astro.46.060407.145222}

\bibitem[{{Balam} \& {Graham}(2015)}]{Class:ASASSN-15no:Early}
{Balam}, D.~D., \& {Graham}, M.~L. 2015, The Astronomer's Telegram, 7931, 1

\bibitem[{{Balcon}(2020)}]{Class:2020tlf}
{Balcon}, C. 2020, Transient Name Server Classification Report, 2020-2839, 1

\bibitem[{{Bauer} {et~al.}(2008){Bauer}, {Dwarkadas}, {Brandt}, {Immler},
  {Smartt}, {Bartel}, \& {Bietenholz}}]{Class:1996cr}
{Bauer}, F.~E., {Dwarkadas}, V.~V., {Brandt}, W.~N., {et~al.} 2008, \apj, 688,
  1210, \dodoi{10.1086/589761}

\bibitem[{{Ben-Ami} {et~al.}(2012){Ben-Ami}, {Gal-Yam}, \&
  {Quimby}}]{Class:2010mb}
{Ben-Ami}, S., {Gal-Yam}, A., \& {Quimby}, R. 2012, Central Bureau Electronic
  Telegrams, 3309, 1

\bibitem[{{Ben-Ami} {et~al.}(2014){Ben-Ami}, {Gal-Yam}, {Mazzali}, {Gnat},
  {Modjaz}, {Rabinak}, {Sullivan}, {Bildsten}, {Poznanski}, {Yaron}, {Arcavi},
  {Bloom}, {Horesh}, {Kasliwal}, {Kulkarni}, {Nugent}, {Ofek}, {Perley},
  {Quimby}, \& {Xu}}]{Ben-Ami14}
{Ben-Ami}, S., {Gal-Yam}, A., {Mazzali}, P.~A., {et~al.} 2014, \apj, 785, 37,
  \dodoi{10.1088/0004-637X/785/1/37}

\bibitem[{{Benetti}(1994)}]{Class:1994aj}
{Benetti}, S. 1994, \iaucirc, 6122, 2

\bibitem[{{Benetti} {et~al.}(2010){Benetti}, {Bufano}, {Vinko}, {Marion},
  {Pritchard}, {Wheeler}, {Chatzopoulos}, \& {Shetrone}}]{Class:2010jl}
{Benetti}, S., {Bufano}, F., {Vinko}, J., {et~al.} 2010, Central Bureau
  Electronic Telegrams, 2536, 1

\bibitem[{{Benetti} {et~al.}(2018){Benetti}, {Zampieri}, {Pastorello},
  {Cappellaro}, {Pumo}, {Elias-Rosa}, {Ochner}, {Terreran}, {Tomasella},
  {Taubenberger}, {Turatto}, {Morales-Garoffolo}, {Harutyunyan}, \&
  {Tartaglia}}]{Benetti18}
{Benetti}, S., {Zampieri}, L., {Pastorello}, A., {et~al.} 2018, \mnras, 476,
  261, \dodoi{10.1093/mnras/sty166}

\bibitem[{{Bhirombhakdi} {et~al.}(2019){Bhirombhakdi}, {Chornock}, {Miller},
  {Filippenko}, {Cenko}, \& {Smith}}]{Bhirombhakdi19}
{Bhirombhakdi}, K., {Chornock}, R., {Miller}, A.~A., {et~al.} 2019, \mnras,
  488, 3783, \dodoi{10.1093/mnras/stz1928}

\bibitem[{{Bietenholz} {et~al.}(2021{\natexlab{a}}){Bietenholz}, {Bartel},
  {Argo}, {Dua}, {Ryder}, \& {Soderberg}}]{Bietenholz20a}
{Bietenholz}, M.~F., {Bartel}, N., {Argo}, M., {et~al.} 2021{\natexlab{a}},
  \apj, 908, 75, \dodoi{10.3847/1538-4357/abccd9}

\bibitem[{{Bietenholz} {et~al.}(2021{\natexlab{b}}){Bietenholz}, {Bartel},
  {Kamble}, {Margutti}, {Matthews}, \& {Milisavljevic}}]{Bietenholz21}
{Bietenholz}, M.~F., {Bartel}, N., {Kamble}, A., {et~al.} 2021{\natexlab{b}},
  \mnras, 502, 1694, \dodoi{10.1093/mnras/staa4003}

\bibitem[{{Bietenholz} {et~al.}(2014){Bietenholz}, {De Colle}, {Granot},
  {Bartel}, \& {Soderberg}}]{Bietenholz14}
{Bietenholz}, M.~F., {De Colle}, F., {Granot}, J., {Bartel}, N., \&
  {Soderberg}, A.~M. 2014, \mnras, 440, 821, \dodoi{10.1093/mnras/stu246}

\bibitem[{Bietenholz {et~al.}(2017)Bietenholz, Kamble, Margutti, Milisavljevic,
  \& Soderberg}]{Bietenholz17}
Bietenholz, M.~F., Kamble, A., Margutti, R., Milisavljevic, D., \& Soderberg,
  A. 2017, Monthly Notices of the Royal Astronomical Society, 475, 1756,
  \dodoi{10.1093/mnras/stx3194}

\bibitem[{{Blanc} {et~al.}(2005){Blanc}, {Bongard}, {Copin}, {Gangler},
  {Sauge}, {Smadja}, {Antilogus}, {Garavini}, {Gilles}, {Pain}, {Pereira},
  {Aldering}, {Bailey}, {Kocevski}, {Lee}, {Loken}, {Nugent}, {Perlmutter},
  {Scalzo}, {Thomas}, {Wang}, {Weaver}, {Bonnaud}, {Pecontal}, {Kessler},
  {Baltay}, {Rabinowitz}, \& {Bauer}}]{Class:2005gl}
{Blanc}, N., {Bongard}, S., {Copin}, Y., {et~al.} 2005, The Astronomer's
  Telegram, 630, 1

\bibitem[{{Blondin} \& {Calkins}(2007)}]{Class:2007od}
{Blondin}, S., \& {Calkins}, M. 2007, Central Bureau Electronic Telegrams,
  1119, 1

\bibitem[{{Bose} {et~al.}(2017){Bose}, {Monard}, {Seidel}, {Dong}, {Hsiao},
  {Shappee}, {Bond}, {Prieto}, {Morrell}, \& {Phillips}}]{Class:2016jbu}
{Bose}, S., {Monard}, L.~A.~G., {Seidel}, M.~K., {et~al.} 2017, The
  Astronomer's Telegram, 9937, 1

\bibitem[{{Bostroem} {et~al.}(2019){Bostroem}, {Valenti}, {Horesh}, {Morozova},
  {Kuin}, {Wyatt}, {Jerkstrand}, {Sand}, {Lundquist}, {Smith}, {Sullivan},
  {Hosseinzadeh}, {Arcavi}, {Callis}, {Cartier}, {Gal-Yam}, {Galbany},
  {Guti{\'e}rrez}, {Howell}, {Inserra}, {Kankare}, {L{\'o}pez}, {McCully},
  {Pignata}, {Piro}, {Rodr{\'\i}guez}, {Smartt}, {Smith}, {Yaron}, \&
  {Young}}]{Bostroem19}
{Bostroem}, K.~A., {Valenti}, S., {Horesh}, A., {et~al.} 2019, \mnras, 485,
  5120, \dodoi{10.1093/mnras/stz570}

\bibitem[{{Bragaglia} {et~al.}(1994){Bragaglia}, {Munari}, {Barbon}, {Mikuz},
  {Spratt}, \& {Vanmunster}}]{Class:1994W}
{Bragaglia}, A., {Munari}, U., {Barbon}, R., {et~al.} 1994, \iaucirc, 6044, 1

\bibitem[{{Brennan} {et~al.}(2021{\natexlab{a}}){Brennan}, {Fraser},
  {Johansson}, {Pastorello}, {Kotak}, {Stevance}, {Chen}, {Eldridge}, {Bose},
  {Brown}, {Callis}, {Cartier}, {Dennefeld}, {Dong}, {Duffy}, {Elias-Rosa},
  {Hosseinzadeh}, {Hsiao}, {Kuncarayakti}, {Martin-Carrillo}, {Monard},
  {Nyholm}, {Pignata}, {Sand}, {Shappee}, {Smartt}, {Tucker}, {Wyrzykowski},
  {Abbot}, {Benetti}, {Blondin}, {Chen}, {Bento}, {Delgado}, {Galbany},
  {Gromadzki}, {Guti{\'e}rrez}, {Hanlon}, {Harrison}, {Hiramatsu}, {Hodgkin},
  {Holoien}, {Howell}, {Inserra}, {Kankare}, {Kozlowski}, {Maguire},
  {M{\"u}ller-Bravo}, {McCully}, {Meintjes}, {Morrell}, {Nicholl}, {O'Neill},
  {Pietrukowicz}, {Poleski}, {Prieto}, {Rau}, {Reichart}, {Schweyer},
  {Shahbandeh}, {Skowron}, {Sollerman}, {Sosz{\'n}yski}, {Stritzinger},
  {Szyma{\'n}ski}, {Tartaglia}, {Udalski}, {Ulaczyk}, {Young}, {van Leeuwen},
  \& {van Soelen}}]{Brennan21a}
{Brennan}, S.~J., {Fraser}, M., {Johansson}, J., {et~al.} 2021{\natexlab{a}},
  arXiv e-prints, arXiv:2102.09572.
\newblock \doarXiv{2102.09572}

\bibitem[{{Brennan} {et~al.}(2021{\natexlab{b}}){Brennan}, {Fraser},
  {Johansson}, {Pastorello}, {Kotak}, {Stevance}, {Chen}, {Eldridge}, {Bose},
  {Brown}, {Callis}, {Cartier}, {Dennefeld}, {Dong}, {Duffy}, {Elias-Rosa},
  {Hosseinzadeh}, {Hsiao}, {Kuncarayakti}, {Martin-Carrillo}, {Monard},
  {Pignata}, {Sand}, {Shappee}, {Smartt}, {Tucker}, {Wyrzykowski}, {Abbot},
  {Benetti}, {Blondin}, {Chen}, {Bento}, {Delgado}, {Galbany}, {Gromadzki},
  {Guti{\'e}rrez}, {Hanlon}, {Harrison}, {Hiramatsu}, {Hodgkin}, {Holoien},
  {Howell}, {Inserra}, {Kankare}, {Kozlowski}, {Maguire}, {M{\"u}ller-Bravo},
  {McCully}, {Meintjes}, {Morrell}, {Nicholl}, {O'Neill}, {Pietrukowicz},
  {Poleski}, {Prieto}, {Rau}, {Reichart}, {Schweyer}, {Shahbandeh}, {Skowron},
  {Sollerman}, {Sosz{\'n}yski}, {Stritzinger}, {Szyma{\'n}ski}, {Tartaglia},
  {Udalski}, {Ulaczyk}, {Young}, {van Leeuwen}, \& {van Soelen}}]{Brennan21b}
---. 2021{\natexlab{b}}, arXiv e-prints, arXiv:2102.09576.
\newblock \doarXiv{2102.09576}

\bibitem[{{Brethauer} {et~al.}(2020){Brethauer}, {Margutti}, {Milisavljevic},
  \& {Bietenholz}}]{Brethauer20}
{Brethauer}, D., {Margutti}, R., {Milisavljevic}, D., \& {Bietenholz}, M. 2020,
  Research Notes of the American Astronomical Society, 4, 235,
  \dodoi{10.3847/2515-5172/abd252}

\bibitem[{{Bruch} {et~al.}(2020){Bruch}, {Schulze}, \&
  {Gal-Yam}}]{Class:2020pni}
{Bruch}, R., {Schulze}, S., \& {Gal-Yam}, A. 2020, Transient Name Server
  Classification Report, 2020-2170, 1

\bibitem[{{Burke} {et~al.}(2019){Burke}, {Hiramatsu}, {Arcavi}, {Howell},
  {Mccully}, \& {Valenti}}]{Class:2019cad}
{Burke}, J., {Hiramatsu}, D., {Arcavi}, I., {et~al.} 2019, Transient Name
  Server Classification Report, 2019-427, 1

\bibitem[{{Cartier} {et~al.}(2017){Cartier}, {Gutierrez}, \&
  {Yaron}}]{Class:2017dio1}
{Cartier}, R., {Gutierrez}, C., \& {Yaron}, O. 2017, Transient Name Server
  Classification Report, 2017-500, 1

\bibitem[{{Cenko} {et~al.}(2012){Cenko}, {Li}, {Filippenko}, {Silverman},
  {Clubb}, \& {Blanchard}}]{Class:2012aa}
{Cenko}, S.~B., {Li}, W., {Filippenko}, A.~V., {et~al.} 2012, Central Bureau
  Electronic Telegrams, 3015, 1

\bibitem[{{Chandra} {et~al.}(2012){Chandra}, {Chevalier}, {Chugai}, {Fransson},
  {Irwin}, {Soderberg}, {Chakraborti}, \& {Immler}}]{Chandra12b}
{Chandra}, P., {Chevalier}, R.~A., {Chugai}, N., {et~al.} 2012, \apj, 755, 110,
  \dodoi{10.1088/0004-637X/755/2/110}

\bibitem[{{Chandra} {et~al.}(2020){Chandra}, {Chevalier}, {Chugai},
  {Milisavljevic}, \& {Fransson}}]{Chandra20}
{Chandra}, P., {Chevalier}, R.~A., {Chugai}, N., {Milisavljevic}, D., \&
  {Fransson}, C. 2020, \apj, 902, 55, \dodoi{10.3847/1538-4357/abb460}

\bibitem[{{Chen} {et~al.}(2018{\natexlab{a}}){Chen}, {Inserra}, {Fraser},
  {Moriya}, {Schady}, {Schweyer}, {Filippenko}, {Perley}, {Ruiter},
  {Seitenzahl}, {Sollerman}, {Taddia}, {Anderson}, {Foley}, {Jerkstrand},
  {Ngeow}, {Pan}, {Pastorello}, {Points}, {Smartt}, {Smith}, {Taubenberger},
  {Wiseman}, {Young}, {Benetti}, {Berton}, {Bufano}, {Clark}, {Della Valle},
  {Galbany}, {Gal-Yam}, {Gromadzki}, {Guti{\'e}rrez}, {Heinze}, {Kankare},
  {Kilpatrick}, {Kuncarayakti}, {Leloudas}, {Lin}, {Maguire}, {Mazzali},
  {McBrien}, {Prentice}, {Rau}, {Rest}, {Siebert}, {Stalder}, {Tonry}, \&
  {Yu}}]{Chen18}
{Chen}, T.~W., {Inserra}, C., {Fraser}, M., {et~al.} 2018{\natexlab{a}}, \apjl,
  867, L31, \dodoi{10.3847/2041-8213/aaeb2e}

\bibitem[{{Chen} {et~al.}(2018{\natexlab{b}}){Chen}, {Inserra}, {Fraser},
  {Moriya}, {Schady}, {Schweyer}, {Filippenko}, {Perley}, {Ruiter},
  {Seitenzahl}, {Sollerman}, {Taddia}, {Anderson}, {Foley}, {Jerkstrand},
  {Ngeow}, {Pan}, {Pastorello}, {Points}, {Smartt}, {Smith}, {Taubenberger},
  {Wiseman}, {Young}, {Benetti}, {Berton}, {Bufano}, {Clark}, {Della Valle},
  {Galbany}, {Gal-Yam}, {Gromadzki}, {Guti{\'e}rrez}, {Heinze}, {Kankare},
  {Kilpatrick}, {Kuncarayakti}, {Leloudas}, {Lin}, {Maguire}, {Mazzali},
  {McBrien}, {Prentice}, {Rau}, {Rest}, {Siebert}, {Stalder}, {Tonry}, \&
  {Yu}}]{Class:2017ens}
---. 2018{\natexlab{b}}, \apjl, 867, L31, \dodoi{10.3847/2041-8213/aaeb2e}

\bibitem[{{Chevalier} \& {Fransson}(2006)}]{Chevalier06}
{Chevalier}, R.~A., \& {Fransson}, C. 2006, \apj, 651, 381,
  \dodoi{10.1086/507606}

\bibitem[{Chevalier \& Fransson(2017)}]{Chevalier17}
Chevalier, R.~A., \& Fransson, C. 2017, Handbook of Supernovae, 875–937,
  \dodoi{10.1007/978-3-319-21846-5_34}

\bibitem[{{Chomiuk} {et~al.}(2011){Chomiuk}, {Chornock}, {Soderberg}, {Berger},
  {Chevalier}, {Foley}, {Huber}, {Narayan}, {Rest}, {Gezari}, {Kirshner},
  {Riess}, {Rodney}, {Smartt}, {Stubbs}, {Tonry}, {Wood-Vasey}, {Burgett},
  {Chambers}, {Czekala}, {Flewelling}, {Forster}, {Kaiser}, {Kudritzki},
  {Magnier}, {Martin}, {Morgan}, {Neill}, {Price}, {Roth}, {Sanders}, \&
  {Wainscoat}}]{Chomiuk11}
{Chomiuk}, L., {Chornock}, R., {Soderberg}, A.~M., {et~al.} 2011, \apj, 743,
  114, \dodoi{10.1088/0004-637X/743/2/114}

\bibitem[{{Chornock} {et~al.}(2016){Chornock}, {Bhirombhakdi}, {Katebi},
  {Blanchard}, {Nicholl}, \& {Berger}}]{Class:2016aps}
{Chornock}, R., {Bhirombhakdi}, K., {Katebi}, R., {et~al.} 2016, The
  Astronomer's Telegram, 8790, 1

\bibitem[{{Chugai} \& {Utrobin}(2022)}]{Chugai22}
{Chugai}, N., \& {Utrobin}, V. 2022, arXiv e-prints, arXiv:2205.07749.
\newblock \doarXiv{2205.07749}

\bibitem[{{Chugai} \& {Chevalier}(2006)}]{Chugai06}
{Chugai}, N.~N., \& {Chevalier}, R.~A. 2006, \apj, 641, 1051,
  \dodoi{10.1086/500539}

\bibitem[{{Ciabattari} {et~al.}(2013){Ciabattari}, {Mazzoni}, {Donati},
  {Petroni}, {Foglia}, {Galli}, {Cenko}, {Clubb}, {Zheng}, {Kelly},
  {Filippenko}, \& {Van Dyk}}]{Class:2013df}
{Ciabattari}, F., {Mazzoni}, E., {Donati}, S., {et~al.} 2013, Central Bureau
  Electronic Telegrams, 3557, 1

\bibitem[{{Clark} {et~al.}(2020){Clark}, {Maguire}, {Inserra}, {Prentice},
  {Smartt}, {Contreras}, {Hossenizadeh}, {Hsiao}, {Kankare}, {Kasliwal},
  {Nugent}, {Shahbandeh}, {Baltay}, {Rabinowitz}, {Arcavi}, {Ashall}, {Burns},
  {Callis}, {Chen}, {Diamond}, {Fraser}, {Howell}, {Karamehmetoglu}, {Kotak},
  {Lyman}, {Morrell}, {Phillips}, {Pignata}, {Pursiainen}, {Sollerman},
  {Stritzinger}, {Sullivan}, \& {Young}}]{Clark20}
{Clark}, P., {Maguire}, K., {Inserra}, C., {et~al.} 2020, \mnras, 492, 2208,
  \dodoi{10.1093/mnras/stz3598}

\bibitem[{{Corsi} {et~al.}(2012){Corsi}, {Kasliwal}, {Ofek}, {Gal-Yam},
  {Kulkarni}, \& {Helou}}]{Class:PTF11qcj}
{Corsi}, A., {Kasliwal}, M., {Ofek}, E., {et~al.} 2012, {PTF11qcj: First
  Discovery of a Radio Luminous Ibn Supernova}, Spitzer Proposal

\bibitem[{{Corsi} {et~al.}(2014){Corsi}, {Ofek}, {Gal-Yam}, {Frail},
  {Kulkarni}, {Fox}, {Kasliwal}, {Sullivan}, {Horesh}, {Carpenter}, {Maguire},
  {Arcavi}, {Cenko}, {Cao}, {Mooley}, {Pan}, {Sesar}, {Sternberg}, {Xu},
  {Bersier}, {James}, {Bloom}, \& {Nugent}}]{Corsi14}
{Corsi}, A., {Ofek}, E.~O., {Gal-Yam}, A., {et~al.} 2014, \apj, 782, 42,
  \dodoi{10.1088/0004-637X/782/1/42}

\bibitem[{{Costantin} {et~al.}(2018){Costantin}, {Avramova-Bonche}, {Pinter},
  {Fiorellino}, {Cabello}, {Smith}, {Benetti}, {Berton}, {Tomasella}, {Ivanov},
  {Korhonen}, \& {Pizzella}}]{Class:2018gep}
{Costantin}, L., {Avramova-Bonche}, A., {Pinter}, V., {et~al.} 2018, The
  Astronomer's Telegram, 12047, 1

\bibitem[{{Crotts} {et~al.}(2006){Crotts}, {Eastman}, {Depoy}, {Prieto}, \&
  {Garnavich}}]{Class:2006jc}
{Crotts}, A., {Eastman}, J., {Depoy}, D., {Prieto}, J.~L., \& {Garnavich}, P.
  2006, Central Bureau Electronic Telegrams, 672, 1

\bibitem[{{Crowther}(2007)}]{Crowther07}
{Crowther}, P.~A. 2007, \araa, 45, 177,
  \dodoi{10.1146/annurev.astro.45.051806.110615}

\bibitem[{{de Jager} {et~al.}(1988){de Jager}, {Nieuwenhuijzen}, \& {van der
  Hucht}}]{deJager88}
{de Jager}, C., {Nieuwenhuijzen}, H., \& {van der Hucht}, K.~A. 1988, \aaps,
  72, 259

\bibitem[{{DeMarchi} {et~al.}(2022){DeMarchi}, {Margutti}, {Dittman},
  {Brunthaler}, {Milisavljevic}, {Bietenholz}, {Stauffer}, {Brethauer},
  {Coppejans}, {Auchettl}, {Alexander}, {Kilpatrick}, {Bright}, {Kelley},
  {Stroh}, \& {Jacobson-Galan}}]{DeMarchi22}
{DeMarchi}, L., {Margutti}, R., {Dittman}, J., {et~al.} 2022, arXiv e-prints,
  arXiv:2203.07388.
\newblock \doarXiv{2203.07388}

\bibitem[{{Dessart} \& {John Hillier}(2022)}]{Dessart22}
{Dessart}, L., \& {John Hillier}, D. 2022, \aap, 660, L9,
  \dodoi{10.1051/0004-6361/202243372}

\bibitem[{{Dewey} {et~al.}(2011){Dewey}, {Bauer}, \& {Dwarkadas}}]{Dewey11}
{Dewey}, D., {Bauer}, F.~E., \& {Dwarkadas}, V.~V. 2011, in American Institute
  of Physics Conference Series, Vol. 1358, American Institute of Physics
  Conference Series, ed. J.~E. {McEnery}, J.~L. {Racusin}, \& N.~{Gehrels},
  289--292, \dodoi{10.1063/1.3621791}

\bibitem[{{Dimitriadis} {et~al.}(2019){Dimitriadis}, {Foley}, {Siebert},
  {Kilpatrick}, \& {Corbett}}]{Class:2019yvr1}
{Dimitriadis}, G., {Foley}, R.~J., {Siebert}, M.~R., {Kilpatrick}, C.~D., \&
  {Corbett}, H.~T. 2019, The Astronomer's Telegram, 13375, 1

\bibitem[{{Dong} {et~al.}(2021){Dong}, {Hallinan}, {Nakar}, {Ho}, {Hughes},
  {Hotokezaka}, {Myers}, {De}, {Mooley}, {Ravi}, {Horesh}, {Kasliwal}, \&
  {Kulkarni}}]{Dong21}
{Dong}, D.~Z., {Hallinan}, G., {Nakar}, E., {et~al.} 2021, Science, 373, 1125,
  \dodoi{10.1126/science.abg6037}

\bibitem[{{Dwarkadas} {et~al.}(2010){Dwarkadas}, {Dewey}, \&
  {Bauer}}]{Dwarkadas10}
{Dwarkadas}, V.~V., {Dewey}, D., \& {Bauer}, F. 2010, \mnras, 407, 812,
  \dodoi{10.1111/j.1365-2966.2010.16966.x}

\bibitem[{{Dyson}(1989)}]{Dyson89}
{Dyson}, J.~E. 1989, {Interstellar Wind-Blown Bubbles}, ed. G.~{Tenorio-Tagle},
  M.~{Moles}, \& J.~{Melnick}, Vol. 350, 137, \dodoi{10.1007/BFb0114858}

\bibitem[{{Elias-Rosa}(2017)}]{Class:2017gmr}
{Elias-Rosa}, N. 2017, Transient Name Server Classification Report, 2017-1015,
  1

\bibitem[{{Filippenko}(1997)}]{Filippenko97}
{Filippenko}, A.~V. 1997, \araa, 35, 309,
  \dodoi{10.1146/annurev.astro.35.1.309}

\bibitem[{{Filippenko}(1999)}]{Class:1999cq}
---. 1999, \iaucirc, 7220, 1

\bibitem[{{Filippenko} \& {Barth}(1997)}]{Class:1997eg}
{Filippenko}, A.~V., \& {Barth}, A.~J. 1997, \iaucirc, 6794, 1

\bibitem[{{Filippenko} \& {Chornock}(2001)}]{Class:2001em1}
{Filippenko}, A.~V., \& {Chornock}, R. 2001, \iaucirc, 7737, 3

\bibitem[{{Filippenko} \& {Foley}(2004)}]{Class:2004gq}
{Filippenko}, A.~V., \& {Foley}, R.~J. 2004, \iaucirc, 8452, 3

\bibitem[{{Filippenko} {et~al.}(2004){Filippenko}, {Ganeshalingam}, {Serduke},
  \& {Hoffman}}]{Class:2004dk1}
{Filippenko}, A.~V., {Ganeshalingam}, M., {Serduke}, F.~J.~D., \& {Hoffman},
  J.~L. 2004, \iaucirc, 8404, 1

\bibitem[{{Filippenko} \& {Schlegel}(1995)}]{Class:1995G}
{Filippenko}, A.~V., \& {Schlegel}, D. 1995, \iaucirc, 6139, 2

\bibitem[{{Foley} {et~al.}(2006){Foley}, {Li}, {Moore}, {Wong}, {Pooley}, \&
  {Filippenko}}]{Class:2006gy}
{Foley}, R.~J., {Li}, W., {Moore}, M., {et~al.} 2006, Central Bureau Electronic
  Telegrams, 695, 1

\bibitem[{{Foley} {et~al.}(2007){Foley}, {Smith}, {Ganeshalingam}, {Li},
  {Chornock}, \& {Filippenko}}]{Foley07}
{Foley}, R.~J., {Smith}, N., {Ganeshalingam}, M., {et~al.} 2007, \apjl, 657,
  L105, \dodoi{10.1086/513145}

\bibitem[{{Foley} {et~al.}(2004){Foley}, {Wong}, {Moore}, \&
  {Filippenko}}]{Class:2004cc}
{Foley}, R.~J., {Wong}, D.~S., {Moore}, M., \& {Filippenko}, A.~V. 2004,
  \iaucirc, 8353, 3

\bibitem[{{Fox} {et~al.}(2015){Fox}, {Smith}, {Ammons}, {Andrews}, {Bostroem},
  {Cenko}, {Clayton}, {Dwek}, {Filippenko}, {Gallagher}, {Kelly}, {Mauerhan},
  {Miller}, \& {Van Dyk}}]{Fox15}
{Fox}, O.~D., {Smith}, N., {Ammons}, S.~M., {et~al.} 2015, \mnras, 454, 4366,
  \dodoi{10.1093/mnras/stv2270}

\bibitem[{{Fransson} {et~al.}(1996){Fransson}, {Lundqvist}, \&
  {Chevalier}}]{Fransson96}
{Fransson}, C., {Lundqvist}, P., \& {Chevalier}, R.~A. 1996, \apj, 461, 993,
  \dodoi{10.1086/177119}

\bibitem[{{Fransson} {et~al.}(2014){Fransson}, {Ergon}, {Challis}, {Chevalier},
  {France}, {Kirshner}, {Marion}, {Milisavljevic}, {Smith}, {Bufano},
  {Friedman}, {Kangas}, {Larsson}, {Mattila}, {Benetti}, {Chornock}, {Czekala},
  {Soderberg}, \& {Sollerman}}]{Fransson14}
{Fransson}, C., {Ergon}, M., {Challis}, P.~J., {et~al.} 2014, \apj, 797, 118,
  \dodoi{10.1088/0004-637X/797/2/118}

\bibitem[{{Fraser} {et~al.}(2016){Fraser}, {Reynolds}, {Inserra}, \&
  {Yaron}}]{Class:2016egz}
{Fraser}, M., {Reynolds}, T., {Inserra}, C., \& {Yaron}, O. 2016, Transient
  Name Server Classification Report, 2016-490, 1

\bibitem[{{Fraser} {et~al.}(2021){Fraser}, {Stritzinger}, {Brennan},
  {Pastorello}, {Cai}, {Piro}, {Ashall}, {Brown}, {Burns}, {Elias-Rosa},
  {Kotak}, {Filippenko}, {Galbany}, {Hsiao}, {Jha}, {Reguitti}, {Zhang},
  {Moran}, {Morrell}, {Shappee}, {Tomasella}, {Anderson}, {Barna}, {Ochner},
  {Phillips}, {Tucker}, {Wang}, {Baron}, {Benetti}, {Bersten}, {Brink},
  {Camacho-Neves}, {Davis}, {Dettman}, {Folatelli}, {Gutierrez}, {Hoflich},
  {Holoien}, {Kankare}, {Kumar}, {Lu}, {Mazzali}, {Taubenberger}, {Tinyanont},
  {Kuncarayakti}, {Kwok}, {Shahbandeh}, {Suntzeff}, {Yan}, {Yang}, \&
  {Zheng}}]{Fraser21}
{Fraser}, M., {Stritzinger}, M.~D., {Brennan}, S.~J., {et~al.} 2021, arXiv
  e-prints, arXiv:2108.07278.
\newblock \doarXiv{2108.07278}

\bibitem[{{Freedman} {et~al.}(2001){Freedman}, {Madore}, {Gibson}, {Ferrarese},
  {Kelson}, {Sakai}, {Mould}, {Kennicutt}, {Ford}, {Graham}, {Huchra},
  {Hughes}, {Illingworth}, {Macri}, \& {Stetson}}]{Freedman01}
{Freedman}, W.~L., {Madore}, B.~F., {Gibson}, B.~K., {et~al.} 2001, \apj, 553,
  47, \dodoi{10.1086/320638}

\bibitem[{{Fremling} \& {Dahiwale}(2019)}]{Class:2019oys1}
{Fremling}, C., \& {Dahiwale}, A. 2019, Transient Name Server Classification
  Report, 2019-1846, 1

\bibitem[{{Fremling} {et~al.}(2018){Fremling}, {Dugas}, \&
  {Sharma}}]{Class:2018ijp}
{Fremling}, C., {Dugas}, A., \& {Sharma}, Y. 2018, Transient Name Server
  Classification Report, 2018-1877, 1

\bibitem[{{Fremling} {et~al.}(2019){Fremling}, {Dugas}, \&
  {Sharma}}]{Class:2019uo}
---. 2019, Transient Name Server Classification Report, 2019-188, 1

\bibitem[{{Gagliano} {et~al.}(2022){Gagliano}, {Izzo}, {Kilpatrick}, {Mockler},
  {Jacobson-Gal{\'a}n}, {Terreran}, {Dimitriadis}, {Zenati}, {Auchettl},
  {Drout}, {Narayan}, {Foley}, {Margutti}, {Rest}, {Jones}, {Aganze}, {Aleo},
  {Burgasser}, {Coulter}, {Gerasimov}, {Gall}, {Hjorth}, {Hsu}, {Magnier},
  {Mandel}, {Piro}, {Rojas-Bravo}, {Siebert}, {Stacey}, {Stroh}, {Swift},
  {Taggart}, {Tinyanont}, \& {Tinyanont}}]{Gagliano22}
{Gagliano}, A., {Izzo}, L., {Kilpatrick}, C.~D., {et~al.} 2022, \apj, 924, 55,
  \dodoi{10.3847/1538-4357/ac35ec}

\bibitem[{{Gal-Yam}(2019)}]{Gal-Yam19}
{Gal-Yam}, A. 2019, \araa, 57, 305, \dodoi{10.1146/annurev-astro-081817-051819}

\bibitem[{{Gal-Yam}(2021)}]{Class:2019hgp}
---. 2021, Transient Name Server Classification Report, 2021-547, 1

\bibitem[{{Gal-Yam} {et~al.}(2002){Gal-Yam}, {Shemmer}, \&
  {Dann}}]{Class:2002ao}
{Gal-Yam}, A., {Shemmer}, O., \& {Dann}, J. 2002, \iaucirc, 7810, 3

\bibitem[{{Gal-Yam} {et~al.}(2021){Gal-Yam}, {Yaron}, {Pastorello},
  {Taubenberger}, {Fraser}, \& {Perley}}]{Gal-Yam21Astronote}
{Gal-Yam}, A., {Yaron}, O., {Pastorello}, A., {et~al.} 2021, Transient Name
  Server AstroNote, 76, 1

\bibitem[{{Gal-Yam} {et~al.}(2007){Gal-Yam}, {Leonard}, {Fox}, {Cenko},
  {Soderberg}, {Moon}, {Sand}, {Caltech Core Collapse Program}, {Li},
  {Filippenko}, {Aldering}, \& {Copin}}]{Gal-Yam07}
{Gal-Yam}, A., {Leonard}, D.~C., {Fox}, D.~B., {et~al.} 2007, \apj, 656, 372,
  \dodoi{10.1086/510523}

\bibitem[{{Gal-Yam} {et~al.}(2014){Gal-Yam}, {Arcavi}, {Ofek}, {Ben-Ami},
  {Cenko}, {Kasliwal}, {Cao}, {Yaron}, {Tal}, {Silverman}, {Horesh}, {De Cia},
  {Taddia}, {Sollerman}, {Perley}, {Vreeswijk}, {Kulkarni}, {Nugent},
  {Filippenko}, \& {Wheeler}}]{Class:2013cu}
{Gal-Yam}, A., {Arcavi}, I., {Ofek}, E.~O., {et~al.} 2014, \nat, 509, 471,
  \dodoi{10.1038/nature13304}

\bibitem[{{Gal-Yam} {et~al.}(2022){Gal-Yam}, {Bruch}, {Schulze}, {Yang},
  {Perley}, {Irani}, {Sollerman}, {Kool}, {Soumagnac}, {Yaron}, {Strotjohann},
  {Zimmerman}, {Barbarino}, {Kulkarni}, {Kasliwal}, {De}, {Yao}, {Fremling},
  {Yan}, {Ofek}, {Fransson}, {Filippenko}, {Zheng}, {Brink}, {Copperwheat},
  {Foley}, {Brown}, {Siebert}, {Leloudas}, {Cabrera-Lavers}, {Garcia-Alvarez},
  {Marante-Barreto}, {Frederick}, {Hung}, {Wheeler}, {Vink{\'o}}, {Thomas},
  {Graham}, {Duev}, {Drake}, {Dekany}, {Bellm}, {Rusholme}, {Shupe},
  {Andreoni}, {Sharma}, {Riddle}, {van Roestel}, \& {Knezevic}}]{Gal-Yam22}
{Gal-Yam}, A., {Bruch}, R., {Schulze}, S., {et~al.} 2022, \nat, 601, 201,
  \dodoi{10.1038/s41586-021-04155-1}

\bibitem[{{Gezari} {et~al.}(2015){Gezari}, {Jones}, {Sanders}, {Soderberg},
  {Hung}, {Heinis}, {Smartt}, {Rest}, {Scolnic}, {Chornock}, {Berger}, {Foley},
  {Huber}, {Price}, {Stubbs}, {Riess}, {Kirshner}, {Smith}, {Wood-Vasey},
  {Schiminovich}, {Martin}, {Burgett}, {Chambers}, {Flewelling}, {Kaiser},
  {Tonry}, \& {Wainscoat}}]{Class:PS1-13arp2}
{Gezari}, S., {Jones}, D.~O., {Sanders}, N.~E., {et~al.} 2015, \apj, 804, 28,
  \dodoi{10.1088/0004-637X/804/1/28}

\bibitem[{{Gomez} {et~al.}(2019){Gomez}, {Berger}, {Nicholl}, {Blanchard},
  {Villar}, {Patton}, {Chornock}, {Leja}, {Hosseinzadeh}, \&
  {Cowperthwaite}}]{Gomez19}
{Gomez}, S., {Berger}, E., {Nicholl}, M., {et~al.} 2019, \apj, 881, 87,
  \dodoi{10.3847/1538-4357/ab2f92}

\bibitem[{{Green}(2006)}]{Class:2006Y/2006ai}
{Green}, D.~W.~E. 2006, \iaucirc, 8689, 3

\bibitem[{{Green}(2009)}]{Class:2008iy}
---. 2009, Central Bureau Electronic Telegrams, 1780, 2

\bibitem[{{Groh}(2014)}]{Groh14}
{Groh}, J.~H. 2014, \aap, 572, L11, \dodoi{10.1051/0004-6361/201424852}

\bibitem[{{Guti{\'e}rrez} {et~al.}(2021){Guti{\'e}rrez}, {Bersten}, {Orellana},
  {Pastorello}, {Ertini}, {Folatelli}, {Pignata}, {Anderson}, {Smartt},
  {Sullivan}, {Pursiainen}, {Inserra}, {Elias-Rosa}, {Fraser}, {Kankare},
  {Moran}, {Reguitti}, {Reynolds}, {Stritzinger}, {Burke}, {Frohmaier},
  {Galbany}, {Hiramatsu}, {Howell}, {Kuncarayakti}, {Mattila},
  {M{\"u}ller-Bravo}, {Pellegrino}, \& {Smith}}]{Gutierrez21}
{Guti{\'e}rrez}, C.~P., {Bersten}, M.~C., {Orellana}, M., {et~al.} 2021,
  \mnras, 504, 4907, \dodoi{10.1093/mnras/stab1009}

\bibitem[{{Hagen} \& {Reimers}(1997)}]{Class:1997ab}
{Hagen}, H.~J., \& {Reimers}, D. 1997, \iaucirc, 6589, 1

\bibitem[{{Harris} \& {Nugent}(2020)}]{Harris20}
{Harris}, C.~E., \& {Nugent}, P.~E. 2020, \apj, 894, 122,
  \dodoi{10.3847/1538-4357/ab879e}

\bibitem[{{Haynie} \& {Piro}(2021)}]{Haynie20}
{Haynie}, A., \& {Piro}, A.~L. 2021, \apj, 910, 128,
  \dodoi{10.3847/1538-4357/abe938}

\bibitem[{{Heathcote} {et~al.}(1988){Heathcote}, {Cowley}, \&
  {Hartwick}}]{Class:1988Z}
{Heathcote}, S., {Cowley}, A., \& {Hartwick}, D. 1988, \iaucirc, 4693, 1

\bibitem[{{Hiramatsu} {et~al.}(2021{\natexlab{a}}){Hiramatsu}, {Howell},
  {Moriya}, {Goldberg}, {Hosseinzadeh}, {Arcavi}, {Anderson}, {Guti{\'e}rrez},
  {Burke}, {McCully}, {Valenti}, {Galbany}, {Fang}, {Maeda}, {Folatelli},
  {Hsiao}, {Morrell}, {Phillips}, {Stritzinger}, {Suntzeff}, {Gromadzki},
  {Maguire}, {M{\"u}ller-Bravo}, \& {Young}}]{Hiramatsu20b}
{Hiramatsu}, D., {Howell}, D.~A., {Moriya}, T.~J., {et~al.} 2021{\natexlab{a}},
  \apj, 913, 55, \dodoi{10.3847/1538-4357/abf6d6}

\bibitem[{{Hiramatsu} {et~al.}(2021{\natexlab{b}}){Hiramatsu}, {Howell}, {Van
  Dyk}, {Goldberg}, {Maeda}, {Moriya}, {Tominaga}, {Nomoto}, {Hosseinzadeh},
  {Arcavi}, {McCully}, {Burke}, {Bostroem}, {Valenti}, {Dong}, {Brown},
  {Andrews}, {Bilinski}, {Williams}, {Smith}, {Smith}, {Sand}, {Anand}, {Xu},
  {Filippenko}, {Bersten}, {Folatelli}, {Kelly}, {Noguchi}, \&
  {Itagaki}}]{Hiramatsu20}
{Hiramatsu}, D., {Howell}, D.~A., {Van Dyk}, S.~D., {et~al.}
  2021{\natexlab{b}}, Nature Astronomy, 5, 903,
  \dodoi{10.1038/s41550-021-01384-2}

\bibitem[{{Ho} {et~al.}(2019){Ho}, {Goldstein}, {Schulze}, {Khatami}, {Perley},
  {Ergon}, {Gal-Yam}, {Corsi}, {Andreoni}, {Barbarino}, {Bellm},
  {Blagorodnova}, {Bright}, {Burns}, {Cenko}, {Cunningham}, {De}, {Dekany},
  {Dugas}, {Fender}, {Fransson}, {Fremling}, {Goldstein}, {Graham}, {Hale},
  {Horesh}, {Hung}, {Kasliwal}, {Kuin}, {Kulkarni}, {Kupfer}, {Lunnan},
  {Masci}, {Ngeow}, {Nugent}, {Ofek}, {Patterson}, {Petitpas}, {Rusholme},
  {Sai}, {Sfaradi}, {Shupe}, {Sollerman}, {Soumagnac}, {Tachibana}, {Taddia},
  {Walters}, {Wang}, {Yao}, \& {Zhang}}]{Ho19}
{Ho}, A. Y.~Q., {Goldstein}, D.~A., {Schulze}, S., {et~al.} 2019, \apj, 887,
  169, \dodoi{10.3847/1538-4357/ab55ec}

\bibitem[{{Holoien} {et~al.}(2017){Holoien}, {Stanek}, {Kochanek}, {Shappee},
  {Prieto}, {Brimacombe}, {Bersier}, {Bishop}, {Dong}, {Brown}, {Danilet},
  {Simonian}, {Basu}, {Beacom}, {Falco}, {Pojmanski}, {Skowron}, {Wo{\'z}niak},
  {{\'A}vila}, {Conseil}, {Contreras}, {Cruz}, {Fern{\'a}ndez}, {Koff}, {Guo},
  {Herczeg}, {Hissong}, {Hsiao}, {Jose}, {Kiyota}, {Long}, {Monard},
  {Nicholls}, {Nicolas}, \& {Wiethoff}}]{Class:ASASSN-14ms}
{Holoien}, T.~W.~S., {Stanek}, K.~Z., {Kochanek}, C.~S., {et~al.} 2017, \mnras,
  464, 2672, \dodoi{10.1093/mnras/stw2273}

\bibitem[{{Hosseinzadeh} {et~al.}(2017){Hosseinzadeh}, {Valenti}, {Arcavi},
  {Howell}, {McCully}, {Sand}, \& {Tartaglia}}]{Class:2017ahn}
{Hosseinzadeh}, G., {Valenti}, S., {Arcavi}, I., {et~al.} 2017, The
  Astronomer's Telegram, 10059, 1

\bibitem[{{Howell} \& {Murray}(2012)}]{Class:2010mc}
{Howell}, D.~A., \& {Murray}, D. 2012, Central Bureau Electronic Telegrams,
  3313, 2

\bibitem[{{Hung} {et~al.}(2019){Hung}, {Dimitriadis}, \&
  {Foley}}]{Class:2019ehk}
{Hung}, T., {Dimitriadis}, G., \& {Foley}, R.~J. 2019, The Astronomer's
  Telegram, 12734, 1

\bibitem[{{Ivanova} {et~al.}(2013){Ivanova}, {Justham}, {Chen}, {De Marco},
  {Fryer}, {Gaburov}, {Ge}, {Glebbeek}, {Han}, {Li}, {Lu}, {Marsh},
  {Podsiadlowski}, {Potter}, {Soker}, {Taam}, {Tauris}, {van den Heuvel}, \&
  {Webbink}}]{Ivanova13}
{Ivanova}, N., {Justham}, S., {Chen}, X., {et~al.} 2013, \aapr, 21, 59,
  \dodoi{10.1007/s00159-013-0059-2}

\bibitem[{{Jacobson-Gal{\'a}n}
  {et~al.}(2022{\natexlab{a}}){Jacobson-Gal{\'a}n}, {Venkatraman}, {Margutti},
  {Khatami}, {Terreran}, {Foley}, {Angulo}, {Angus}, {Auchettl}, {Blanchard},
  {Bobrick}, {Bright}, {Couch}, {Coulter}, {Clever}, {Davis}, {de Boer},
  {DeMarchi}, {Dodd}, {Jones}, {Johnson}, {Kilpatrick}, {Khetan}, {Lai},
  {Langeroodi}, {Lin}, {Magnier}, {Milisavljevic}, {Perets}, {Pierel},
  {Raymond}, {Rest}, {Rest}, {Ridden-Harper}, {Shen}, {Siebert}, {Smith},
  {Taggart}, {Tinyanont}, {Valdes}, {Villar}, {Wang}, {Karthik Yadavalli},
  {Zenati}, \& {Zenteno}}]{Jacobson-Galan22b}
{Jacobson-Gal{\'a}n}, W., {Venkatraman}, P., {Margutti}, R., {et~al.}
  2022{\natexlab{a}}, arXiv e-prints, arXiv:2203.03785.
\newblock \doarXiv{2203.03785}

\bibitem[{{Jacobson-Gal{\'a}n} {et~al.}(2020){Jacobson-Gal{\'a}n}, {Margutti},
  {Kilpatrick}, {Hiramatsu}, {Perets}, {Khatami}, {Foley}, {Raymond}, {Yoon},
  {Bobrick}, {Zenati}, {Galbany}, {Andrews}, {Brown}, {Cartier}, {Coppejans},
  {Dimitriadis}, {Dobson}, {Hajela}, {Howell}, {Kuncarayakti}, {Milisavljevic},
  {Rahman}, {Rojas-Bravo}, {Sand}, {Shepherd}, {Smartt}, {Stacey}, {Stroh},
  {Swift}, {Terreran}, {Vinko}, {Wang}, {Anderson}, {Baron}, {Berger},
  {Blanchard}, {Burke}, {Coulter}, {DeMarchi}, {DerKacy}, {Fremling}, {Gomez},
  {Gromadzki}, {Hosseinzadeh}, {Kasen}, {Kriskovics}, {McCully},
  {M{\"u}ller-Bravo}, {Nicholl}, {Ordasi}, {Pellegrino}, {Piro}, {P{\'a}l},
  {Ren}, {Rest}, {Rich}, {Sai}, {S{\'a}rneczky}, {Shen}, {Short}, {Siebert},
  {Stauffer}, {Szak{\'a}ts}, {Zhang}, {Zhang}, \& {Zhang}}]{Jacobson-Galan20}
{Jacobson-Gal{\'a}n}, W.~V., {Margutti}, R., {Kilpatrick}, C.~D., {et~al.}
  2020, \apj, 898, 166, \dodoi{10.3847/1538-4357/ab9e66}

\bibitem[{{Jacobson-Gal{\'a}n}
  {et~al.}(2022{\natexlab{b}}){Jacobson-Gal{\'a}n}, {Dessart}, {Jones},
  {Margutti}, {Coppejans}, {Dimitriadis}, {Foley}, {Kilpatrick}, {Matthews},
  {Rest}, {Terreran}, {Aleo}, {Auchettl}, {Blanchard}, {Coulter}, {Davis}, {de
  Boer}, {DeMarchi}, {Drout}, {Earl}, {Gagliano}, {Gall}, {Hjorth}, {Huber},
  {Ibik}, {Milisavljevic}, {Pan}, {Rest}, {Ridden-Harper}, {Rojas-Bravo},
  {Siebert}, {Smith}, {Taggart}, {Tinyanont}, {Wang}, \&
  {Zenati}}]{Jacobson-Galan22}
{Jacobson-Gal{\'a}n}, W.~V., {Dessart}, L., {Jones}, D.~O., {et~al.}
  2022{\natexlab{b}}, \apj, 924, 15, \dodoi{10.3847/1538-4357/ac3f3a}

\bibitem[{{Jin} {et~al.}(2021){Jin}, {Yoon}, \& {Blinnikov}}]{Jin21}
{Jin}, H., {Yoon}, S.-C., \& {Blinnikov}, S. 2021, \apj, 910, 68,
  \dodoi{10.3847/1538-4357/abe0b1}

\bibitem[{{Jin} \& {Kong}(2019)}]{Jin19}
{Jin}, R., \& {Kong}, A. K.~H. 2019, \apj, 879, 112,
  \dodoi{10.3847/1538-4357/ab2461}

\bibitem[{{Kalberla} {et~al.}(2005){Kalberla}, {Burton}, {Hartmann}, {Arnal},
  {Bajaja}, {Morras}, \& {P{\"o}ppel}}]{Kalberla05}
{Kalberla}, P.~M.~W., {Burton}, W.~B., {Hartmann}, D., {et~al.} 2005, \aap,
  440, 775, \dodoi{10.1051/0004-6361:20041864}

\bibitem[{{Kamble} {et~al.}(2016){Kamble}, {Margutti}, {Soderberg},
  {Chakraborti}, {Fransson}, {Chevalier}, {Powell}, {Milisavljevic}, {Parrent},
  \& {Bietenholz}}]{Kamble16}
{Kamble}, A., {Margutti}, R., {Soderberg}, A.~M., {et~al.} 2016, \apj, 818,
  111, \dodoi{10.3847/0004-637X/818/2/111}

\bibitem[{{Kiewe} {et~al.}(2012){Kiewe}, {Gal-Yam}, {Arcavi}, {Leonard},
  {Emilio Enriquez}, {Cenko}, {Fox}, {Moon}, {Sand}, {Soderberg}, \&
  {CCCP}}]{Kiewe12}
{Kiewe}, M., {Gal-Yam}, A., {Arcavi}, I., {et~al.} 2012, \apj, 744, 10,
  \dodoi{10.1088/0004-637X/744/1/10}

\bibitem[{{Kilpatrick} {et~al.}(2021){Kilpatrick}, {Drout}, {Auchettl},
  {Dimitriadis}, {Foley}, {Jones}, {DeMarchi}, {French}, {Gall}, {Hjorth},
  {Jacobson-Gal{\'a}n}, {Margutti}, {Piro}, {Ramirez-Ruiz}, {Rest}, \&
  {Rojas-Bravo}}]{Kilpatrick21}
{Kilpatrick}, C.~D., {Drout}, M.~R., {Auchettl}, K., {et~al.} 2021, \mnras,
  504, 2073, \dodoi{10.1093/mnras/stab838}

\bibitem[{{Kuncarayakti} {et~al.}(2018){Kuncarayakti}, {Maeda}, {Ashall},
  {Prentice}, {Mattila}, {Kankare}, {Fransson}, {Lundqvist}, {Pastorello},
  {Leloudas}, {Anderson}, {Benetti}, {Bersten}, {Cappellaro}, {Cartier},
  {Denneau}, {Della Valle}, {Elias-Rosa}, {Folatelli}, {Fraser}, {Galbany},
  {Gall}, {Gal-Yam}, {Guti{\'e}rrez}, {Hamanowicz}, {Heinze}, {Inserra},
  {Kangas}, {Mazzali}, {Melandri}, {Pignata}, {Rest}, {Reynolds}, {Roy},
  {Smartt}, {Smith}, {Sollerman}, {Somero}, {Stalder}, {Stritzinger}, {Taddia},
  {Tomasella}, {Tonry}, {Weiland}, \& {Young}}]{Kuncarayakti18}
{Kuncarayakti}, H., {Maeda}, K., {Ashall}, C.~J., {et~al.} 2018, \apjl, 854,
  L14, \dodoi{10.3847/2041-8213/aaaa1a}

\bibitem[{{Kundu} {et~al.}(2019){Kundu}, {Lundqvist}, {Sorokina},
  {P{\'e}rez-Torres}, {Blinnikov}, {O'Connor}, {Ergon}, {Chandra}, \&
  {Das}}]{Kundu19}
{Kundu}, E., {Lundqvist}, P., {Sorokina}, E., {et~al.} 2019, \apj, 875, 17,
  \dodoi{10.3847/1538-4357/ab0d81}

\bibitem[{{Leung} {et~al.}(2021{\natexlab{a}}){Leung}, {Fuller}, \&
  {Nomoto}}]{Leung21}
{Leung}, S.-C., {Fuller}, J., \& {Nomoto}, K. 2021{\natexlab{a}}, \apj, 915,
  80, \dodoi{10.3847/1538-4357/abfcbe}

\bibitem[{{Leung} {et~al.}(2021{\natexlab{b}}){Leung}, {Wu}, \&
  {Fuller}}]{Leung21b}
{Leung}, S.-C., {Wu}, S., \& {Fuller}, J. 2021{\natexlab{b}}, \apj, 923, 41,
  \dodoi{10.3847/1538-4357/ac2c63}

\bibitem[{{Li} {et~al.}(2015){Li}, {Wang}, \& {Zhang}}]{Class:iPTF14hls}
{Li}, W., {Wang}, X., \& {Zhang}, T. 2015, The Astronomer's Telegram, 6898, 1

\bibitem[{{Li} {et~al.}(1998){Li}, {Li}, {Filippenko}, \&
  {Moran}}]{Class:1998S}
{Li}, W.~D., {Li}, C., {Filippenko}, A.~V., \& {Moran}, E.~C. 1998, \iaucirc,
  6829, 1

\bibitem[{{Liedahl} {et~al.}(1995){Liedahl}, {Osterheld}, \&
  {Goldstein}}]{Liedahl95}
{Liedahl}, D.~A., {Osterheld}, A.~L., \& {Goldstein}, W.~H. 1995, \apjl, 438,
  L115, \dodoi{10.1086/187729}

\bibitem[{{Lunnan} {et~al.}(2018){Lunnan}, {Fransson}, {Vreeswijk}, {Woosley},
  {Leloudas}, {Perley}, {Quimby}, {Yan}, {Blagorodnova}, {Bue}, {Cenko}, {De
  Cia}, {Cook}, {Fremling}, {Gatkine}, {Gal-Yam}, {Kasliwal}, {Kulkarni},
  {Masci}, {Nugent}, {Nyholm}, {Rubin}, {Suzuki}, \& {Wozniak}}]{Lunnan18}
{Lunnan}, R., {Fransson}, C., {Vreeswijk}, P.~M., {et~al.} 2018, Nature
  Astronomy, 2, 887, \dodoi{10.1038/s41550-018-0568-z}

\bibitem[{Maeda {et~al.}(2021)Maeda, Chandra, Matsuoka, Ryder, Moriya,
  Kuncarayakti, Lee, Kundu, Patnaude, Saito, \& et~al.}]{Maeda21}
Maeda, K., Chandra, P., Matsuoka, T., {et~al.} 2021, The Astrophysical Journal,
  918, 34, \dodoi{10.3847/1538-4357/ac0dbc}

\bibitem[{{Margalit} {et~al.}(2022){Margalit}, {Quataert}, \&
  {Ho}}]{Margalit22}
{Margalit}, B., {Quataert}, E., \& {Ho}, A. Y.~Q. 2022, \apj, 928, 122,
  \dodoi{10.3847/1538-4357/ac53b0}

\bibitem[{{Margutti} {et~al.}(2014){Margutti}, {Milisavljevic}, {Soderberg},
  {Chornock}, {Zauderer}, {Murase}, {Guidorzi}, {Sanders}, {Kuin}, {Fransson},
  {Levesque}, {Chandra}, {Berger}, {Bianco}, {Brown}, {Challis},
  {Chatzopoulos}, {Cheung}, {Choi}, {Chomiuk}, {Chugai}, {Contreras}, {Drout},
  {Fesen}, {Foley}, {Fong}, {Friedman}, {Gall}, {Gehrels}, {Hjorth}, {Hsiao},
  {Kirshner}, {Im}, {Leloudas}, {Lunnan}, {Marion}, {Martin}, {Morrell},
  {Neugent}, {Omodei}, {Phillips}, {Rest}, {Silverman}, {Strader},
  {Stritzinger}, {Szalai}, {Utterback}, {Vinko}, {Wheeler}, {Arnett},
  {Campana}, {Chevalier}, {Ginsburg}, {Kamble}, {Roming}, {Pritchard}, \&
  {Stringfellow}}]{Margutti14}
{Margutti}, R., {Milisavljevic}, D., {Soderberg}, A.~M., {et~al.} 2014, \apj,
  780, 21, \dodoi{10.1088/0004-637X/780/1/21}

\bibitem[{{Margutti} {et~al.}(2017){Margutti}, {Kamble}, {Milisavljevic},
  {Zapartas}, {de Mink}, {Drout}, {Chornock}, {Risaliti}, {Zauderer},
  {Bietenholz}, {Cantiello}, {Chakraborti}, {Chomiuk}, {Fong}, {Grefenstette},
  {Guidorzi}, {Kirshner}, {Parrent}, {Patnaude}, {Soderberg}, {Gehrels}, \&
  {Harrison}}]{Margutti2017}
{Margutti}, R., {Kamble}, A., {Milisavljevic}, D., {et~al.} 2017, \apj, 835,
  140, \dodoi{10.3847/1538-4357/835/2/140}

\bibitem[{{Marshall} {et~al.}(2004){Marshall}, {van Loon}, {Matsuura}, {Wood},
  {Zijlstra}, \& {Whitelock}}]{Marshall04}
{Marshall}, J.~R., {van Loon}, J.~T., {Matsuura}, M., {et~al.} 2004, \mnras,
  355, 1348, \dodoi{10.1111/j.1365-2966.2004.08417.x}

\bibitem[{{Marston}(1997)}]{Marston97}
{Marston}, A.~P. 1997, \apj, 475, 188, \dodoi{10.1086/303534}

\bibitem[{{Mauerhan} {et~al.}(2018){Mauerhan}, {Filippenko}, {Zheng}, {Brink},
  {Graham}, {Shivvers}, \& {Clubb}}]{Mauerhan18}
{Mauerhan}, J.~C., {Filippenko}, A.~V., {Zheng}, W., {et~al.} 2018, \mnras,
  478, 5050, \dodoi{10.1093/mnras/sty1307}

\bibitem[{{Mauerhan} {et~al.}(2013){Mauerhan}, {Smith}, {Filippenko},
  {Blanchard}, {Blanchard}, {Casper}, {Cenko}, {Clubb}, {Cohen}, {Fuller},
  {Li}, \& {Silverman}}]{Mauerhan13}
{Mauerhan}, J.~C., {Smith}, N., {Filippenko}, A.~V., {et~al.} 2013, \mnras,
  430, 1801, \dodoi{10.1093/mnras/stt009}

\bibitem[{{McNaught} {et~al.}(1996){McNaught}, {Russell}, {James}, {Benetti},
  {Barthel}, \& {de Vries}}]{Class:1996L}
{McNaught}, R.~H., {Russell}, K.~S., {James}, J.~M., {et~al.} 1996, \iaucirc,
  6346, 1

\bibitem[{{Mewe} {et~al.}(1985){Mewe}, {Gronenschild}, \& {van den
  Oord}}]{Mewe85}
{Mewe}, R., {Gronenschild}, E.~H.~B.~M., \& {van den Oord}, G.~H.~J. 1985,
  \aaps, 62, 197

\bibitem[{{Mewe} {et~al.}(1986){Mewe}, {Lemen}, \& {van den Oord}}]{Mewe86}
{Mewe}, R., {Lemen}, J.~R., \& {van den Oord}, G.~H.~J. 1986, \aaps, 65, 511

\bibitem[{{Milisavljevic} {et~al.}(2015){Milisavljevic}, {Margutti}, {Kamble},
  {Patnaude}, {Raymond}, {Eldridge}, {Fong}, {Bietenholz}, {Challis},
  {Chornock}, {Drout}, {Fransson}, {Fesen}, {Grindlay}, {Kirshner}, {Lunnan},
  {Mackey}, {Miller}, {Parrent}, {Sanders}, {Soderberg}, \&
  {Zauderer}}]{Milisavljevic15}
{Milisavljevic}, D., {Margutti}, R., {Kamble}, A., {et~al.} 2015, \apj, 815,
  120, \dodoi{10.1088/0004-637X/815/2/120}

\bibitem[{{Modjaz} {et~al.}(2005){Modjaz}, {Kirshner}, {Challis}, \&
  {Calkins}}]{Class:2005ip}
{Modjaz}, M., {Kirshner}, R., {Challis}, P., \& {Calkins}, M. 2005, Central
  Bureau Electronic Telegrams, 276, 1

\bibitem[{{Monard} {et~al.}(2011{\natexlab{a}}){Monard}, {Prieto}, \&
  {Seth}}]{Class:2011fh}
{Monard}, L.~A.~G., {Prieto}, J.~L., \& {Seth}, K. 2011{\natexlab{a}}, Central
  Bureau Electronic Telegrams, 2799, 1

\bibitem[{{Monard} {et~al.}(2011{\natexlab{b}}){Monard}, {Milisavljevic},
  {Fesen}, {Pickering}, {Romero-Colmenero}, {Turatto}, {Benetti}, {Pastorello},
  {Valenti}, {Bufano}, {Tomasella}, {Ryder}, {Soderberg}, {Stockdale}, {van
  Dyk}, {Immler}, {Weiler}, \& {Panagia}}]{Class:2011ja}
{Monard}, L.~A.~G., {Milisavljevic}, D., {Fesen}, R., {et~al.}
  2011{\natexlab{b}}, Central Bureau Electronic Telegrams, 2946, 1

\bibitem[{{Monard} {et~al.}(2013){Monard}, {Morales Garoffolo}, {Elias-Rosa},
  {Benitez-Herrera}, {Taubenberger}, {Walker}, {Benetti}, {Pastorello},
  {Valenti}, {Smartt}, {Smith}, {Young}, {Sullivan}, {Gal-Yam}, \&
  {Yaron}}]{Class:2013L}
{Monard}, L.~A.~G., {Morales Garoffolo}, A., {Elias-Rosa}, N., {et~al.} 2013,
  Central Bureau Electronic Telegrams, 3392, 1

\bibitem[{{Moriya}(2015)}]{Moriya15}
{Moriya}, T.~J. 2015, \apjl, 803, L26, \dodoi{10.1088/2041-8205/803/2/L26}

\bibitem[{{Moriya} {et~al.}(2013){Moriya}, {Blinnikov}, {Tominaga}, {Yoshida},
  {Tanaka}, {Maeda}, \& {Nomoto}}]{Moriya13}
{Moriya}, T.~J., {Blinnikov}, S.~I., {Tominaga}, N., {et~al.} 2013, \mnras,
  428, 1020, \dodoi{10.1093/mnras/sts075}

\bibitem[{{Morozova} {et~al.}(2020){Morozova}, {Piro}, {Fuller}, \& {Van
  Dyk}}]{Morozova20}
{Morozova}, V., {Piro}, A.~L., {Fuller}, J., \& {Van Dyk}, S.~D. 2020, \apjl,
  891, L32, \dodoi{10.3847/2041-8213/ab77c8}

\bibitem[{{Morozova} {et~al.}(2018){Morozova}, {Piro}, \&
  {Valenti}}]{Morozova18}
{Morozova}, V., {Piro}, A.~L., \& {Valenti}, S. 2018, \apj, 858, 15,
  \dodoi{10.3847/1538-4357/aab9a6}

\bibitem[{{Nicholl} {et~al.}(2020){Nicholl}, {Blanchard}, {Berger}, {Chornock},
  {Margutti}, {Gomez}, {Lunnan}, {Miller}, {Fong}, {Terreran},
  {Vigna-G{\'o}mez}, {Bhirombhakdi}, {Bieryla}, {Challis}, {Laher}, {Masci}, \&
  {Paterson}}]{Nicholl20}
{Nicholl}, M., {Blanchard}, P.~K., {Berger}, E., {et~al.} 2020, Nature
  Astronomy, 4, 893, \dodoi{10.1038/s41550-020-1066-7}

\bibitem[{{Nomoto} {et~al.}(1993){Nomoto}, {Suzuki}, {Shigeyama}, {Kumagai},
  {Yamaoka}, \& {Saio}}]{Class:1993J}
{Nomoto}, K., {Suzuki}, T., {Shigeyama}, T., {et~al.} 1993, \nat, 364, 507,
  \dodoi{10.1038/364507a0}

\bibitem[{{Ofek} {et~al.}(2007){Ofek}, {Cameron}, {Kasliwal}, {Gal-Yam}, {Rau},
  {Kulkarni}, {Frail}, {Chandra}, {Cenko}, {Soderberg}, \& {Immler}}]{Ofek07}
{Ofek}, E.~O., {Cameron}, P.~B., {Kasliwal}, M.~M., {et~al.} 2007, \apjl, 659,
  L13, \dodoi{10.1086/516749}

\bibitem[{{Ofek} {et~al.}(2013){Ofek}, {Sullivan}, {Cenko}, {Kasliwal},
  {Gal-Yam}, {Kulkarni}, {Arcavi}, {Bildsten}, {Bloom}, {Horesh}, {Howell},
  {Filippenko}, {Laher}, {Murray}, {Nakar}, {Nugent}, {Silverman}, {Shaviv},
  {Surace}, \& {Yaron}}]{Ofek13}
{Ofek}, E.~O., {Sullivan}, M., {Cenko}, S.~B., {et~al.} 2013, \nat, 494, 65,
  \dodoi{10.1038/nature11877}

\bibitem[{{Ofek} {et~al.}(2014){Ofek}, {Sullivan}, {Shaviv}, {Steinbok},
  {Arcavi}, {Gal-Yam}, {Tal}, {Kulkarni}, {Nugent}, {Ben-Ami}, {Kasliwal},
  {Cenko}, {Laher}, {Surace}, {Bloom}, {Filippenko}, {Silverman}, \&
  {Yaron}}]{Ofek14}
{Ofek}, E.~O., {Sullivan}, M., {Shaviv}, N.~J., {et~al.} 2014, \apj, 789, 104,
  \dodoi{10.1088/0004-637X/789/2/104}

\bibitem[{{Pastorello} {et~al.}(2006){Pastorello}, {Sauer}, {Taubenberger},
  {Mazzali}, {Nomoto}, {Kawabata}, {Benetti}, {Elias-Rosa}, {Harutyunyan},
  {Navasardyan}, {Zampieri}, {Iijima}, {Botticella}, {di Rico}, {Del Principe},
  {Dolci}, {Gagliardi}, {Ragni}, \& {Valentini}}]{Pastorello06}
{Pastorello}, A., {Sauer}, D., {Taubenberger}, S., {et~al.} 2006, \mnras, 370,
  1752, \dodoi{10.1111/j.1365-2966.2006.10587.x}

\bibitem[{{Pastorello} {et~al.}(2007){Pastorello}, {Smartt}, {Mattila},
  {Eldridge}, {Young}, {Itagaki}, {Yamaoka}, {Navasardyan}, {Valenti}, {Patat},
  {Agnoletto}, {Augusteijn}, {Benetti}, {Cappellaro}, {Boles}, {Bonnet-Bidaud},
  {Botticella}, {Bufano}, {Cao}, {Deng}, {Dennefeld}, {Elias-Rosa},
  {Harutyunyan}, {Keenan}, {Iijima}, {Lorenzi}, {Mazzali}, {Meng}, {Nakano},
  {Nielsen}, {Smoker}, {Stanishev}, {Turatto}, {Xu}, \&
  {Zampieri}}]{Pastorello07}
{Pastorello}, A., {Smartt}, S.~J., {Mattila}, S., {et~al.} 2007, \nat, 447,
  829, \dodoi{10.1038/nature05825}

\bibitem[{{Pastorello} {et~al.}(2008){Pastorello}, {Mattila}, {Zampieri},
  {Della Valle}, {Smartt}, {Valenti}, {Agnoletto}, {Benetti}, {Benn}, {Branch},
  {Cappellaro}, {Dennefeld}, {Eldridge}, {Gal-Yam}, {Harutyunyan}, {Hunter},
  {Kjeldsen}, {Lipkin}, {Mazzali}, {Milne}, {Navasardyan}, {Ofek}, {Pian},
  {Shemmer}, {Spiro}, {Stathakis}, {Taubenberger}, {Turatto}, \&
  {Yamaoka}}]{Pastorello08}
{Pastorello}, A., {Mattila}, S., {Zampieri}, L., {et~al.} 2008, \mnras, 389,
  113, \dodoi{10.1111/j.1365-2966.2008.13602.x}

\bibitem[{{Pastorello} {et~al.}(2013){Pastorello}, {Cappellaro}, {Inserra},
  {Smartt}, {Pignata}, {Benetti}, {Valenti}, {Fraser}, {Tak{\'a}ts}, {Benitez},
  {Botticella}, {Brimacombe}, {Bufano}, {Cellier-Holzem}, {Costado}, {Cupani},
  {Curtis}, {Elias-Rosa}, {Ergon}, {Fynbo}, {Hambsch}, {Hamuy}, {Harutyunyan},
  {Ivarson}, {Kankare}, {Martin}, {Kotak}, {LaCluyze}, {Maguire}, {Mattila},
  {Maza}, {McCrum}, {Miluzio}, {Norgaard-Nielsen}, {Nysewander}, {Ochner},
  {Pan}, {Pumo}, {Reichart}, {Tan}, {Taubenberger}, {Tomasella}, {Turatto}, \&
  {Wright}}]{Pastorello13}
{Pastorello}, A., {Cappellaro}, E., {Inserra}, C., {et~al.} 2013, \apj, 767, 1,
  \dodoi{10.1088/0004-637X/767/1/1}

\bibitem[{{Pastorello} {et~al.}(2015{\natexlab{a}}){Pastorello}, {Benetti},
  {Brown}, {Tsvetkov}, {Inserra}, {Taubenberger}, {Tomasella}, {Fraser},
  {Rich}, {Botticella}, {Bufano}, {Cappellaro}, {Ergon}, {Gorbovskoy},
  {Harutyunyan}, {Huang}, {Kotak}, {Lipunov}, {Magill}, {Miluzio}, {Morrell},
  {Ochner}, {Smartt}, {Sollerman}, {Spiro}, {Stritzinger}, {Turatto},
  {Valenti}, {Wang}, {Wright}, {Yurkov}, {Zampieri}, \&
  {Zhang}}]{Pastorello15a}
{Pastorello}, A., {Benetti}, S., {Brown}, P.~J., {et~al.} 2015{\natexlab{a}},
  \mnras, 449, 1921, \dodoi{10.1093/mnras/stu2745}

\bibitem[{{Pastorello} {et~al.}(2015{\natexlab{b}}){Pastorello}, {Wyrzykowski},
  {Valenti}, {Prieto}, {Koz{\l}owski}, {Udalski}, {Elias-Rosa},
  {Morales-Garoffolo}, {Anderson}, {Benetti}, {Bersten}, {Botticella},
  {Cappellaro}, {Fasano}, {Fraser}, {Gal-Yam}, {Gillone}, {Graham}, {Greiner},
  {Hachinger}, {Howell}, {Inserra}, {Parrent}, {Rau}, {Schulze}, {Smartt},
  {Smith}, {Turatto}, {Yaron}, {Young}, {Kubiak}, {Szyma{\'n}ski},
  {Pietrzy{\'n}ski}, {Soszy{\'n}ski}, {Ulaczyk}, {Poleski}, {Pietrukowicz},
  {Skowron}, \& {Mr{\'o}z}}]{Pastorello15b}
{Pastorello}, A., {Wyrzykowski}, {\L}., {Valenti}, S., {et~al.}
  2015{\natexlab{b}}, \mnras, 449, 1941, \dodoi{10.1093/mnras/stu2621}

\bibitem[{{Pastorello} {et~al.}(2016){Pastorello}, {Wang}, {Ciabattari},
  {Bersier}, {Mazzali}, {Gao}, {Xu}, {Zhang}, {Tokuoka}, {Benetti},
  {Cappellaro}, {Elias-Rosa}, {Harutyunyan}, {Huang}, {Miluzio}, {Mo},
  {Ochner}, {Tartaglia}, {Terreran}, {Tomasella}, \& {Turatto}}]{Pastorello16}
{Pastorello}, A., {Wang}, X.~F., {Ciabattari}, F., {et~al.} 2016, \mnras, 456,
  853, \dodoi{10.1093/mnras/stv2634}

\bibitem[{{Pastorello} {et~al.}(2018){Pastorello}, {Kochanek}, {Fraser},
  {Dong}, {Elias-Rosa}, {Filippenko}, {Benetti}, {Cappellaro}, {Tomasella},
  {Drake}, {Harmanen}, {Reynolds}, {Shappee}, {Smartt}, {Chambers}, {Huber},
  {Smith}, {Stanek}, {Christensen}, {Denneau}, {Djorgovski}, {Flewelling},
  {Gall}, {Gal-Yam}, {Geier}, {Heinze}, {Holoien}, {Isern}, {Kangas},
  {Kankare}, {Koff}, {Llapasset}, {Lowe}, {Lundqvist}, {Magnier}, {Mattila},
  {Morales-Garoffolo}, {Mutel}, {Nicolas}, {Ochner}, {Ofek}, {Prosperi},
  {Rest}, {Sano}, {Stalder}, {Stritzinger}, {Taddia}, {Terreran}, {Tonry},
  {Wainscoat}, {Waters}, {Weiland}, {Willman}, {Young}, \&
  {Zheng}}]{Pastorello18}
{Pastorello}, A., {Kochanek}, C.~S., {Fraser}, M., {et~al.} 2018, \mnras, 474,
  197, \dodoi{10.1093/mnras/stx2668}

\bibitem[{{Pellegrino} {et~al.}(2022){Pellegrino}, {Howell}, {Terreran},
  {Arcavi}, {Bostroem}, {Brown}, {Burke}, {Dong}, {Gilkis}, {Hiramatsu},
  {Hosseinzadeh}, {McCully}, {Modjaz}, {Newsome}, {Padilla Gonzalez},
  {Pritchard}, {Sand}, {Valenti}, \& {Williamson}}]{Pellegrino22}
{Pellegrino}, C., {Howell}, D.~A., {Terreran}, G., {et~al.} 2022, arXiv
  e-prints, arXiv:2205.07894.
\newblock \doarXiv{2205.07894}

\bibitem[{{Perley}(2021)}]{Class:2021csp}
{Perley}, D. 2021, Transient Name Server Classification Report, 2021-570, 1

\bibitem[{{Perley} {et~al.}(2020){Perley}, {Taggart}, {Dahiwale}, \&
  {Fremling}}]{Class:2020faa}
{Perley}, D.~A., {Taggart}, K., {Dahiwale}, A., \& {Fremling}, C. 2020,
  Transient Name Server Classification Report, 2020-987, 1

\bibitem[{{Perley} {et~al.}(2022){Perley}, {Sollerman}, {Schulze}, {Yao},
  {Fremling}, {Gal-Yam}, {Ho}, {Yang}, {Kool}, {Irani}, {Yan}, {Andreoni},
  {Baade}, {Bellm}, {Brink}, {Chen}, {Cikota}, {Coughlin}, {Dahiwale},
  {Dekany}, {Duev}, {Filippenko}, {Hoeflich}, {Kasliwal}, {Kulkarni}, {Lunnan},
  {Masci}, {Maund}, {Medford}, {Riddle}, {Rosnet}, {Shupe}, {Strotjohann},
  {Tzanidakis}, \& {Zheng}}]{Perley22}
{Perley}, D.~A., {Sollerman}, J., {Schulze}, S., {et~al.} 2022, \apj, 927, 180,
  \dodoi{10.3847/1538-4357/ac478e}

\bibitem[{{Pessi} {et~al.}(2022){Pessi}, {Prieto}, {Monard}, {Kochanek},
  {Bock}, {Drake}, {Fox}, {Parker}, \& {Stevance}}]{Pessi21}
{Pessi}, T., {Prieto}, J.~L., {Monard}, B., {et~al.} 2022, \apj, 928, 138,
  \dodoi{10.3847/1538-4357/ac562d}

\bibitem[{{Podsiadlowski} {et~al.}(1992){Podsiadlowski}, {Joss}, \&
  {Hsu}}]{Podsiadlowski92}
{Podsiadlowski}, P., {Joss}, P.~C., \& {Hsu}, J.~J.~L. 1992, \apj, 391, 246,
  \dodoi{10.1086/171341}

\bibitem[{{Pollas} {et~al.}(1995){Pollas}, {Albanese}, {Benetti}, {Bouchet}, \&
  {Schwarz}}]{Class:1995N}
{Pollas}, C., {Albanese}, D., {Benetti}, S., {Bouchet}, P., \& {Schwarz}, H.
  1995, \iaucirc, 6170, 1

\bibitem[{Prentice {et~al.}(2020)Prentice, Maguire, Boian, Groh, Anderson,
  Barbarino, Bostroem, Burke, Clark, Dong, Fraser, Galbany, Gromadzki,
  Gutiérrez, Howell, Hiramatsu, Inserra, James, Kankare, Kuncarayakti,
  Mazzali, McCully, Müller-Bravo, Nichol, Pellegrino, Smartt, Sollerman,
  Tartaglia, Valenti, \& Young}]{Prentice20}
Prentice, S.~J., Maguire, K., Boian, I., {et~al.} 2020, Monthly Notices of the
  Royal Astronomical Society, \dodoi{10.1093/mnras/staa2947}

\bibitem[{{Prieto} {et~al.}(2011){Prieto}, {McMillan}, {Bakos}, \&
  {Grennan}}]{Class:2011ht}
{Prieto}, J.~L., {McMillan}, R., {Bakos}, G., \& {Grennan}, D. 2011, Central
  Bureau Electronic Telegrams, 2903, 1

\bibitem[{{Prieto} {et~al.}(2017){Prieto}, {Chen}, {Dong}, {Shappee},
  {Seibert}, {Bersier}, {Holoien}, {Kochanek}, {Stanek}, \&
  {Thompson}}]{Class:2017hcc}
{Prieto}, J.~L., {Chen}, P., {Dong}, S., {et~al.} 2017, Research Notes of the
  American Astronomical Society, 1, 28, \dodoi{10.3847/2515-5172/aa9c46}

\bibitem[{{Quataert} \& {Shiode}(2012)}]{Quataert12}
{Quataert}, E., \& {Shiode}, J. 2012, \mnras, 423, L92,
  \dodoi{10.1111/j.1745-3933.2012.01264.x}

\bibitem[{{Quimby} {et~al.}(2007){Quimby}, {Castro}, {Mondol}, {Caldwell}, \&
  {Terrazas}}]{Class:2006tf}
{Quimby}, R., {Castro}, F., {Mondol}, P., {Caldwell}, J., \& {Terrazas}, E.
  2007, Central Bureau Electronic Telegrams, 793, 1

\bibitem[{{Quimby}(2012)}]{Quimby12}
{Quimby}, R.~M. 2012, in Death of Massive Stars: Supernovae and Gamma-Ray
  Bursts, ed. P.~{Roming}, N.~{Kawai}, \& E.~{Pian}, Vol. 279, 22--28,
  \dodoi{10.1017/S174392131201263X}

\bibitem[{{Rodney} {et~al.}(2009){Rodney}, {Trundle}, {Valenti}, \&
  {Pastorello}}]{Class:2009kf}
{Rodney}, S., {Trundle}, C., {Valenti}, S., \& {Pastorello}, A. 2009, Central
  Bureau Electronic Telegrams, 1988, 3

\bibitem[{{Roming} {et~al.}(2012){Roming}, {Pritchard}, {Prieto}, {Kochanek},
  {Fryer}, {Davidson}, {Humphreys}, {Bayless}, {Beacom}, {Brown}, {Holland},
  {Immler}, {Kuin}, {Oates}, {Pogge}, {Pojmanski}, {Stoll}, {Shappee},
  {Stanek}, \& {Szczygiel}}]{Roming12}
{Roming}, P.~W.~A., {Pritchard}, T.~A., {Prieto}, J.~L., {et~al.} 2012, \apj,
  751, 92, \dodoi{10.1088/0004-637X/751/2/92}

\bibitem[{{Roy} {et~al.}(2016){Roy}, {Sollerman}, {Silverman}, {Pastorello},
  {Fransson}, {Drake}, {Taddia}, {Fremling}, {Kankare}, {Kumar}, {Cappellaro},
  {Bose}, {Benetti}, {Filippenko}, {Valenti}, {Nyholm}, {Ergon}, {Sutaria},
  {Kumar}, {Pandey}, {Nicholl}, {Garcia-{\'A}lvarez}, {Tomasella},
  {Karamehmetoglu}, \& {Migotto}}]{Roy16}
{Roy}, R., {Sollerman}, J., {Silverman}, J.~M., {et~al.} 2016, \aap, 596, A67,
  \dodoi{10.1051/0004-6361/201527947}

\bibitem[{{Salas} {et~al.}(2013){Salas}, {Bauer}, {Stockdale}, \&
  {Prieto}}]{Salas13}
{Salas}, P., {Bauer}, F.~E., {Stockdale}, C., \& {Prieto}, J.~L. 2013, \mnras,
  428, 1207, \dodoi{10.1093/mnras/sts104}

\bibitem[{{Sana} {et~al.}(2012){Sana}, {de Mink}, {de Koter}, {Langer},
  {Evans}, {Gieles}, {Gosset}, {Izzard}, {Le Bouquin}, \& {Schneider}}]{Sana12}
{Sana}, H., {de Mink}, S.~E., {de Koter}, A., {et~al.} 2012, Science, 337, 444,
  \dodoi{10.1126/science.1223344}

\bibitem[{{Schinzel} {et~al.}(2009){Schinzel}, {Taylor}, {Stockdale}, {Granot},
  \& {Ramirez-Ruiz}}]{Schinzel09}
{Schinzel}, F.~K., {Taylor}, G.~B., {Stockdale}, C.~J., {Granot}, J., \&
  {Ramirez-Ruiz}, E. 2009, \apj, 691, 1380,
  \dodoi{10.1088/0004-637X/691/2/1380}

\bibitem[{{Schlegel}(1990)}]{Schlegel90}
{Schlegel}, E.~M. 1990, \mnras, 244, 269

\bibitem[{{Shiode}(2013)}]{Shiode13b}
{Shiode}, J.~H. 2013, PhD thesis, University of California, Berkeley

\bibitem[{{Shiode} \& {Quataert}(2014)}]{Shiode14}
{Shiode}, J.~H., \& {Quataert}, E. 2014, \apj, 780, 96,
  \dodoi{10.1088/0004-637X/780/1/96}

\bibitem[{{Shiode} {et~al.}(2013){Shiode}, {Quataert}, {Cantiello}, \&
  {Bildsten}}]{Shiode13}
{Shiode}, J.~H., {Quataert}, E., {Cantiello}, M., \& {Bildsten}, L. 2013,
  \mnras, 430, 1736, \dodoi{10.1093/mnras/sts719}

\bibitem[{{Shivvers} {et~al.}(2019){Shivvers}, {Filippenko}, {Silverman},
  {Zheng}, {Foley}, {Chornock}, {Barth}, {Cenko}, {Clubb}, {Fox},
  {Ganeshalingam}, {Graham}, {Kelly}, {Kleiser}, {Leonard}, {Li}, {Matheson},
  {Mauerhan}, {Modjaz}, {Serduke}, {Shields}, {Steele}, {Swift}, {Wong}, \&
  {Yuk}}]{Shivver19}
{Shivvers}, I., {Filippenko}, A.~V., {Silverman}, J.~M., {et~al.} 2019, \mnras,
  482, 1545, \dodoi{10.1093/mnras/sty2719}

\bibitem[{{Siebert} {et~al.}(2020){Siebert}, {Kilpatrick}, {Foley}, \&
  {Cartier}}]{Class:2020oi}
{Siebert}, M.~R., {Kilpatrick}, C.~D., {Foley}, R.~J., \& {Cartier}, R. 2020,
  The Astronomer's Telegram, 13393, 1

\bibitem[{{Smartt}(2015)}]{Smartt15}
{Smartt}, S.~J. 2015, \pasa, 32, e016, \dodoi{10.1017/pasa.2015.17}

\bibitem[{{Smartt, S. J.} {et~al.}(2015){Smartt, S. J.}, {Valenti, S.},
  {Fraser, M.}, {Inserra, C.}, {Young, D. R.}, {Sullivan, M.}, {Pastorello,
  A.}, {Benetti, S.}, {Gal-Yam, A.}, {Knapic, C.}, {Molinaro, M.}, {Smareglia,
  R.}, {Smith, K. W.}, {Taubenberger, S.}, {Yaron, O.}, {Anderson, J. P.},
  {Ashall, C.}, {Balland, C.}, {Baltay, C.}, {Barbarino, C.}, {Bauer, F. E.},
  {Baumont, S.}, {Bersier, D.}, {Blagorodnova, N.}, {Bongard, S.}, {Botticella,
  M. T.}, {Bufano, F.}, {Bulla, M.}, {Cappellaro, E.}, {Campbell, H.},
  {Cellier-Holzem, F.}, {Chen, T.-W.}, {Childress, M. J.}, {Clocchiatti, A.},
  {Contreras, C.}, {Dall\'{}Ora, M.}, {Danziger, J.}, {de Jaeger, T.}, {De Cia,
  A.}, {Della Valle, M.}, {Dennefeld, M.}, {Elias-Rosa, N.}, {Elman, N.},
  {Feindt, U.}, {Fleury, M.}, {Gall, E.}, {Gonzalez-Gaitan, S.}, {Galbany, L.},
  {Morales Garoffolo, A.}, {Greggio, L.}, {Guillou, L. L.}, {Hachinger, S.},
  {Hadjiyska, E.}, {Hage, P. E.}, {Hillebrandt, W.}, {Hodgkin, S.}, {Hsiao, E.
  Y.}, {James, P. A.}, {Jerkstrand, A.}, {Kangas, T.}, {Kankare, E.}, {Kotak,
  R.}, {Kromer, M.}, {Kuncarayakti, H.}, {Leloudas, G.}, {Lundqvist, P.},
  {Lyman, J. D.}, {Hook, I. M.}, {Maguire, K.}, {Manulis, I.}, {Margheim, S.
  J.}, {Mattila, S.}, {Maund, J. R.}, {Mazzali, P. A.}, {McCrum, M.},
  {McKinnon, R.}, {Moreno-Raya, M. E.}, {Nicholl, M.}, {Nugent, P.}, {Pain,
  R.}, {Pignata, G.}, {Phillips, M. M.}, {Polshaw, J.}, {Pumo, M. L.},
  {Rabinowitz, D.}, {Reilly, E.}, {Romero-Ca\~nizales, C.}, {Scalzo, R.},
  {Schmidt, B.}, {Schulze, S.}, {Sim, S.}, {Sollerman, J.}, {Taddia, F.},
  {Tartaglia, L.}, {Terreran, G.}, {Tomasella, L.}, {Turatto, M.}, {Walker,
  E.}, {Walton, N. A.}, {Wyrzykowski, L.}, {Yuan, F.}, \& {Zampieri,
  L.}}]{Smartt14}
{Smartt, S. J.}, {Valenti, S.}, {Fraser, M.}, {et~al.} 2015, A\&A, 579, A40,
  \dodoi{10.1051/0004-6361/201425237}

\bibitem[{{Smith}(2014)}]{Smith14}
{Smith}, N. 2014, \araa, 52, 487, \dodoi{10.1146/annurev-astro-081913-040025}

\bibitem[{{Smith} \& {Andrews}(2020)}]{Smith20}
{Smith}, N., \& {Andrews}, J.~E. 2020, \mnras, 499, 3544,
  \dodoi{10.1093/mnras/staa3047}

\bibitem[{{Smith} {et~al.}(2016){Smith}, {Andrews}, \& {Mauerhan}}]{Smith16}
{Smith}, N., {Andrews}, J.~E., \& {Mauerhan}, J.~C. 2016, \mnras, 463, 2904,
  \dodoi{10.1093/mnras/stw2190}

\bibitem[{{Smith} \& {Arnett}(2014)}]{Smith14b}
{Smith}, N., \& {Arnett}, W.~D. 2014, \apj, 785, 82,
  \dodoi{10.1088/0004-637X/785/2/82}

\bibitem[{{Smith} {et~al.}(2008{\natexlab{a}}){Smith}, {Chornock}, {Li},
  {Ganeshalingam}, {Silverman}, {Foley}, {Filippenko}, \& {Barth}}]{Smith08b}
{Smith}, N., {Chornock}, R., {Li}, W., {et~al.} 2008{\natexlab{a}}, \apj, 686,
  467, \dodoi{10.1086/591021}

\bibitem[{{Smith} \& {Owocki}(2006)}]{Smith06}
{Smith}, N., \& {Owocki}, S.~P. 2006, \apjl, 645, L45, \dodoi{10.1086/506523}

\bibitem[{{Smith} {et~al.}(2008{\natexlab{b}}){Smith}, {Foley}, {Bloom}, {Li},
  {Filippenko}, {Gavazzi}, {Ghez}, {Konopacky}, {Malkan}, {Marshall}, {Pooley},
  {Treu}, \& {Woo}}]{Smith08}
{Smith}, N., {Foley}, R.~J., {Bloom}, J.~S., {et~al.} 2008{\natexlab{b}}, \apj,
  686, 485, \dodoi{10.1086/590141}

\bibitem[{{Smith} {et~al.}(2015){Smith}, {Mauerhan}, {Cenko}, {Kasliwal},
  {Silverman}, {Filippenko}, {Gal-Yam}, {Clubb}, {Graham}, {Leonard}, {Horst},
  {Williams}, {Andrews}, {Kulkarni}, {Nugent}, {Sullivan}, {Maguire}, {Xu}, \&
  {Ben-Ami}}]{Smith15}
{Smith}, N., {Mauerhan}, J.~C., {Cenko}, S.~B., {et~al.} 2015, \mnras, 449,
  1876, \dodoi{10.1093/mnras/stv354}

\bibitem[{{Soderberg} {et~al.}(2006){Soderberg}, {Chevalier}, {Kulkarni}, \&
  {Frail}}]{Soderberg062003bg}
{Soderberg}, A.~M., {Chevalier}, R.~A., {Kulkarni}, S.~R., \& {Frail}, D.~A.
  2006, \apj, 651, 1005, \dodoi{10.1086/507571}

\bibitem[{{Soderberg} {et~al.}(2004){Soderberg}, {Gal-Yam}, \&
  {Kulkarni}}]{Class:2001em2}
{Soderberg}, A.~M., {Gal-Yam}, A., \& {Kulkarni}, S.~R. 2004, GRB Coordinates
  Network, 2586, 1

\bibitem[{{Soderberg} {et~al.}(2012){Soderberg}, {Margutti}, {Zauderer},
  {Krauss}, {Katz}, {Chomiuk}, {Dittmann}, {Nakar}, {Sakamoto}, {Kawai},
  {Hurley}, {Barthelmy}, {Toizumi}, {Morii}, {Chevalier}, {Gurwell},
  {Petitpas}, {Rupen}, {Alexander}, {Levesque}, {Fransson}, {Brunthaler},
  {Bietenholz}, {Chugai}, {Grindlay}, {Copete}, {Connaughton}, {Briggs},
  {Meegan}, {von Kienlin}, {Zhang}, {Rau}, {Golenetskii}, {Mazets}, \&
  {Cline}}]{Soderberg12}
{Soderberg}, A.~M., {Margutti}, R., {Zauderer}, B.~A., {et~al.} 2012, \apj,
  752, 78, \dodoi{10.1088/0004-637X/752/2/78}

\bibitem[{{Sollerman} {et~al.}(2003){Sollerman}, {Andersson}, {Gustafsson},
  {Jakobsson}, {Oye}, \& {Patat}}]{Class:2003gk}
{Sollerman}, J., {Andersson}, J., {Gustafsson}, M., {et~al.} 2003, \iaucirc,
  8164, 3

\bibitem[{{Sollerman} {et~al.}(2020){Sollerman}, {Fransson}, {Barbarino},
  {Fremling}, {Horesh}, {Kool}, {Schulze}, {Sfaradi}, {Yang}, {Bellm},
  {Burruss}, {Cunningham}, {De}, {Drake}, {Golkhou}, {Green}, {Kasliwal},
  {Kulkarni}, {Kupfer}, {Laher}, {Masci}, {Rodriguez}, {Rusholme}, {Williams},
  {Yan}, \& {Zolkower}}]{Sollerman20}
{Sollerman}, J., {Fransson}, C., {Barbarino}, C., {et~al.} 2020, \aap, 643,
  A79, \dodoi{10.1051/0004-6361/202038960}

\bibitem[{{Stroh} {et~al.}(2021){Stroh}, {Terreran}, {Coppejans}, {Bright},
  {Margutti}, {Bietenholz}, {De Colle}, {DeMarchi}, {Duran}, {Milisavljevic},
  {Murase}, {Paterson}, \& {Williams}}]{Stroh21}
{Stroh}, M.~C., {Terreran}, G., {Coppejans}, D.~L., {et~al.} 2021, \apjl, 923,
  L24, \dodoi{10.3847/2041-8213/ac375e}

\bibitem[{{Strotjohann} {et~al.}(2021){Strotjohann}, {Ofek}, {Gal-Yam},
  {Bruch}, {Schulze}, {Shaviv}, {Sollerman}, {Filippenko}, {Yaron}, {Fremling},
  {Nordin}, {Kool}, {Perley}, {Ho}, {Yang}, {Yao}, {Soumagnac}, {Graham},
  {Barbarino}, {Tartaglia}, {De}, {Goldstein}, {Cook}, {Brink}, {Taggart},
  {Yan}, {Lunnan}, {Kasliwal}, {Kulkarni}, {Nugent}, {Masci}, {Rosnet},
  {Adams}, {Andreoni}, {Bagdasaryan}, {Bellm}, {Burdge}, {Duev}, {Dugas},
  {Frederick}, {Goldwasser}, {Hankins}, {Irani}, {Karambelkar}, {Kupfer},
  {Liang}, {Neill}, {Porter}, {Riddle}, {Sharma}, {Short}, {Taddia},
  {Tzanidakis}, {van Roestel}, {Walters}, \& {Zhuang}}]{Strotjohann21}
{Strotjohann}, N.~L., {Ofek}, E.~O., {Gal-Yam}, A., {et~al.} 2021, \apj, 907,
  99, \dodoi{10.3847/1538-4357/abd032}

\bibitem[{Sun {et~al.}(2020)Sun, Maund, \& Crowther}]{Sun20}
Sun, N.-C., Maund, J.~R., \& Crowther, P.~A. 2020, Monthly Notices of the Royal
  Astronomical Society, \dodoi{10.1093/mnras/staa2277}

\bibitem[{{Suzuki} {et~al.}(2021){Suzuki}, {Nicholl}, {Moriya}, \&
  {Takiwaki}}]{Suzuki21}
{Suzuki}, A., {Nicholl}, M., {Moriya}, T.~J., \& {Takiwaki}, T. 2021, \apj,
  908, 99, \dodoi{10.3847/1538-4357/abd6ce}

\bibitem[{{Taddia} {et~al.}(2013){Taddia}, {Stritzinger}, {Sollerman},
  {Phillips}, {Anderson}, {Boldt}, {Campillay}, {Castell{\'o}n}, {Contreras},
  {Folatelli}, {Hamuy}, {Heinrich-Josties}, {Krzeminski}, {Morrell}, {Burns},
  {Freedman}, {Madore}, {Persson}, \& {Suntzeff}}]{Taddia13}
{Taddia}, F., {Stritzinger}, M.~D., {Sollerman}, J., {et~al.} 2013, \aap, 555,
  A10, \dodoi{10.1051/0004-6361/201321180}

\bibitem[{{Taddia} {et~al.}(2020){Taddia}, {Stritzinger}, {Fransson}, {Brown},
  {Contreras}, {Holmbo}, {Moriya}, {Phillips}, {Sollerman}, {Suntzeff},
  {Ashall}, {Burns}, {Busta}, {Campillay}, {Castell{\'o}n}, {Corco}, {Di
  Mille}, {Gall}, {Gonz{\'a}lez}, {Hsiao}, {Morrell}, {Nyholm}, {Simon}, \&
  {Ser{\'o}n}}]{Taddia20}
{Taddia}, F., {Stritzinger}, M.~D., {Fransson}, C., {et~al.} 2020, \aap, 638,
  A92, \dodoi{10.1051/0004-6361/201936654}

\bibitem[{{Tak{\'a}ts} {et~al.}(2014){Tak{\'a}ts}, {Pumo}, {Elias-Rosa},
  {Pastorello}, {Pignata}, {Paillas}, {Zampieri}, {Anderson}, {Vink{\'o}},
  {Benetti}, {Botticella}, {Bufano}, {Campillay}, {Cartier}, {Ergon},
  {Folatelli}, {Foley}, {F{\"o}rster}, {Hamuy}, {Hentunen}, {Kankare},
  {Leloudas}, {Morrell}, {Nissinen}, {Phillips}, {Smartt}, {Stritzinger},
  {Taubenberger}, {Valenti}, {Van Dyk}, {Haislip}, {LaCluyze}, {Moore}, \&
  {Reichart}}]{Takats14}
{Tak{\'a}ts}, K., {Pumo}, M.~L., {Elias-Rosa}, N., {et~al.} 2014, \mnras, 438,
  368, \dodoi{10.1093/mnras/stt2203}

\bibitem[{{Tartaglia}(2019)}]{Class:2015da}
{Tartaglia}, L. 2019, Transient Name Server Discovery Report, 2019-1671, 1

\bibitem[{{Tartaglia} {et~al.}(2020){Tartaglia}, {Pastorello}, {Sollerman},
  {Fransson}, {Mattila}, {Fraser}, {Taddia}, {Tomasella}, {Turatto},
  {Morales-Garoffolo}, {Elias-Rosa}, {Lundqvist}, {Harmanen}, {Reynolds},
  {Cappellaro}, {Barbarino}, {Nyholm}, {Kool}, {Ofek}, {Gao}, {Jin}, {Tan},
  {Sand}, {Ciabattari}, {Wang}, {Zhang}, {Huang}, {Li}, {Mo}, {Rui}, {Xiang},
  {Zhang}, {Hosseinzadeh}, {Howell}, {McCully}, {Valenti}, {Benetti}, {Callis},
  {Carracedo}, {Fremling}, {Kangas}, {Rubin}, {Somero}, \&
  {Terreran}}]{Tartaglia20a}
{Tartaglia}, L., {Pastorello}, A., {Sollerman}, J., {et~al.} 2020, \aap, 635,
  A39, \dodoi{10.1051/0004-6361/201936553}

\bibitem[{{Tartaglia} {et~al.}(2021{\natexlab{a}}){Tartaglia}, {Sand}, {Groh},
  {Valenti}, {Wyatt}, {Bostroem}, {Brown}, {Yang}, {Burke}, {Chen}, {Davis},
  {F{\"o}rster}, {Galbany}, {Haislip}, {Hiramatsu}, {Hosseinzadeh}, {Howell},
  {Hsiao}, {Jha}, {Kouprianov}, {Kuncarayakti}, {Lyman}, {McCully}, {Phillips},
  {Rau}, {Reichart}, {Shahbandeh}, \& {Strader}}]{Tartaglia21}
{Tartaglia}, L., {Sand}, D.~J., {Groh}, J.~H., {et~al.} 2021{\natexlab{a}},
  \apj, 907, 52, \dodoi{10.3847/1538-4357/abca8a}

\bibitem[{{Tartaglia} {et~al.}(2021{\natexlab{b}}){Tartaglia}, {Sollerman},
  {Barbarino}, {Taddia}, {Mason}, {Berton}, {Taggart}, {Bellm}, {De},
  {Frederick}, {Fremling}, {Gal-Yam}, {Golkhou}, {Graham}, {Ho}, {Hung},
  {Kaye}, {Kim}, {Laher}, {Masci}, {Perley}, {Porter}, {Reiley}, {Riddle},
  {Rusholme}, {Soumagnac}, \& {Walters}}]{Tartaglia20b}
{Tartaglia}, L., {Sollerman}, J., {Barbarino}, C., {et~al.} 2021{\natexlab{b}},
  \aap, 650, A174, \dodoi{10.1051/0004-6361/202039068}

\bibitem[{{Terreran} {et~al.}(2022){Terreran}, {Jacobson-Gal{\'a}n}, {Groh},
  {Margutti}, {Coppejans}, {Dimitriadis}, {Kilpatrick}, {Matthews}, {Siebert},
  {Angus}, {Brink}, {Filippenko}, {Foley}, {Jones}, {Tinyanont}, {Gall},
  {Pfister}, {Zenati}, {Ansari}, {Auchettl}, {El-Badry}, {Magnier}, \&
  {Zheng}}]{Terreran22}
{Terreran}, G., {Jacobson-Gal{\'a}n}, W.~V., {Groh}, J.~H., {et~al.} 2022,
  \apj, 926, 20, \dodoi{10.3847/1538-4357/ac3820}

\bibitem[{{Thomas} {et~al.}(2022){Thomas}, {Wheeler}, {Dwarkadas}, {Stockdale},
  {Vinko}, {Pooley}, {Xu}, {Zeimann}, \& {MacQueen}}]{Thomas22}
{Thomas}, B.~P., {Wheeler}, J.~C., {Dwarkadas}, V.~V., {et~al.} 2022, arXiv
  e-prints, arXiv:2203.12747.
\newblock \doarXiv{2203.12747}

\bibitem[{{Tinyanont} {et~al.}(2016){Tinyanont}, {Kasliwal}, {Fox}, {Lau},
  {Smith}, {Williams}, {Jencson}, {Perley}, {Dykhoff}, {Gehrz}, {Johansson},
  {Masci}, {Cody}, \& {Prince}}]{Tinyanont16}
{Tinyanont}, S., {Kasliwal}, M.~M., {Fox}, O.~D., {et~al.} 2016, ArXiv
  e-prints.
\newblock \doarXiv{1601.03440}

\bibitem[{{Tinyanont} {et~al.}(2019){Tinyanont}, {Lau}, {Kasliwal}, {Maeda},
  {Smith}, {Fox}, {Gehrz}, {De}, {Jencson}, {Bally}, \& {Masci}}]{Tinyanont19}
{Tinyanont}, S., {Lau}, R.~M., {Kasliwal}, M.~M., {et~al.} 2019, \apj, 887, 75,
  \dodoi{10.3847/1538-4357/ab521b}

\bibitem[{{Valenti} {et~al.}(2011){Valenti}, {Pastorello}, {Benetti},
  {Tomasella}, {Bufano}, \& {Ochner}}]{Class:2011hw}
{Valenti}, S., {Pastorello}, A., {Benetti}, S., {et~al.} 2011, Central Bureau
  Electronic Telegrams, 2906, 2

\bibitem[{{Vallely} {et~al.}(2018){Vallely}, {Prieto}, {Stanek}, {Kochanek},
  {Sukhbold}, {Bersier}, {Brown}, {Chen}, {Dong}, {Falco}, {Berlind},
  {Calkins}, {Koff}, {Kiyota}, {Brimacombe}, {Shappee}, {Holoien}, {Thompson},
  \& {Stritzinger}}]{Vallely18}
{Vallely}, P.~J., {Prieto}, J.~L., {Stanek}, K.~Z., {et~al.} 2018, \mnras, 475,
  2344, \dodoi{10.1093/mnras/stx3303}

\bibitem[{{van Loon} {et~al.}(2005){van Loon}, {Cioni}, {Zijlstra}, \&
  {Loup}}]{vanLoon05}
{van Loon}, J.~T., {Cioni}, M.-R.~L., {Zijlstra}, A.~A., \& {Loup}, C. 2005,
  \aap, 438, 273, \dodoi{10.1051/0004-6361:20042555}

\bibitem[{{Vargas} {et~al.}(2021){Vargas}, {De Colle}, {Brethauer}, {Margutti},
  \& {Bernal}}]{Vargas21}
{Vargas}, F., {De Colle}, F., {Brethauer}, D., {Margutti}, R., \& {Bernal},
  C.~G. 2021, arXiv e-prints, arXiv:2102.12581.
\newblock \doarXiv{2102.12581}

\bibitem[{{Wegner} \& {Swanson}(1996)}]{Class:1987F}
{Wegner}, G., \& {Swanson}, S.~R. 1996, \mnras, 278, 22,
  \dodoi{10.1093/mnras/278.1.22}

\bibitem[{{Weil} {et~al.}(2020){Weil}, {Fesen}, {Patnaude}, \&
  {Milisavljevic}}]{Weil20}
{Weil}, K.~E., {Fesen}, R.~A., {Patnaude}, D.~J., \& {Milisavljevic}, D. 2020,
  \apj, 900, 11, \dodoi{10.3847/1538-4357/aba4b1}

\bibitem[{{Wellons} {et~al.}(2012){Wellons}, {Soderberg}, \&
  {Chevalier}}]{Wellons12}
{Wellons}, S., {Soderberg}, A.~M., \& {Chevalier}, R.~A. 2012, \apj, 752, 17,
  \dodoi{10.1088/0004-637X/752/1/17}

\bibitem[{{Woosley} \& {Heger}(2015)}]{Woosley15}
{Woosley}, S.~E., \& {Heger}, A. 2015, \apj, 810, 34,
  \dodoi{10.1088/0004-637X/810/1/34}

\bibitem[{{Wu} \& {Fuller}(2021)}]{Wu20}
{Wu}, S., \& {Fuller}, J. 2021, \apj, 906, 3, \dodoi{10.3847/1538-4357/abc87c}

\bibitem[{{Wu} \& {Fuller}(2022)}]{Wu22}
---. 2022, arXiv e-prints, arXiv:2205.03319.
\newblock \doarXiv{2205.03319}

\bibitem[{{Xiang} {et~al.}(2017){Xiang}, {Rui}, {Wang}, {Fu}, {Xiao}, {Zhang},
  \& {Zhang}}]{Class:2017eaw}
{Xiang}, D., {Rui}, L., {Wang}, X., {et~al.} 2017, The Astronomer's Telegram,
  10376, 1

\bibitem[{{Yan} {et~al.}(2015){Yan}, {Quimby}, {Ofek}, {Gal-Yam}, {Mazzali},
  {Perley}, {Vreeswijk}, {Leloudas}, {de Cia}, {Masci}, {Cenko}, {Cao},
  {Kulkarni}, {Nugent}, {Rebbapragada}, {Wo{\'z}niak}, \& {Yaron}}]{Yan15}
{Yan}, L., {Quimby}, R., {Ofek}, E., {et~al.} 2015, ArXiv e-prints.
\newblock \doarXiv{1508.04420}

\bibitem[{{Yan} {et~al.}(2017){Yan}, {Lunnan}, {Perley}, {Gal-Yam}, {Yaron},
  {Roy}, {Quimby}, {Sollerman}, {Fremling}, {Leloudas}, {Cenko}, {Vreeswijk},
  {Graham}, {Howell}, {De Cia}, {Ofek}, {Nugent}, {Kulkarni}, {Hosseinzadeh},
  {Masci}, {McCully}, {Rebbapragada}, \& {Wo{\'z}niak}}]{Yan17}
{Yan}, L., {Lunnan}, R., {Perley}, D.~A., {et~al.} 2017, \apj, 848, 6,
  \dodoi{10.3847/1538-4357/aa8993}

\bibitem[{{Yang} {et~al.}(2021){Yang}, {Sollerman}, {Chen}, {Kool}, {Lunnan},
  {Schulze}, {Strotjohann}, {Horesh}, {Kasliwal}, {Kupfer}, {Mahabal}, {Masci},
  {Nugent}, {Perley}, {Riddle}, {Rusholme}, \& {Sharma}}]{Yang21}
{Yang}, S., {Sollerman}, J., {Chen}, T.~W., {et~al.} 2021, \aap, 646, A22,
  \dodoi{10.1051/0004-6361/202039440}

\bibitem[{{Yaron} {et~al.}(2017){Yaron}, {Perley}, {Gal-Yam}, {Groh}, {Horesh},
  {Ofek}, {Kulkarni}, {Sollerman}, {Fransson}, {Rubin}, {Szabo}, {Sapir},
  {Taddia}, {Cenko}, {Valenti}, {Arcavi}, {Howell}, {Kasliwal}, {Vreeswijk},
  {Khazov}, {Fox}, {Cao}, {Gnat}, {Kelly}, {Nugent}, {Filippenko}, {Laher},
  {Wozniak}, {Lee}, {Rebbapragada}, {Maguire}, {Sullivan}, \&
  {Soumagnac}}]{Yaron17}
{Yaron}, O., {Perley}, D.~A., {Gal-Yam}, A., {et~al.} 2017, Nature Physics, 13,
  510, \dodoi{10.1038/nphys4025}

\bibitem[{{Yuan} {et~al.}(2008){Yuan}, {Quimby}, {McKay}, {Chamarro}, {Sisson},
  {Akerlof}, {Wheeler}, {Odewahn}, {Rileyand}, {Rostopchin}, {Westfall}, \&
  {Cenko}}]{Class:2008es}
{Yuan}, F., {Quimby}, R., {McKay}, T., {et~al.} 2008, Central Bureau Electronic
  Telegrams, 1462, 1

\end{thebibliography}
\bibliographystyle{aasjournal}

\end{document}